\def\ltsima{$\; \buildrel < \over \sim \;$}
\def\simlt{\lower.5ex\hbox{\ltsima}}
\def\gtsima{$\; \buildrel > \over \sim \;$}
\def\simgt{\lower.5ex\hbox{\gtsima}}

\def\arcsec{\hbox{$^{\prime\prime}$}}
\def\arcmin{\hbox{$^\prime$}}

\def\spicapol{B-BOP}
\newcommand{\spica}{{\it SPICA}}
\newcommand\micron{\mbox{$\mu$m}}
\documentclass{pasa}%

\usepackage{graphicx}
\usepackage{subcaption}  
\usepackage{xcolor}  

\title[The cold magnetized Universe with {\spicapol}]
{Probing the cold magnetized Universe with SPICA-POL (B-BOP)}

\author[Andr\'e et al.]{Ph. Andr\'e$^1$, A. Hughes$^{2}$, V. Guillet$^{3,4}$, F. Boulanger$^{3,5}$, 
A. Bracco$^{1,5,6}$, E. Ntormousi$^{1,7}$, D.~Arzoumanian$^{1,8}$, A.J. Maury$^{1,9}$, 
J.-Ph. Bernard$^{2}$, S. Bontemps$^{10}$, I. Ristorcelli$^{2}$, J.M. Girart$^{11}$, F.~Motte$^{12,1}$, 
K. Tassis$^{7}$, E. Pantin$^1$, T. Montmerle$^{13}$, D. Johnstone$^{14}$, S. Gabici$^{15}$, A. Efstathiou$^{16}$,
S.~Basu$^{17}$, M. B\'ethermin$^{18}$, 
H. Beuther$^{19}$, J. Braine$^9$, J. Di Francesco$^{14}$, E. Falgarone$^{5}$,  K.~Ferri\`ere$^{2}$, A. Fletcher$^{20}$,
M. Galametz$^{1}$, 
M.~Giard$^{2}$, P. Hennebelle$^{1}$, A. Jones$^{3}$, A.A. Kepley$^{21}$,  
J.Kwon$^{22}$, G.~Lagache$^{18}$, P.  Lesaffre$^{5}$, F. Levrier$^{5}$, D.~Li$^{23}$, 
Z.-Y. Li$^{24}$,  S.A. Mao$^{25}$, T. Nakagawa$^{22}$, T.~Onaka$^{26}$,  R.~Paladino$^{27}$, N. Peretto$^{28}$
A. Poglitsch$^{1,29}$, V.~Rev\'eret$^1$, L. Rodriguez$^1$, 
M. Sauvage$^1$, J.D.Soler$^{1,19}$, L.~Spinoglio$^{30}$, F. Tabatabaei$^{31}$, A. Tritsis$^{32}$,  F.~van~der~Tak$^{33}$, 
D. Ward-Thompson$^{34}$, H.~Wiesemeyer$^{25}$, N. Ysard$^{3}$, H. Zhang$^{35}$
}

\jid{PASA}
\doi{10.1017/pas.\the\year.xxx}
\jyear{\the\year}

\usepackage{aas_macros}
\usepackage{hyperref} 
\hypersetup{colorlinks,citecolor=blue,linkcolor=blue,urlcolor=blue}

\hypersetup{draft}

\begin{document}

\begin{frontmatter}
\maketitle

\begin{abstract}
SPICA, the cryogenic infrared space telescope recently pre-selected for a ``Phase A'' concept study
as one of the three remaining candidates for ESA's fifth medium class (M5) mission, 
is foreseen to include a far-infrared polarimetric imager (SPICA-POL, now called {\spicapol}),  
which would offer a unique opportunity to resolve major issues in our understanding of the nearby, cold magnetized Universe.
This paper presents an overview of the main science drivers for {\spicapol}, including high dynamic range polarimetric imaging 
of the cold interstellar medium (ISM) in both our Milky Way and nearby galaxies.
Thanks to a cooled telescope, {\spicapol} will 
deliver wide-field 100--350$\, \mu$m images of linearly polarized dust emission in Stokes Q and U 
with a resolution, signal-to-noise ratio, and both intensity and spatial dynamic ranges comparable 
to those achieved by {\it Herschel} images of the cold ISM in total intensity 
(Stokes I).  
The {\spicapol} 200$\, \mu$m images will also have a factor $\sim \,$30 higher resolution than {\it Planck} polarization data. 
This will make {\spicapol} a unique tool for characterizing the statistical properties of the magnetized interstellar medium
and probing the role of magnetic fields in the formation and evolution 
of the interstellar web of dusty molecular filaments giving birth to most stars in our Galaxy. 
{\spicapol} will also be a powerful instrument for studying the magnetism of nearby galaxies and testing galactic dynamo models, 
constraining the physics of dust grain alignment, informing the problem of the interaction of cosmic rays with molecular clouds, 
tracing magnetic fields in the inner layers of protoplanetary disks, 
and monitoring accretion bursts in embedded protostars.
\end{abstract}

\vspace{-0.5cm}
\begin{keywords}
observations: submillimeter -- space missions -- interstellar medium: structure -- stars: formation -- magnetic fields
\end{keywords}
\end{frontmatter}

{\bf Preface}

\vspace{0.5cm}
\noindent
The following set of articles describe in detail the science goals of
the future Space Infrared telescope for Cosmology and Astrophysics
(\spica).  The \spica\ satellite will employ a 2.5-m telescope,
actively cooled to below 8\,K, and a suite of mid- to far-infrared
spectrometers and photometric cameras, equipped with state-of-the-art
detectors.  In particular, the \spica\ Far Infrared Instrument
(SAFARI) will be a grating spectrograph with low ($R$\,$=$\,300) and
medium ($R$\,$=$\,3000--11000) resolution observing modes
instantaneously covering the 35--230\,$\mu$m wavelength range.  The
\spica\ Mid-Infrared Instrument (SMI) will have three operating modes:
a large field of view (12\arcmin$\times$10\arcmin) low-resolution
17--36\,$\mu$m spectroscopic ($R$\,$=$\,50--120) and photometric
camera at 34\,$\mu$m, a medium resolution ($R$\,$=$\,2000) grating
spectrometer covering wavelengths of 18--36\,\micron\ and a
high-resolution echelle module ($R$\,$=$\,28000) for the
12--18\,\micron\ domain.  A large-field-of-view 
  (160\arcsec$\times$160\arcsec)\footnote{Some other
  \spica\ papers refer to this field of view as
  80\arcsec$\times$80\arcsec, but it is
  160\arcsec$\times$160\arcsec\ according to the latest design.},
three-channel (100, 200, and 350\,\micron) polarimetric camera (B-BOP~\footnote{B-BOP stands for ``B-fields with BOlometers and Polarizers''.}) 
will also be part of the instrument complement.  These articles will
focus on some of the major scientific questions that the
\spica\ mission aims to address; more details about the mission and
instruments can be found in \citet{Roelfsema+2018}.

\section{INTRODUCTION: SPICA AND THE NATURE OF COSMIC MAGNETISM}
\label{sec:intro}

Alongside gravity, magnetic fields play a key role in the formation and evolution of a wide range of structures in the Universe, 
from galaxies to stars and planets.
They simultaneously are an actor, an outcome, and a tracer of cosmic evolution. These three facets of cosmic magnetism are intertwined and 
must be thought of together.  On one hand, the role magnetic fields play in the formation of stars and galaxies 
results from and traces their interplay with gas dynamics. On the other hand, 
turbulence is central  to the dynamo processes that  initially amplified cosmic magnetic fields and have since maintained their strength in galaxies across  time \citep{Brandenburg05}. 
A transfer from gas kinetic  to magnetic energy inevitably takes place in turbulent cosmic flows, while   
magnetic fields act on gas dynamics through the Lorentz force. These physical couplings relate cosmic magnetism to structure formation in 
the Universe across time and scales, and make the observation of magnetic fields a tracer of cosmic evolution, which is today 
yet to be disclosed. 
Improving our observational understanding of cosmic magnetism on a broad range of physical scales is  
thus at the heart of the ``Origins''  big question and is an integral part of one of ESA's four Grand Science Themes 
(``Cosmic Radiation and Magnetism'') as defined by the ESA High-level Science Policy Advisory Committee (HISPAC) in 2013.

As often in Astrophysics, our understanding of the Universe is rooted in 
observations of the very local universe: the Milky Way and nearby galaxies. 
In the interstellar medium (ISM) of these galaxies, the magnetic energy is observed to be in rough equipartition 
with the kinetic (e.g. turbulent), radiative, and cosmic ray energies, all on the order of $\sim 1\, {\rm eV\, cm^{-3}} $, 
suggesting that magnetic fields are a key player in the dynamics of the ISM \citep[e.g.][]{Draine2011}. 
Their exact role in the formation of molecular clouds and stellar systems is not well understood, however, 
and remains highly debated \citep[e.g.][]{Crutcher2012}. 
Interstellar magnetic fields also hold the key for making 
headway on other main issues in Astrophysics, including 
the dynamics and energetics of the multiphase ISM, 
the acceleration and propagation of cosmic rays, and the physics of stellar and back-hole feedback. 
Altogether, a broad range of science topics call for progress in our understanding of interstellar magnetic fields, which in turn 
motivates ambitious efforts to obtain relevant data  \citep[cf.][]{Imagine18}. 

Observations of Galactic polarization are a highlight and a lasting legacy of the {\it Planck} space mission. 
Spectacular images combining the intensity of dust emission with the texture derived from polarization data 
have received world-wide attention and have become part of the general scientific culture \citep{planck2014-a01}. 
Beyond their popular impact, the {\it Planck} polarization maps represented an 
immense step forward for Galactic astrophysics \citep{PlanckXII2018}. 
{\it Planck} has paved the way for statistical studies of the structure of the Galactic magnetic field 
and its coupling with interstellar matter and turbulence, in the diffuse ISM and star-forming molecular clouds. 

SPICA, the {\it Space Infrared Telescope for Cosmology and Astrophysics} proposed to ESA as an M5 mission concept \citep{Roelfsema+2018}, 
provides one of the best opportunities to take the next big leap forward and 
gain fundamental insight into the role of magnetic fields in structure formation 
in the cold Universe, 
thanks to the unprecedented sensitivity, angular resolution, and dynamic range of 
its far-infrared (far-IR) imaging polarimeter, {\spicapol}$^2$ (previously called SPICA-POL, for ``SPICA polarimeter'').
The baseline {\spicapol} instrument will allow simultaneous imaging observations in three bands, $100\, \mu$m, $200\, \mu$m, 
and $350\, \mu$m, with an individual pixel ${\rm NEP} < 3 \times 10^{-18}\, {\rm W\, Hz}^{-1/2}$,  
over an instantaneous field of view of $\sim 2.7' \times 2.7'$ at resolutions of 9\arcsec, 18\arcsec, and 32\arcsec, respectively 
\citep{Rodriguez+2018}. 
Benefiting from a 2.5-m space telescope cooled to 
$< 8\, $K,  
{\spicapol} will be two to three orders of magnitude more sensitive 
than current or planned far-IR/submillimeter polarimeters (see \S ~\ref{subsec:spica-adv} below) 
and will produce far-IR dust polarization images at a factor 20--30 higher resolution than the {\it Planck} satellite. 
It will provide wide-field 100--350$\, \mu$m polarimetric images 
in Stokes Q and U 
of comparable quality (in terms of resolution, signal-to-noise ratio, and both intensity and spatial dynamic ranges)  
to  {\it Herschel} images in Stokes I. 

The present paper gives an overview of the main science drivers for the {\spicapol} polarimeter 
and is complementary to the papers by, e.g., \citet{Spinoglio+2017} 
and \citet{vandertak+2018} 
which discuss the science questions addressed by the other two instruments of SPICA, 
SMI \citep{Kaneda+2016}
and SAFARI \citep{Roelfsema+2014},
mainly through highly sensitive spectroscopy. 
The outline is as follows: Section~\ref{sec:filaments} describes the prime science driver for {\spicapol}, 
namely high dynamic range polarimetric mapping of Galactic filamentary structures to unravel the role of magnetic fields 
in the star formation process. 
Section~\ref{sec:turbulence} introduces the contribution of  {\spicapol} to 
the statistical characterization of magnetized interstellar turbulence. 
Section~\ref{sec:protostars} and Section~\ref{sec:massive-sf} emphasize the importance of {\spicapol} polarization observations for 
our understanding of the physics of protostellar dense cores and high-mass star protoclusters, respectively.  
Section~\ref{sec:galaxies} discusses dust polarization observations of galaxies, focusing mainly on nearby galaxies. 
Section~\ref{sec:dust-physics} describes how multi-wavelength polarimetry with {\spicapol} can constrain 
dust models and the physics of dust grain alignment. 
Finally, Sections~\ref{sec:cosmic-rays}, ~\ref{sec:disks}, and ~\ref{sec:proto-var} discuss three topics which, although not among the main drivers 
of the {\spicapol} instrument, will significantly benefit from {\spicapol} observations, 
namely the study of the origin of cosmic rays and of their interaction with molecular clouds (Sect.~\ref{sec:cosmic-rays}), 
the detection of polarized far-IR dust emission from protoplanetary disks, thereby tracing magnetic fields 
in the inner layers of the disks (Sect.~\ref{sec:disks}), and  
the (non-polarimetric) monitoring of protostars in the far-IR, 
i.e., close to the peak of their spectral energy distributions (SEDs), 
to provide direct constraints on the process of episodic protostellar accretion (Sect.~\ref{sec:proto-var}). 
Section~\ref{sec:conclusions} concludes the paper.

\begin{table}
	\caption{B-BOP performance parameters}
	\setlength{\tabcolsep}{3pt}
	\centering
	\begin{tabular}{lccc}
		\hline\hline
		Band 			& 100 $\mu$m  	& 200 $\mu$m  		& 350 $\mu$m \rule{0pt}{10pt} \\ 
		\hline 
		$\lambda$ range & 75-125 $\mu$m	& 150-250 $\mu$m	& 280-420 $\mu$m \rule{0pt}{10pt} \\ 
		Array size 		& 32$\times$32  & 16$\times$16 		& 8$\times$8 		\\ 
		Pixel size 		& 5"$\times$5" 	& 10"$\times$10" 	& 20"$\times$20" 	\\ 
		FWHM			& 9"			& 18"				& 32"				\\
		\hline 
		\multicolumn{4}{l}{Point source sensitivity 2.5'$\times$2.5' 5$\sigma$-1hr \rule{0pt}{10pt} } \\
		Unpol.		& 21 $\mu$Jy	& 42 $\mu$Jy		& 85 $\mu$Jy		\\
		 Q, U 	& 30 $\mu$Jy	& 60 $\mu$Jy 		& 120 $\mu$Jy		\\
		\hline 
		\multicolumn{4}{l}{Point source sensitivity  1 deg$^2$ 5$\sigma$-10hr \rule{0pt}{10pt} } \\
		Unpol.			& 160 $\mu$Jy	& 320 $\mu$Jy		& 650 $\mu$Jy		\\
		 Q, U 	& 230 $\mu$Jy	& 460 $\mu$Jy		& 920 $\mu$Jy		\\
		\hline
		\multicolumn{4}{l}{Surface brightness sensitivity 1 deg$^2$ 5$\sigma$-10hr \rule{0pt}{10pt} } \\
		Unpol.			& 0.09 MJy/sr	& 0.045 MJy/sr		& 0.025 MJy/sr		\\
		5\% Q, U $^\dag$	& 2.5 MJy/sr	& 1.25 MJy/sr		& 0.7 MJy/sr		\\
		\hline
		\multicolumn{4}{l}{Dynamic range in I for accurate I, Q, U measurements$^\ddag$  \rule{0pt}{10pt} } \\
			& $\geq 100$	&    $\geq 100$		& $\geq 100$		\\
		\hline
		\multicolumn{4}{l}{Maximum scanning speed for full resolution imaging  \rule{0pt}{10pt} } \\
			& $\geq 20\arcsec$/sec	&    $\geq 20\arcsec$/sec  & $\geq 20\arcsec$/sec		\\			
		\hline\hline
	\end{tabular} 

 \smallskip
 \raggedright
 {$^\dag$} Surface brightness level in I to map Q, U at 5$\sigma$ over 1~deg$^2$ in 10hr 
assuming 5\% fractional polarization.

\raggedright
 {$^\ddag$} Assuming $\geq 1\% $ fractional polarization.
	\label{tab:POL}
\end{table}

\section{Magnetic fields and star formation in filamentary clouds}
\label{sec:filaments} 

Understanding how stars form in the cold ISM of galaxies is central in Astrophysics. 
Star formation is both one of the main factors that drive the evolution of galaxies 
on global scales and the process that sets the physical conditions for planet formation on local scales. 
Star formation is also a complex, multi-scale process, involving a subtle interplay between gravity, 
turbulence, magnetic fields, feedback mechanisms. 
As a consequence, and despite recent progress, the basic questions of what regulates star formation in galaxies 
and what determines the mass distribution of forming stars (i,e. the stellar initial mass function or IMF) 
remain two of the most debated problems in Astronomy. 
Today, a popular  school of thought for understanding star formation and these two big questions
is the gravo-turbulent paradigm \citep[e.g.][]{MacLowKlessen2004, McKee07, Padoan+2014}, 
whereby magnetized supersonic turbulence creates structure and seeds 
in interstellar clouds, which subsequently grow and collapse under the primary influence of gravity.
A variation on this scenario is that of dominant magnetic fields in cloud envelopes, and a turbulence-enhanced 
ambipolar diffusion leading to gravity-dominated subregions \citep[e.g.,][]{Li04,KudohBasu2008}.

Moreover, while the global rate of star formation in galaxies and the positions of galaxies 
in the Schmidt-Kennicutt diagram \citep[e.g.][]{Kennicutt+2012}
are likely controlled by macroscopic phenomena such as cosmic accretion, large-scale feedback, 
and large-scale turbulence \citep{SanchezAlmeida+2014}, there 
is some evidence that the star formation efficiency in the dense molecular gas of galaxies is 
nearly universal\footnote{With the possible exception 
of extreme star-forming environments like the central molecular zone (CMZ)  
of our Milky Way \citep{Longmore+2013} or extreme starburst galaxies \citep[e.g.][]{Garcia-Burillo+2012}. 
See other caveats for galaxies in \citet{Bigiel+2016}.} 
\citep[e.g.][]{Gao+2004,Lada+2012} 
and primarily 
governed by the physics of filamentary cloud fragmentation on much smaller scales \citep[e.g.][]{Andre+2014, Padoan+2014}. 
As argued in \S ~\ref{subsec:fil-paradigm} and \S ~\ref{subsec:filaments} below, magnetic fields are likely a key element 
of the physics behind the formation and fragmentation of filamentary structures in interstellar clouds. 

Often ignored,
strong, organized magnetic fields, in rough equipartition with the turbulent and 
cosmic ray energy densities, have been detected in the ISM of a large number of galaxies out to 
$z = 2$ \citep[e.g.][]{Beck2015,Bernet+2008}. 
Recent cosmological magneto-hydrodynamic (MHD) 
simulations of structure formation in the Universe suggest 
that magnetic-field strengths comparable to those measured in nearby galaxies ($\simlt 10\, \mu$G)
can be quickly built up in high-redshift galaxies (in $<< 1\,$ Gyr), through the dynamo amplification 
of initially weak seed fields  \citep[e.g.][]{RiederTeyssier2017,Marinacci+2018}. 
Magnetic fields are therefore expected to play a dynamically important role 
in the formation of giant molecular clouds (GMCs) on kpc scales 
within galaxies \citep[e.g.][]{Inoue+2012} 
and in the formation of filamentary structures leading to individual 
star formation on $\sim$1--10$\,$pc scales within GMCs \citep[e.g.][see \S ~\ref{subsec:filaments} below]{Inutsuka+2015,Inoue+2018}. 
On dense core ($\leq 0.1\, $pc) scales, the magnetic field and angular momentum of most protostellar systems  
are likely inherited from the processes of filament formation and fragmentation 
(cf. Misugi et al., in prep.). 
On even smaller ($< 0.01\, $pc or $< 2000\, $au) scales, magnetic fields are essential 
to solve the angular momentum problem of star formation, 
generate protostellar outflows, and control the formation of protoplanetary disks 
\citep[e.g.][]{Pudritz+2007,Machida+2008,LiPPVI}.

In this context, {\spicapol} will be a unique tool for characterizing the morphology 
of magnetic fields on scales ranging from $\sim 0.01\,$pc to $\sim 1\,$kpc 
in Milky Way like galaxies.
In particular, a key science driver for {\spicapol} is to clarify the role of magnetic fields 
in shaping the rich web of filamentary structures
pervading the cold ISM, from the low-density striations seen in HI clouds and the outskirts of CO clouds 
\citep[e.g.][]{Clark+2014,Kalberla+2016,Goldsmith+2008} 
to the denser molecular filaments 
within which most prestellar cores and protostars are forming 
according to {\it Herschel} results 
(see Fig.~\ref{taurus_planck} and \S ~\ref{subsec:fil-paradigm} below).

\subsection{Dust polarization observations: A probe of magnetic fields in star-forming clouds}
\label{subsec:dustpol}

\subsubsection{Dust grain alignment}
\label{subsubsec:alignment}

Polarization of background starlight from dichroic extinction produced by 
intervening interstellar dust has been known since
the late 1940s
\citep{Hall1949,Hiltner1949}. The analysis of the
extinction data in polarization, in particular its variation with
wavelength in the visible to near-UV, has allowed major discoveries
regarding dust properties, in particular regarding the
size distribution of dust.  Like the first large-scale total intensity
mapping
in the far-IR that was provided by the 
Infrared Astronomical Satellite (IRAS) satellite data,
extensive studies of polarized far-IR emission today bring the prospect of a new
revolution in our understanding of dust physics. 
This endeavor includes pioneering observations with ground-based, 
balloon-borne, and space-borne facilities  such as the very
recent all-sky observations by the {\it Planck} satellite at $850\, \mu$m and
beyond \citep[e.g.][]{PlanckXII2018}.
However, polarimetric imaging of polarized dust continuum emission  is
still in its infancy and amazing improvements are expected in the next
decades from instruments such as  the Atacama Large Millimeter Array (ALMA) 
in the submillimeter and {\spicapol} in the far-IR.

The initial discovery that starlight extinguished by intervening dust
is polarized led to the
conclusion that dust grains must be somewhat elongated and globally
aligned in space in order to produce the observed polarized
extinction. While the elongation of dust grains was not unexpected,
coherent grain alignment over large spatial scales has been more difficult to
explain. 
A very important constraint has come from recent measurements
in emission with, e.g.,  
the Archeops balloon-borne experiment \citep{Benoit+2004} and the
{\it Planck} satellite \citep{PlanckXIX2015} which indicated that the polarization degree of dust emission 
can be as high as $20\%$ in some regions of the diffuse ISM in the solar neighborhood.
This requires more efficient dust alignment processes than
previously anticipated \citep{PlanckXII2018}. 

The most widely accepted dust grain alignment theories, 
already alluded to by
\cite{Hiltner1949}, propose that alignment is with respect to the
magnetic field that pervades the ISM.
Rapidly spinning grains will naturally align their angular momentum
with the magnetic field direction
\citep{Purcell1979,Lazarian_Draine1999}, but the mechanism leading to
such rapid spin remains a mystery.  The formation of molecular
hydrogen at the surface of dust grains could provide the required momentum 
\citep{Purcell1979}. 
Today's leading grain alignment theory
is Radiative Alignment Torques (RATs) \citep[][and references
therein]{Dolginov_Mitrofanov1976,Draine_Weingartner1996,Lazarian_Hoang2007, Hoang_Lazarian2016}, 
where supra-thermal spinup of irregularly-shaped dust grains results 
from their irradiation by an anisotropic radiation field \citep[a process experimentally confirmed, see ][]{Abbas+2004}. 

\subsubsection{Probing magnetic fields with imaging polarimetry}
\label{subsubsec:polarimetry}

In the conventional picture 
that the minor axis of elongated dust grains is 
aligned with the local direction of the magnetic field, 
mapping observations of linearly-polarized continuum emission 
at far-IR and submillimeter wavelengths 
are a powerful tool to measure the morphology and structure of magnetic field lines in star-forming clouds and dense cores 
\citep[cf.][]{Matthews+2009,Crutcher+2004, Crutcher2012}.
A key advantage of this technique is that it images the structure of magnetic fields through an emission process that traces the {\it mass} of cold interstellar matter, 
i.e., the reservoir of gas directly involved in star formation. 
Indirect estimates of the plane-of-sky magnetic field strength $B_{POS}$ can also be obtained using the Davis-Chandrasekhar-Fermi method \citep{Davis1951, CF1953}:
$B_{POS} = \alpha_{corr}\, \sqrt{4\pi \rho}\, \delta V / \delta \Phi $, where $\rho $ is the gas density (which can be estimated to reasonable 
accuracy from $Herschel$ column density maps, especially in the case of resolved filaments and cores -- cf. \citealp{Palmeirim+2013, Roy+2014}), 
$ \delta V $ is the one-dimensional velocity dispersion (which can be estimated from line observations in an appropriate  tracer such as 
N$_2$H$^+$ for star-forming filaments and dense cores-- e.g. \citealp{Andre+2007, Tafalla+2015}),  
$ \delta \Phi $ is the dispersion in polarization position angles directly measured in a dust polarization map, 
and $\alpha_{corr} \approx 0.5 $ is a correction factor obtained through numerical simulations 
\citep[cf.][]{Ostriker+2001}. 
Large-scale maps that resolve the above quantities over a large dynamic range of densities
can be used to estimate the mass-to-flux ratio in different parts of a molecular cloud. 
This can test the idea that cloud envelopes may be magnetically supported 
and have a subcritical mass-to-flux ratio \citep{MouschoviasCiolek1999,Shu+1999}.
Recent applications of the Davis-Chandrasekhar-Fermi method using SCUBA2-POL $850\, \mu$m data taken as part of the BISTRO survey 
\citep{Ward+2017} toward dusty molecular 
clumps in the Orion and Ophiuchus clouds are presented in \citet{Pattle+2017}, \citet{Kwon+2018}, and \citet{Soam+2018}. 
Refined estimates of both the mean 
and the turbulent component of $B_{POS}$ can be derived from an analysis of the 
second-order angular structure function (or angular dispersion function) of observed polarization position angles 
$<\Delta\Phi^2 (l)>  = \frac{1}{N(l)} \Sigma [\Phi (r) - \Phi (r+l)]^2 $ \citep{Hildebrand+2009,Houde+2009}.
Alternatively, in localized regions where gravity dominates over MHD turbulence, the polarization-intensity gradient method can be used 
to obtain maps of the local magnetic field strength from maps of the misalignment angle $\delta $ between the local magnetic field 
(estimated from observed polarization position angles) and the local column density gradient (estimated from maps of total dust emission). 
Indeed, such $\delta $ maps provide information on the local ratio between the magnetic field tension force and the gravitational force 
\citep{Koch+2012, Koch+2014}. 
Additionally, the paradigm of Alfv\'enic turbulence can be tested in dense regions where gravity dominates, in which the 
observed angular dispersion $\Delta \Phi$ 
is expected to decrease in amplitude toward the center of dense cores where $\delta V$ also decreases 
\citep{Auddy+2019}. 

\begin{figure*}
\begin{center}
\includegraphics[width=42pc]{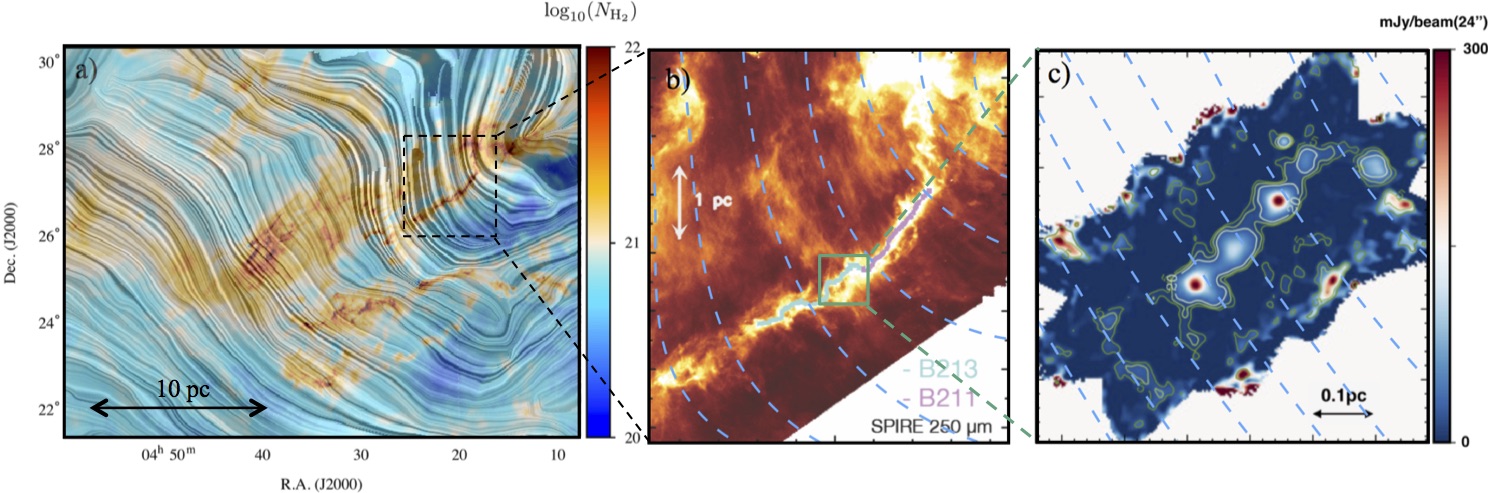}
\caption{{\bf(a)} Multi-resolution column density map of the Taurus molecular cloud as derived 
from a combination of high-resolution (18\arcsec--36\arcsec $\,$HPBW) observations from the 
{\it Herschel} Gould Belt survey and low-resolution (5\arcmin $\,$HPBW) {\it Planck} data. 
The superimposed  ``drapery'' pattern traces the magnetic-field orientation projected 
on the plane of the sky, as inferred from {\it Planck} polarization data at 850 $\mu $m \citep{PlanckXXXV2016}. 
{\bf(b)} {\it Herschel}/SPIRE 250~$\mu$m dust continuum image of the B211/B213 filament in the Taurus cloud 
\citep{Palmeirim+2013,Marsh+2016}.
The superimposed blue dashed curves trace the magnetic field orientation projected on the plane 
of the sky, as inferred from $Planck$ dust polarization data at 850~$\mu$m \citep{PlanckXXXV2016}.
Note the presence of faint striations oriented roughly perpendicular to the main filament 
and parallel to the plane-of-sky magnetic field. 
{\bf(c)} IRAM/NIKA1 1.2~mm dust continuum image of the central part of the {\it Herschel} field shown in (b) (effective HPBW resolution of 20\arcsec ), 
showing a chain of at least four equally-spaced dense cores along the B211/B213 filament \citep[from][]{Bracco+2017}.
{\spicapol} can image the magnetic field lines at a factor 30 better resolution than {\it Planck} over the entire Taurus cloud (cf. panel {\bf a}), probing scales from $\sim 0.01$ pc to $> 10$ pc.}
\label{taurus_planck}
\end{center}
\end{figure*}

Because the typical degree of polarized dust continuum emission is low ($\sim \, $2\%--5\%  -- e.g. \citealp{Matthews+2009}) 
and the range of relevant column densities spans three orders of magnitude from equivalent visual extinctions\footnote{In Galactic molecular clouds, 
a visual extinction $A_V = 1$ roughly corresponds to a column density of H$_2$ molecules $N_{H_2} \sim 10^{21}\, {\rm cm}^{-2}$ \citep[cf.][]{Bohlin+1978}.} 
$A_V \sim 0.1$ in the atomic medium to  $A_V  > 100$ 
in the densest molecular filaments/cores, 
a systematic dust polarization study of the rich filamentary networks pervading nearby interstellar clouds and their connection to star formation
requires a large improvement in sensitivity, mapping speed, and dynamic range over existing far-IR/submillimeter polarimeters. 
A big improvement in polarimetric mapping speed is also needed for statistical reasons.
As only the plane-of-sky component of the magnetic field 
is directly accessible to dust continuum polarimetry, 
a large number of systems must be imaged in various Galactic environments before physically meaningful conclusions 
can be drawn statistically on the role of magnetic fields.
As shown in \S ~\ref{subsec:spica-adv} below, 
the required step forward in performance can be uniquely provided by a large, cryogenically cooled space-borne telescope such as SPICA, 
which can do in far-IR polarimetric imaging what {\it Herschel} achieved in total-power continuum imaging. 

\subsection{Insights from {\it Herschel} and {\it Planck}: A filamentary paradigm for star formation?}
\label{subsec:fil-paradigm}

The {\it Herschel} mission has led to spectacular advances in our knowledge of the texture of the cold ISM and its link with star formation. 
While 
interstellar clouds have been known to be filamentary for a long time 
\citep[e.g.][and references therein]{Schneider+1979, Bally+1987, Myers2009}, 
{\it Herschel} imaging surveys have established the ubiquity of 
filaments on almost all length scales 
($\sim 0.5\,$pc to $\sim 100\,$pc) in the molecular clouds of the Galaxy and shown that this filamentary structure likely plays a key role in the star formation process 
\citep[e.g.][]{Andre+2010, Henning+2010, Molinari+2010, Hill+2011, Schisano+2014, Wang+2015}.

The interstellar filamentary structures detected with {\it Herschel} span broad ranges in length, 
central column density, and mass per unit length \citep[e.g.][]{Schisano+2014,Arzoumanian+2018}.
In contrast, detailed analysis of the radial column density profiles 
indicates that, at least in the nearby molecular clouds of the Gould Belt, 
 {\it Herschel} filaments are characterized 
by a narrow distribution of inner widths with a typical 
value of $\sim 0.1$~pc and a dispersion of less than a factor of 2, 
when the data are averaged over the filament crests  
\citep[][]{Arzoumanian+2011,Arzoumanian+2018}.
Independent studies 
of filament widths in nearby clouds have generally confirmed
this result 
when using submillimeter continuum data
\citep[e.g.][]{KochRosolowsky2015,Salji+2015, Rivera-Ingraham+2016}, 
even if factor of $\sim \,$2--4
variations around the mean inner width of $\sim 0.1\,$pc 
have been found  
along the main axis of a given filament 
\citep[e.g.][]{Juvela+2012, Ysard+2013}. 
%
Measurements of filament widths obtained in molecular line tracers 
 \citep[e.g.,][]{Pineda+2011,FernandezLopez+2014, Panopoulou+2014, Hacar+2018}
have been less consistent with the {\it Herschel} dust continuum results of \citet[][2018]{Arzoumanian+2011}, 
but this can be attributed to the lower dynamic range achieved by observations in any given molecular line tracer. 
%
\citet{Panopoulou+2017} pointed out an apparent contradiction between the existence of a characteristic filament width 
and the essentially scale-free nature of the power spectrum of interstellar cloud images 
(well described by a single power law 
from $\sim 0.01\,$pc to $\sim 50\,$pc -- \citealp{mamd+2010,mamd+2016}), but 
\citet{Roy+2018} showed that there is no contradiction given the only modest area filling factors ($\simlt 10\% $) 
and column density contrasts ($\leq 100\% $ in most cases) 
derived by \citet{Arzoumanian+2018}
for the filaments seen in {\it Herschel} 
images. 
While further high-resolution submillimeter continuum studies would be required to investigate whether the same result holds beyond the Gould Belt, 
the median inner width of $\sim 0.1\,$pc measured with {\it Herschel} 
appears to reflect the presence of a true common 
scale in the filamentary structure of nearby interstellar clouds. 
If confirmed, this result may have far-reaching consequences as
it introduces a characteristic scale in a system generally thought to be chaotic and turbulent (i.e. largely scale-free -- cf.  \citealp{Guszejnov+2018}).
It may thus present a severe challenge in any attempt to interpret all ISM observations in terms of scale-free
processes.

Another major result from {\it Herschel} studies of nearby 
clouds is that most ($>75\% $) 
prestellar cores and protostars are found to lie in dense,  ``supercritical'' 
filaments above a critical threshold $\sim 16\, M_\odot $/pc in mass per unit length, 
equivalent to a critical threshold $\sim 160\, M_\odot $/pc$^2$ ($A_V \sim 8)$ in column density or $n_{H_2} \sim  2 \times 10^4\, {\rm cm}^{-3} $ in volume density 
\citep{Andre+2010,Konyves+2015,Marsh+2016}.
A similar column density threshold for the formation of prestellar cores (at $A_V \sim \, $5--10) had been suggested earlier based 
on ground-based millimeter and submillimeter studies \citep[e.g.][]{Onishi+1998,Johnstone+2004,Kirk+2006}, but without clear connection to filaments. 
Interestingly, a comparable threshold in extinction (at $A_V \sim \, $8) has also been observed 
in the spatial distribution of young stellar objects (YSOs) with {\it Spitzer} \citep[e.g.][]{Heiderman+2010,Lada+2010,Evans+2014}.

\begin{figure*}
\begin{center}
\includegraphics[width=30pc]{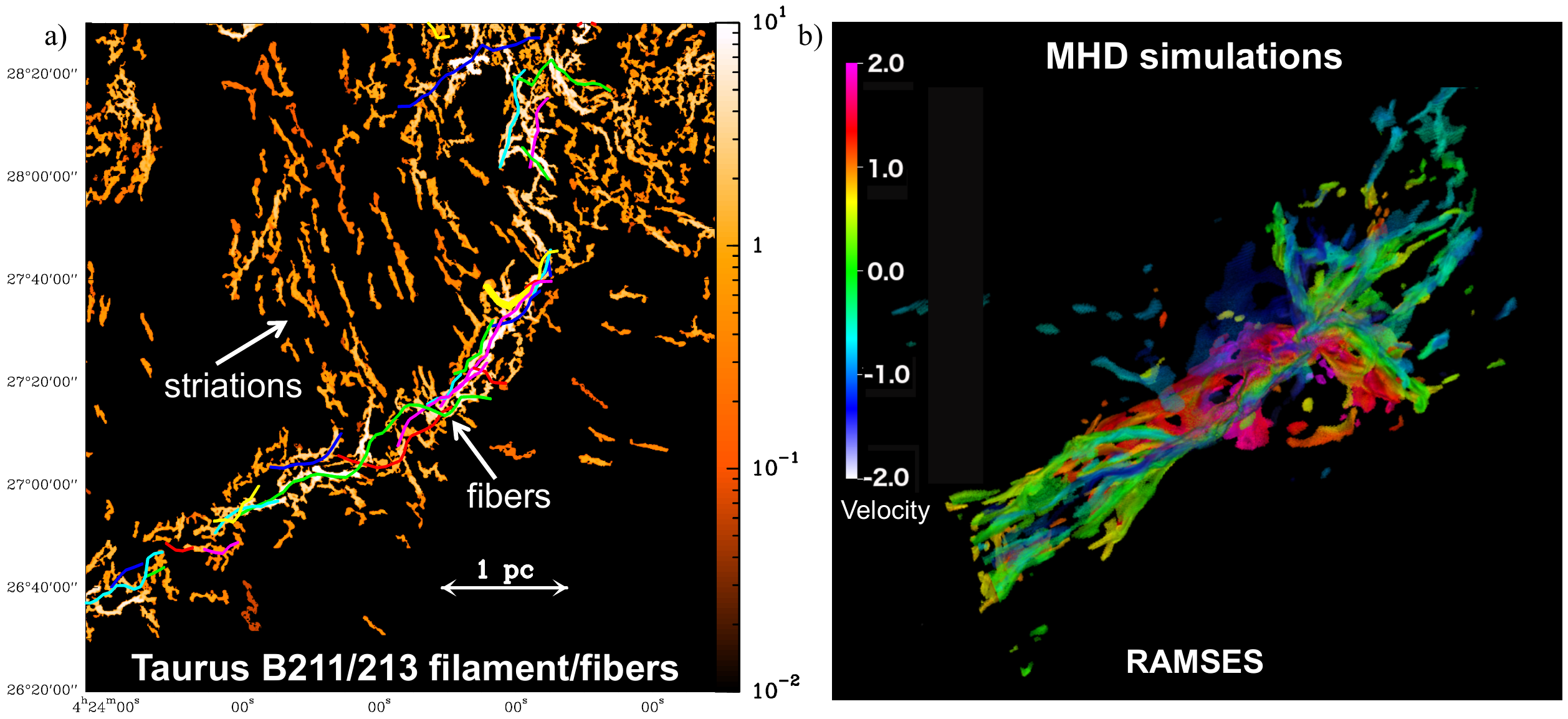}
\caption{{\bf(a)} Fine (column) density structure of the B211/B213 filament based on a filtered version of the $Herschel$ 250 $\mu$m 
    image of \citet{Palmeirim+2013}  
    using the algorithm {\it getfilaments} \citep{Menshchikov2013}. 
    In this view, all transverse angular scales larger than $72\arcsec$ (or $\sim 0.05$~pc) 
    were filtered out to enhance the contrast of the small-scale structure. 
    The color scale 
    is in MJy/sr at 250 $\mu$m. 
    The colored curves display the velocity-coherent fibers independently 
     identified by \citet{Hacar+2013}  
     using N$_2$H$^+$/C$^{18}$O observations.  
 {\bf(b)} MHD simulation of a collapsing/accreting filament performed by E. Ntormousi \& P. Hennebelle with the adaptive mesh refinement (AMR) code RAMSES. 
 Line-of-sight velocities (in km/s) after one free-fall time ($\sim 0.9$~Myr)  
 are coded by  colors. For clarity, only the dense gas with $10^4\, {\rm cm}^{-3} < {n_{\rm H_2}} <  10^5\, {\rm cm}^{-3} $ is shown. 
 Note the braid-like velocity structure and the morphological similarity with the 
 fiber-like pattern seen in the B211/B213 observations on the left. 
 Thanks to its high resolution and dynamic range, 
 {\spicapol} can probe, for the first time, the geometry of the magnetic field {\it within} the dense system of fibers and the connection with the low-density striations 
 in the ambient cloud. 
 }
 \label{taurus_fibers}
\end{center}
\end{figure*}

Overall, the {\it Herschel} results support a filamentary paradigm for star formation 
in two main steps \citep[e.g.][]{Andre+2014,Inutsuka+2015}: 
First, multiple large-scale compressions 
of interstellar material in supersonic turbulent 
MHD flows generate a cobweb of $\sim 0.1$ pc-wide filaments in the cold ISM; 
second, the densest filaments fragment into prestellar cores (and subsequently protostars) by gravitational instability 
above the critical mass per unit length $ M_{\rm line,crit} = 2\, c_s^2/G$ of nearly isothermal, cylinder-like filaments (see Fig.~\ref{taurus_planck}), 
where $c_s$ is the sound speed and $G$ the gravitational constant. 
This paradigm differs from the classical gravo-turbulent picture 
in that it relies on the unique features of filamentary geometry, 
such as the existence of a critical line mass for nearly isothermal filaments  \citep[e.g.][and references therein]{Inutsuka+1997}. 
The validity and details of the filamentary paradigm are strongly debated, however, 
and many issues remain open.
For instance, according to some numerical simulations, the above two steps 
may not occur consecutively but simultaneously, in the sense that both filamentary structures and dense cores may grow in mass at the same time 
\citep[e.g.][]{Gomez+2014,ChenOstriker2015}.
The physical origin of the typical $\sim 0.1\,$pc inner width of molecular filaments is also poorly understood 
and remains a challenge for numerical models \citep[e.g.][]{Padoan+2001,Hennebelle2013, Smith+2014, Federrath2016, Ntormousi+2016}. 
\citet{Auddy+2016} point out that magnetized filaments may actually be ribbon-like and quasi-equilibrium structures supported by the magnetic field, 
and therefore not have cylindrical symmetry.
Regardless of any particular scenario,  
there is nevertheless  little doubt after {\it Herschel} results that 
dense molecular filaments represent an integral part of the initial conditions of the bulk of star formation in our Galaxy. 

As molecular filaments are known to be present in the Large Magellanic Cloud 
(LMC -- \citealp{Fukui+2015}), 
the proposed filamentary paradigm 
may have implications on galaxy-wide scales. 
Assuming that all filaments have similar inner widths, 
it has been argued that they may help to regulate the star formation efficiency in dense molecular gas \citep{Andre+2014}, 
and that they 
may be responsible for a quasi-universal 
star formation law in the dense molecular ISM of galaxies  \citep[cf.][]{Lada+2012,Shimajiri+2017}, 
with possible variations in extreme environments such as the CMZ  \citep{Longmore+2013,Federrath+2016}.

In parallel, the {\it Planck} mission has led to major advances in our knowledge of the geometry of the magnetic field on large scales in the Galactic ISM. 
The first all-sky maps of dust polarization 
provided by {\it Planck} at 850$\, \mu$m have revealed a very organized magnetic field structure on $\simgt \,  $1--10~pc scales in 
Galactic interstellar clouds \citep[][see Fig.~\ref{taurus_planck}a]{PlanckXXXV2016}. 
The large-scale magnetic field tends to be aligned with low-density filamentary structures with subcritical line masses 
such as striations (see Fig.~\ref{taurus_planck}b) 
and perpendicular to dense star-forming filaments with supercritical line masses 
\citep[][see Figs.~\ref{taurus_planck}b \& \ref{taurus_planck}c]{PlanckXXXII2016,PlanckXXXV2016}, 
a trend also seen in optical and near-IR polarization observations 
\citep[][]{Chapman+2011,Palmeirim+2013,Panopoulou+2016,Soler+2016}. 
There is also a hint from {\it Planck}  polarization observations of the nearest 
clouds that the direction of the magnetic field may change {\it within} dense filaments
from nearly perpendicular in the ambient cloud to more parallel in the filament interior 
\citep[cf.][]{PlanckXXXIII2016}. 
These findings suggest that magnetic fields are dynamically important and play a key role in the formation and evolution of filamentary structures in interstellar clouds, 
supporting the view that dense molecular filaments form by accumulation of interstellar matter along field lines. 

The low resolution of {\it Planck} polarization data (10\arcmin$\,$at best or 0.4 pc in nearby clouds) is however insufficient 
to probe the organization of field lines in the $\sim 0.1\,$pc interior of filaments, corresponding both to the characteristic transverse 
scale of filaments \citep[][]{Arzoumanian+2011,Arzoumanian+2018}
and to the scale at which fragmentation into prestellar cores occurs  \citep[cf.][]{Tafalla+2015}. 
Consequently,  the geometry of the magnetic field {\it within} interstellar filaments and its effects on fragmentation and star formation 
are essentially unknown today.

\subsection{Investigating the role of magnetic fields in the formation and evolution of  molecular filaments with {\spicapol}}
\label{subsec:filaments} 
Improving our understanding of the physics and detailed properties of molecular filaments is of paramount importance 
as the latter are representative of the initial conditions of star formation in molecular clouds and GMCs\footnote{There is a whole spectrum 
of molecular clouds in the Galaxy, ranging from individual clouds $\sim \,$2--10$\,$pc in size and $\sim 10^{2-4}\, M_\odot$ in mass 
to GMCs $\sim \,$50$\,$pc or more in diameter and $\sim 10^{5-6}\, M_\odot $ in mass (\citealp{Williams00a}, \citealp{Heyer+2015},  
and references therein).}   
(see \S ~\ref{subsec:fil-paradigm} above).
In particular, 
investigating how dense, ``supercritical'' molecular filaments can maintain a roughly constant $\sim$0.1$\,$pc inner width and fragment into prestellar cores 
instead of collapsing radially to spindles 
is crucial to understanding star formation. The topology of magnetic field lines may be one of the key elements here. 
For instance, a longitudinal magnetic field can support a filament against radial collapse but not against fragmentation along its main axis, 
while a perpendicular magnetic field works against fragmentation and increases the critical mass per unit length 
but cannot prevent the radial collapse of a supercritical filament \citep[e.g.][]{Tomisaka2014,Hanawa+2017}. 
The actual topology of the field within molecular filaments is likely more complex and may be a combination of these two extreme configurations.

One plausible evolutionary scenario, consistent with existing observations, is that star-forming filaments accrete ambient cloud material along field lines 
through a network of  
magnetically-dominated striations \citep[e.g.][see also Figs.~\ref{taurus_planck}b \& \ref{taurus_fibers}a]{Palmeirim+2013, Cox+2016,Shimajiri+2018}. 
Accretion-driven MHD waves may then generate a system of velocity-coherent fibers within dense filaments 
\citep[][cf. Fig.~\ref{taurus_fibers}]{Hacar+2013, Hacar+2018, Arzoumanian+2013, HennebelleAndre2013}
and the corresponding organization of magnetic field lines may play a central role in accounting 
for the roughly constant $\sim 0.1\,$pc inner width of star-forming filaments as measured in {\it Herschel} observations (cf. \S ~\ref{subsec:fil-paradigm}).  
Constraining this process further is key to understanding star formation itself, since filaments with supercritical 
masses per unit length would otherwise undergo rapid radial contraction with time, effectively preventing fragmentation 
into prestellar cores  and the formation of protostars \citep[e.g.][]{Inutsuka+1997}.
Information on the geometry of magnetic field lines {\it within} star-forming filaments at $A_V > 8$ is thus crucially needed, 
and can be obtained through 200--350$\, \mu$m  dust polarimetric imaging at high angular resolution with {\spicapol}. 
Large area coverage and 
both high angular resolution and high spatial dynamic range are needed to resolve the 0.1~pc scale by a factor $\sim \,$ 3--10 on one hand 
and to probe spatial scales from $> 10\,$pc in the low-density striations of the ambient cloud (see Fig.~\ref{taurus_planck}), down to $\sim \, $0.01--0.03 pc 
for the fibers of dense filaments  (see Fig.~\ref{taurus_fibers}).
In nearby Galactic regions (at $d \sim \, $150--500~pc), this corresponds to angular scales from $> 5\, $deg or more down to $\sim 20$\arcsec $\,$ or less.  

\begin{figure}
\begin{center}
\includegraphics[width=\columnwidth]{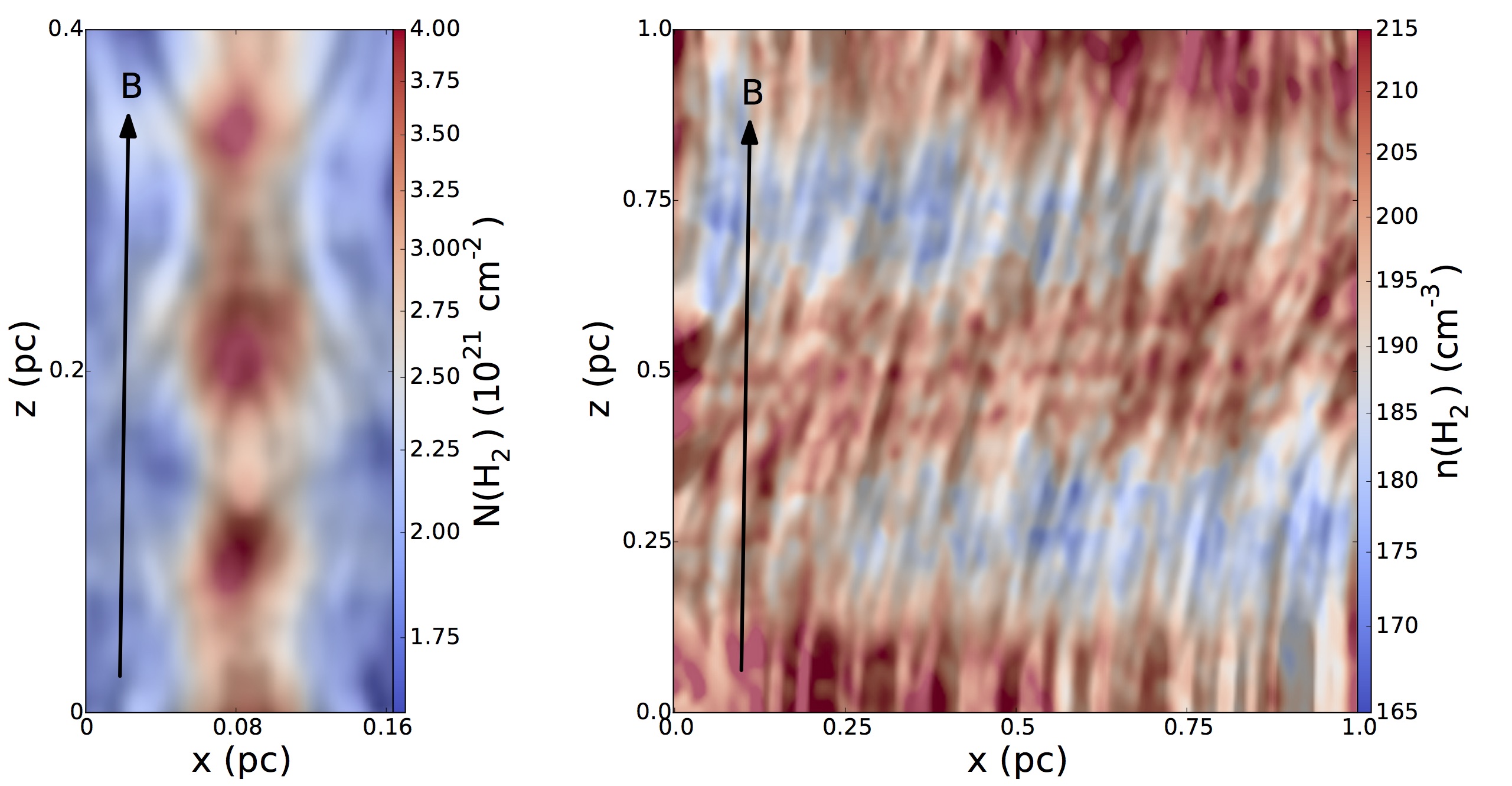}
\caption{Simulated striations (from Tritsis and Tassis 2016). 
Right panel: Volume density image from the simulations.
Left panel: Zoomed-in column density view of a single striation, 
showing the ``sausage''
instability setting in, with characteristic imprints 
in both the magnetic-field and the column-density distribution. 
In both panels, the drapery pattern traces the magnetic field lines and the 
mean direction of the magnetic field is indicated by a black arrow. 
  The passage of Alfv\'{e}n waves excites magnetosonic
  modes that create compressions and rarefactions (colorbar) along
  field lines, giving rise to striations.  
  The simulated data in both panels have been convolved to an effective spatial 
  resolution of 0.012~pc, corresponding to the $18\arcsec $ HPBW of {\spicapol} at $200\, \mu$m.}
 \label{striations_tritsis_sims}
\end{center}
\end{figure}

\begin{figure*}
\begin{center}
\includegraphics[width=40pc]{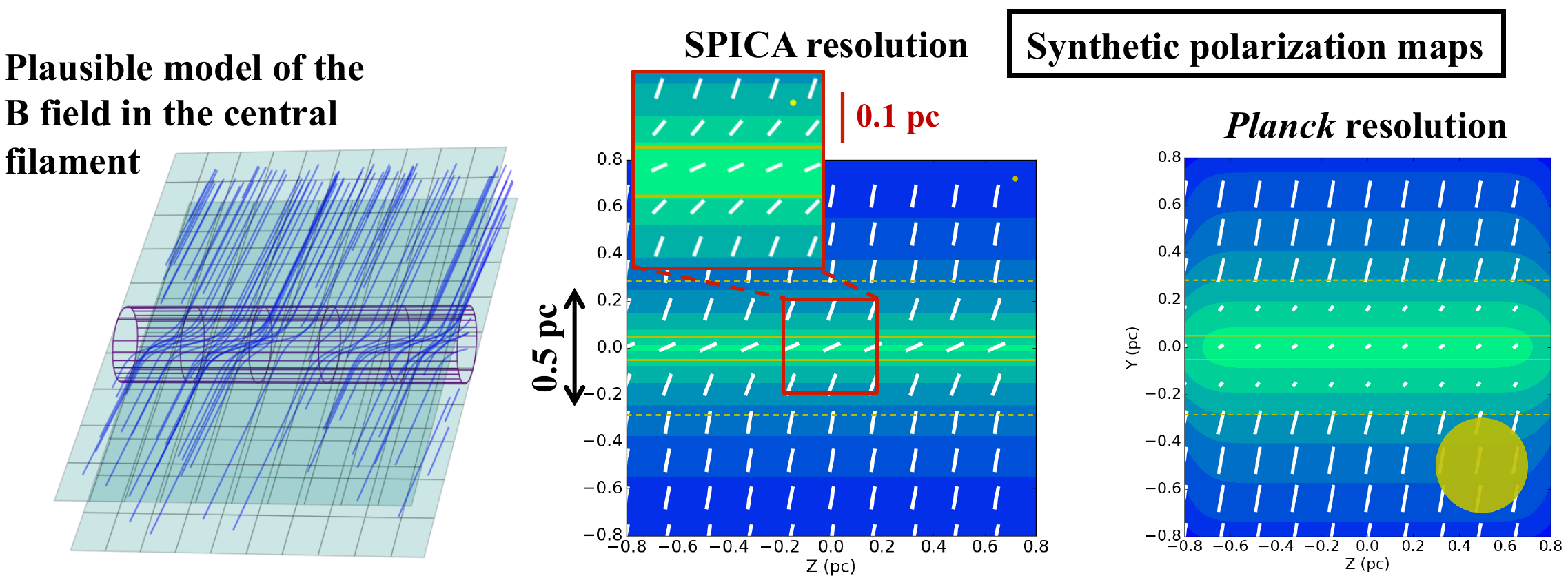}
\caption{{\bf(a)} 3D view of a model filament system similar to Taurus B211/3 and associated magnetic field lines (in blue), 
with a cylindrical filament (red lines) embedded in a sheet-like background cloud (in light green). 
In this model, the magnetic field in the ambient cloud is nearly (but not exactly) perpendicular to the filament axis 
and the axial component is amplified by (gravitational or turbulent) compression in the filament interior. 
 {\bf(b)} Synthetic polarization map expected at the $\sim 20\arcsec $ resolution of {\spicapol} at 200 $\mu$m for the model filament system shown in a). 
 SPICA will follow the magnetic field all the way from the background cloud to the central filament. 
 {\bf(c)} Synthetic polarization map of the same model filament system at the {\it Planck} resolution.
 Note how {\it Planck} data {\it cannot} constrain the geometry of the field lines within the central filament.}
 \label{synthetic_polar_maps}
\end{center}
\end{figure*}

Low-density striations are remarkably ordered structures in an otherwise chaotic-looking turbulent medium.
While the exact physical origin of both low-density striations \citep[][]{Heyer+2016,Tritsis+2016,Tritsis+2018, Chen+2017}
and high-density fibers \citep[e.g.][]{Clarke+2017,Zamora-Aviles+2017} is not well understood and remains highly debated in the literature, 
there is little doubt that magnetic fields are involved. 
For instance, \citet{Tritsis+2016} modeled striations as density fluctuations associated with 
magnetosonic waves in the linear regime (the column density contrast of
observed striations does not exceed $25\%$). These waves are excited
as a result of the passage of Alfv\'{e}n waves, which couple to other
MHD modes through phase mixing 
(see Fig.~\ref{striations_tritsis_sims}, right panel). 
In contrast, \citet{Chen+2017} proposed that striations do not represent real density fluctuations,
but are rather a line-of-sight column density effect in a corrugated layer forming in the dense post-shock region 
of an oblique MHD shock. 
High-resolution polarimetric imaging data would be of great interest to set direct observational constraints 
and discriminate between these possible models. 
Specifically, the magnetosonic wave model predicts
 that a  zoo of MHD wave effects should be observable in these
regions. One of them, that
linear waves in an isolated cloud should establish standing waves
(normal modes) imprinted in the striations pattern, has recently been confirmed in the case of the Musca 
 cloud \citep{Tritsis+2018}.  Other such effects include the
``sausage'' and ``kink'' modes (see Fig.~\ref{striations_tritsis_sims}, left panel), which are studied extensively in the
context of heliophysics \citep[e.g.][]{Nakariakov+2016}, and which
could open a new window to probe the local conditions in molecular clouds  \citep{Tritsis+2018b}.

A first specific objective of {\spicapol} observations will be to test the hypothesis, tentatively suggested by {\it Planck} polarization results \citep[cf.][]{PlanckXXXIII2016}
that the magnetic field may become nearly parallel to the long axis of star-forming filaments in their dense interiors
at scales $< 0.1$~pc, due to, e.g., gravitational or turbulent compression (see Fig.~\ref{synthetic_polar_maps}) and/or reorientation of oblique shocks 
in magnetized colliding flows \citep[][]{Fogerty+2017}.
A change of field orientation inside dense star-forming filaments is also predicted 
by numerical MHD simulations in which gravity dominates and the magnetic field is 
dragged by gas flowing along the filament axis \citep[][]{Gomez+2018,PSLi+2018},  
as observed in the velocity field of some massive infrared dark filaments \citep[][]{Peretto+2014}. 
An alternative topology for the field lines within dense molecular filaments often advocated in the literature 
is that of helical magnetic fields wrapping around the filament axis \citep[e.g.][]{Fiege+2000,StutzGould2016,SchleicherStutz2018,Tahani+2018}. 
As significant degeneracies exist between different models because only the plane-of-sky magnetic field is directly 
accessible to dust polarimetry \citep[cf.][]{Reissl+2018,Tomisaka2015}, discriminating between these various magnetic topologies will require sensitive imaging 
observations of large samples of molecular filaments for which the distribution of viewing angles may be assumed to be essentially random.
One advantage of 
the model of oblique MHD shocks \citep[e.g.][]{ChenOstriker2014,Inoue+2018,Lehmann+2016} 
is that it could potentially 
explain both how dense filaments maintain a roughly constant $\sim 0.1\,$pc width while 
evolving \citep[cf.][]{Seifried+2015} 
and why the observed spacing of prestellar cores along the filaments is significantly shorter than the characteristic fragmentation scale 
of 4$\, \times$ the filament diameter expected in the case of non-magnetized nearly isothermal gas cylinders \citep[e.g.][]{Inutsuka+1992, Nakamura+1993, Kainulainen+2017}. 

A second specific objective of {\spicapol} observations will be to better characterize the transition column density at which a switch occurs 
between filamentary structures primarily parallel to the magnetic field (at low $N_{H_2}$) 
and filamentary structures preferentially perpendicular to the magnetic field (at high $N_{H_2}$) 
\citep[see][and \S ~\ref{subsec:fil-paradigm} above]{PlanckXXXV2016}. 
Based on a detailed analysis of numerical MHD simulations, \citet{Soler+2017} postulated that 
this transition column density depends primarily on the strength of the magnetic field in the parent molecular cloud
and therefore constitutes a key observable piece of information. 
Moreover, \citet{Chen+2016} showed  that, in their colliding flow MHD simulations, the transition occurs where 
the ambient gas is accelerated gravitationally from sub-Alfv\'enic to super-Alfv\'enic speeds. 
They also concluded that the nature of the transition and the 3D magnetic field morphology in the  super-Alfv\'enic region can be constrained 
from the observed polarization fraction and dispersion of polarization angles in the plane of the sky, which provides information 
on the tangledness of the field.

As a practical illustration of what could be achieved with {\spicapol}, a reference polarimetric imaging survey would map, in Stokes I, Q, U 
at $100\, \mu$m, $200\, \mu$m, $350\, \mu$m, the same $\sim 500\,$deg$^2$ area in nearby interstellar clouds 
imaged by {\it Herschel} in Stokes I at 70--500 $\, \mu$m 
as part of the Gould Belt, HOBYS, and Hi-GAL surveys \citep[][]{Andre+2010, Motte+2010, Molinari+2010}. 
To first order, the gain in sensitivity of {\spicapol} over SPIRE \& PACS on {\it Herschel} would compensate for the low degree of polarization (only a few \%) 
and make it possible to obtain Q and U maps of polarized dust emission with a signal-to-noise ratio similar to the {\it Herschel} images in Stokes I. 
Assuming the {\spicapol} performance parameters given in Table~\ref{tab:POL} (see also Table~4 of \citealp{Roelfsema+2018}, and Table~1 of \citealp{Rodriguez+2018})
and an integration time of $\sim 2\,$hr per square degree, 
such a survey would reach a signal-to-noise ratio of 7 
in Q, U intensity at both 200$\, \mu$m and 350$\, \mu$m in low column density areas with $A_V \sim 0.2$ 
(corresponding to the diffuse, cold ISM),  
for a typical polarization fraction of 5\% and a typical dust temperature of $T_d \sim 15\,$K. 
The same survey would reach a signal-to-noise ratio of 5 
in Q, U at 100$\, \mu$m down to $A_V \sim 1$. 
The entire survey of $\sim 500\,$deg$^2$ would require $\sim 1500\,$hr of telescope time, including overheads. 
It would provide key information on the magnetic field geometry for thousands of filamentary structures spanning $\sim 3$ orders of magnitude in column density 
from low-density subcritical filaments in the atomic (HI) medium at $A_V < 0.5$ to star-forming supercritical filaments in the dense inner parts 
of molecular clouds at $A_V > 100$.

\subsection{Key advantages of {\spicapol} over other polarimetric facilities}
\label{subsec:spica-adv} 
Far-IR/submillimeter polarimetric imaging from space with {\spicapol} will have unique advantages, especially in terms of spatial
dynamic range {\it and} surface brightness  dynamic range. 
Studying the multi-scale physics of star formation within molecular filaments requires a spatial dynamic range of $\sim 1000$  
or more to simultaneously probe scales $> 10\,$pc in the parent clouds down to $\sim 0.01\,$pc in the interior of star-forming filaments 
(corresponding to angular scales from $\sim 18$\arcsec $\,$to $> 5 \,$deg in the nearest molecular clouds -- see Fig.~\ref{taurus_planck}). 
Such a high spatial dynamic range was routinely achieved with {\it Herschel} in non-polarized imaging, 
but has never been obtained in ground-based submillimeter continuum observations. It will be achieved for the first time with {\spicapol} 
in polarized far-IR imaging. 

The angular resolution and surface brightness dynamic range of {\spicapol} will make it possible to resolve 0.1$\,$pc-wide filaments 
out to 350 pc 
and to image a few~\% polarized dust emission 
through the entire extent of nearby cloud complexes (cf. Fig.~\ref{taurus_planck}), 
from the low-density outer parts of molecular clouds ($A_V \simlt 0.5$) 
all the way to the densest filaments and cores ($A_V > 100$). 
In comparison, {\it Planck} had far too low resolution (10\arcmin $\,$ at best in polarization) to probe the magnetic field within 
dense 0.1 pc-wide filaments or detect faint 0.1 pc-wide striations. 
Near-infrared polarimetry cannot penetrate the dense inner parts of star-forming filaments, 
and ground-based or air-borne millimeter/submillimeter polarimetric instruments, such as 
SCUBA2-POL (the polarimeter for the Sub-millimetre Common-User Bolometer Array 2 -- \citealp{Friberg+2016}), 
NIKA2-POL (the polarization channel for the New IRAM KID Arrays 2 -- cf. \citealp{Adam+2018}), HAWC$+$ (the High-resolution Airborne Wideband Camera-Plus 
for SOFIA, the Stratospheric Observatory for Infrared Astronomy -- \citealp{Harper+2018}), or 
BLAST-TNG (the next generation Balloon-borne Large Aperture Submillimeter Telescope for Polarimetry -- \citealp{Galitzki+2014}),  
will lack the required sensitivity and dynamic range in both spatial scales and intensity.

\begin{figure}
\begin{center}
\includegraphics[width=\columnwidth]{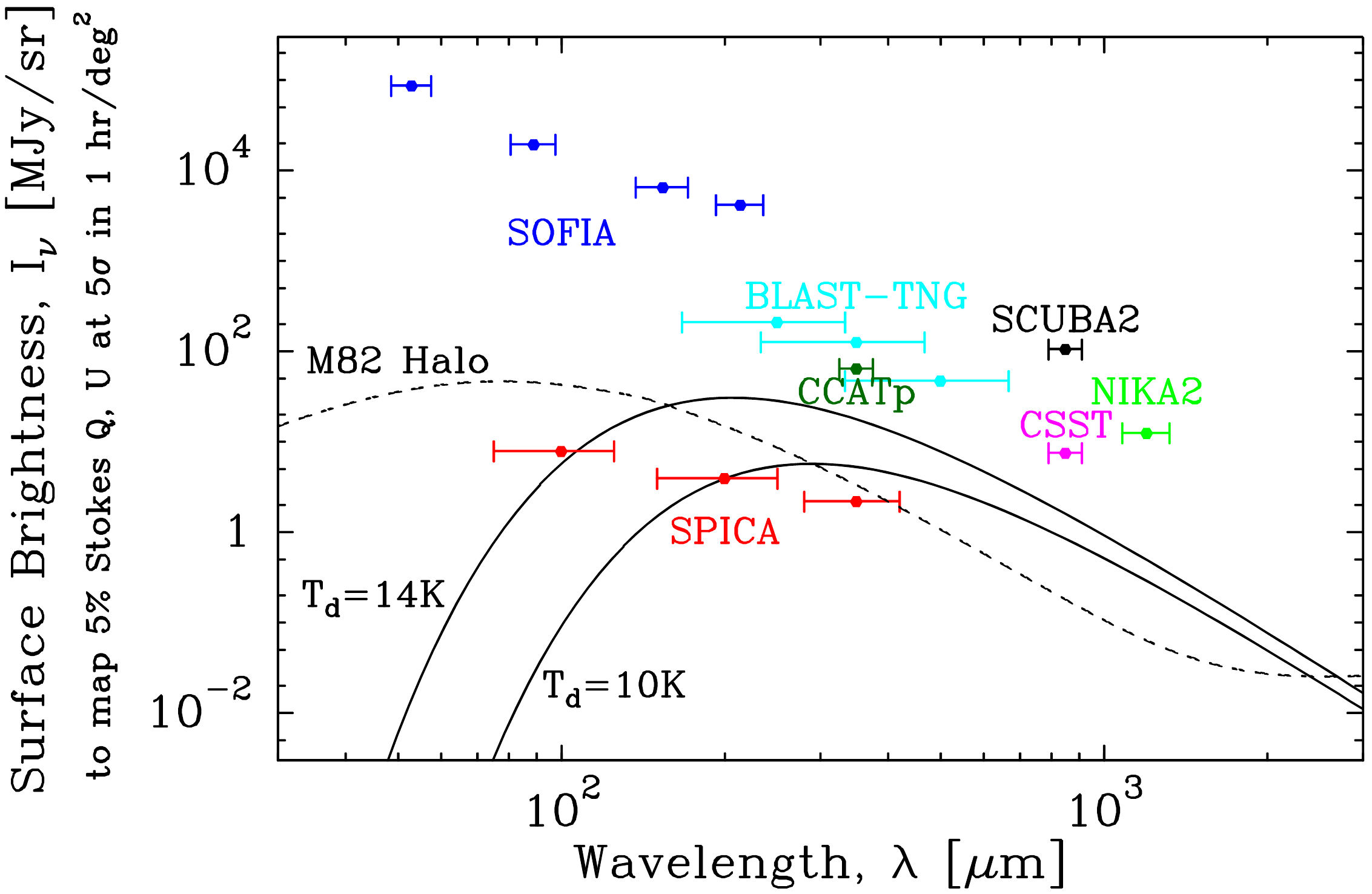}
\caption{Surface-brightness sensitivity of {\spicapol} for wide-field polarimetric imaging compared to other existing or planned polarimetric facilities. 
The total surface-brightness level required to detect polarization (i.e., Stokes parameters Q, U) with a signal-to-noise ratio 
of 7 per resolution element (e.g. $9\arcsec $ pixel at $200\mu $m for {\spicapol}) when mapping 1~deg$^2$ in 2~hr assuming 5\% fractional polarization 
is plotted as a function of wavelength for each instrument (SOFIA-HAWC$+$, {\spicapol}, BLAST-TNG, CCAT-p, SCUBA2-POL, CSST,  NIKA2-POL). 
For comparison, the typical surface-brightness level expected in total intensity from the diffuse outer parts of molecular clouds ($A_V = 1$) is shown   
for two representative dust temperatures ($T_d = 10\,$K and  $T_d = 14\,$K, assuming simple modified blackbody emission 
with a dust emissivity index $\beta = 2$), as well as the SED of the halo of the nearby galaxy M82 \citep[cf.][]{Galliano+2008,Roussel+2010}.
} 
 \label{spica_sensitivity}
\end{center}
\end{figure}

More specifically, BLAST-Pol, the Balloon-borne Large Aperture Submillimeter Telescope for Polarimetry operating at 250, 350, and 500 $\mu$m \citep[][]{Fissel+2010, Fissel+2016}, 
has only modest resolution (30\arcsec --1\arcmin), sensitivity, and dynamic range. 
HAWC$+$, the far-infrared camera and polarimeter for SOFIA \citep{Dowell+2010}; 
and 
BLAST-TNG  \citep[cf.][]{Dober+2014},
both benefit 
from a larger 2.5-m primary mirror equivalent to that of SPICA and thus have comparable angular resolution, but are not cooled and therefore are 
two to three orders of magnitude less sensitive 
(Noise Equivalent Power ${\rm NEP} > 10^{-16}\, {\rm W\, Hz}^{-1/2}$) than {\spicapol}. 
Stated another way, the mapping speed\footnote{The mapping speed is defined as the surface area that can be imaged to a given sensitivity level in a given observing time.}
of {\spicapol} will be four to five 
orders of magnitude higher than that of HAWC$+$ or BLAST-TNG.
%
Future ground-based submillimeter telescopes on high, dry sites such as CCAT-p (the Cerro Chajnantor Atacama Telescope, prime) 
and CSST (the Chajnantor Submillimeter Survey Telescope)
will benefit from larger aperture sizes (6$\,$m and 30$\,$m, respectively) 
and will thus achieve higher angular resolution than SPICA at 350$\, \mu$m, but will be limited in sensitivity by the atmospheric background load on the detectors 
and in spatial dynamic range by the need to remove atmospheric fluctuations. 
The performance and advantage of {\spicapol} over other instruments for wide-field dust polarimetric imaging 
are illustrated in Fig.~\ref{spica_sensitivity}.

Dust polarimetric imaging with ALMA at $\lambda \sim \,$0.8--3$\,$mm will provide excellent sensitivity and resolution, but only on 
small angular scales (from $\sim \,$0.02\arcsec $\,$ to $\sim 20\arcsec $).  
Indeed, even with additional observations with ACA (the ALMA Compact Array),  
the maximum angular scale recoverable by the ALMA interferometer remains smaller 
than $\sim \,$1\arcmin $\,$ in total intensity and $\sim \, 20\arcsec $ in polarized emission 
(see ALMA technical handbook)\footnote{Note that total power ALMA data can only be obtained for spectral line observations 
and are not possible for continuum observations.}.
This implies that ALMA polarimetry is intrinsically insensitive to all angular scales $> 20\arcsec $, 
corresponding to structures larger than 0.015--0.05$\,$pc in nearby clouds.  
Using multi-configuration imaging, ALMA can achieve a spatial dynamic range of $\sim 1000$, comparable to that of {\it Herschel} or {\spicapol}, 
but only for relatively high surface brightness emission. 
Because ALMA can only image the sky at high resolution, 
it is indeed $\sim \,$2--3 orders of magnitude less sensitive to low surface brightness emission than a cooled single-dish space telescope such as SPICA.
Expressed in terms of column density, this means that ALMA can only produce polarized dust continuum images of compact objects 
with $N_{H_2} \simgt 10^{23}\, {\rm cm}^{-2}$ 
(such as substructure in distant, massive supercritical filaments -- see \citealp{Beuther+2018})   
at significantly higher resolution  ($\sim 1$\arcsec $\,$ or better) than SPICA, while polarimetric imaging of extended, low column density structures down 
to $N_{H_2} \simgt 5 \times 10^{20}\, {\rm cm}^{-2}$ (such as subcritical filaments and striations) will be possible with {\spicapol}. 
Furthermore,  the small size of the primary beam ($\sim \,$0.3\arcmin-1\arcmin $\,$at 0.8--3$\,$mm) makes mosaicing of wide ($> 1\, {\rm deg}^2$) fields impractical and prohibitive with ALMA. 
In practice, ALMA polarimetric studies of star-forming molecular clouds will provide invaluable insight into the role of magnetic fields within individual protostellar cores/disks 
and will be very complementary to, but will not compete with, the {\spicapol} observations discussed here which target the role of magnetic fields in the formation and evolution 
of filaments on larger scales.

\begin{figure*} 
\begin{center}
\includegraphics[width=7.5cm]{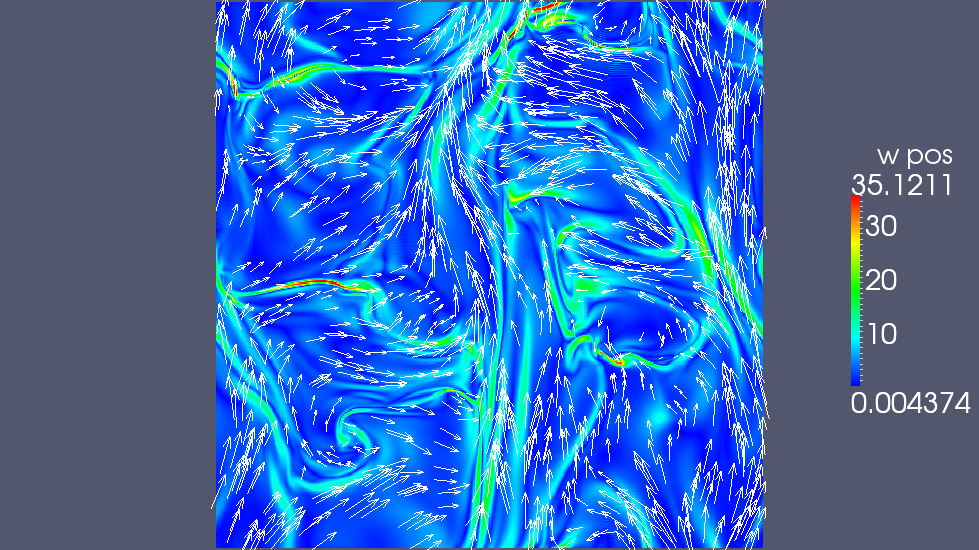}
\includegraphics[width=7.5cm]{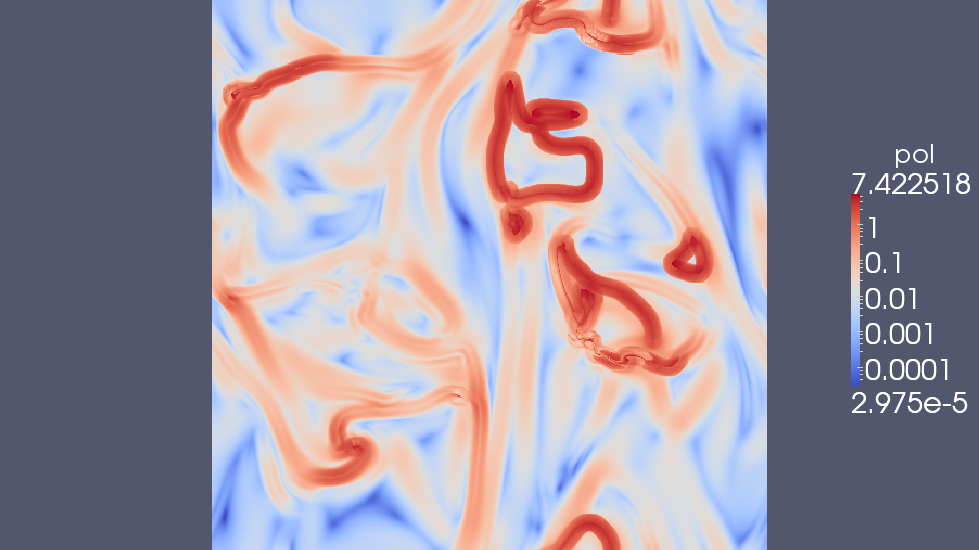}
\end{center}
\caption{
Coherent structures in simulations of 3D decaying incompressible MHD turbulence from \citet{Momferratos14}. 
These simulations resolve the dissipation scales of turbulence; they characterize the morphology of magnetic structures 
formed in magnetized turbulence. For parameters typical of diffuse molecular clouds, the box size is $\approx$ 1 pc.
Left: Projections  of the vorticity (the modulus in color) and of the magnetic field (arrows) on the plane of the sky. 
Right: Small-scale increments of the orientation of the plane of the sky component of the magnetic field, a proxy for the dust polarization angle gradient. 
Figure adapted from \citet{Falgarone15}. } 
\label{fig:simus_intermittency}
\end{figure*}

\section{The turbulent magnetized interstellar medium}
\label{sec:turbulence}

Magnetic fields and turbulence are central to the dynamics and energetics of gas in galactic disks, but also in their halos and possibly in the cosmic web at large.  
These two intertwined actors of cosmic evolution coupled to gravity drive the formation of coherent structures from the warm and hot tenuous gas phases to the onset of star formation in molecular clouds. 

Reaching a statistical description of turbulence in the magnetized ISM 
is an outstanding challenge, because its extreme characteristics may not be reproduced
in laboratory experiments nor in numerical simulations. This challenge is of fundamental importance to Astrophysics, in particular to understand 
how galaxies and stars form, as well as the chemical evolution of matter in space.  This section summarizes 
the contribution we expect {\spicapol} to bring to this ambitious endeavor.

\subsection{Interstellar magnetic fields}
\label{subsec:turbulence_polarization}

Magnetic fields pervade the multi-phase ISM of galaxies. 
In the Milky Way, and more generally in local universe galaxies, 
the ordered (mean) and turbulent (random) components of interstellar magnetic fields are comparable and in near equipartition
with turbulent kinetic energy \citep{Heiles05,Beck2015}. The galactic dynamo amplification is saturated, but exchanges between gas kinetic and magnetic energy 
still occur and are of major importance to gas dynamics. Magnetic fields are involved in the driving of turbulence and in the turbulent energy cascade \citep{Subramanian08}.  
The two facets of interstellar turbulence: gas kinematics 
and magnetic fields, are dynamically so intertwined that they may not be studied independently of each other.

The multiphase magnetized ISM is far too complex to be described by an analytic theory. Our understanding in 
this research field follows from observations, MHD simulations and phenomenological models.
MHD simulations allow us to quantify the non-linear ISM physics but within numerical constraints that limit their scope. 
They may guide the interpretation observations but alone they do not provide conclusive answers
because they are very far from reproducing the high Reynolds ($R_e$) and magnetic Prandtl numbers  ($P_m$)\footnote{The magnetic Prandtl number is the ratio between the kinetic viscosity and the magnetic diffusivity.} of interstellar turbulence \citep{Kritsuk11}. 
The fluctuation dynamo and shock waves contribute to produce highly intermittent magnetic fields where the field strength is enhanced in localized magnetic structures.
The volume filling factor of these structures decreseases for increasing values of the magnetic Reynolds number $R_m = R_e\times P_m$ \citep{Schekochihin02,Brandenburg05}. 
The inhomegeneity in the degree of magnetization of matter associated with intermittency is an essential facet of interstellar turbulence \citep{Falgarone15,Nixon19}, which simulations miss
because they are far from reproducing the interstellar values of $R_m$. 
In this context, to make headway, we must follow an empirical approach where a statistical model of interstellar turbulence is inferred from observations. 

\subsection{The promise of {\spicapol}}
\label{subsec:turbulence_promises}

{\spicapol} will image dust polarization with an unprecedented combination of sensitivity and angular resolution, providing a unique data set (Sect.~\ref{subsec:spica-adv}) to characterize the magnetic facet of interstellar turbulence. 
This leap forward 
will open an immense discovery space, which
will revolutionize our understanding of interstellar magnetic fields, and their correlation with matter and gas kinematics. 

At $200\,\mu$m, 
for the SED of the diffuse ISM \citep{planck2013-XVII}, the surface brightness sensitivity of {\spicapol}  is three orders of magnitude greater than that 
of the {\it Planck} $353\,$ GHz all-sky map for deep imaging (10~hr per square degree) and a few hundred times better for faster mapping (2~hr per square degree). 
The analysis of dust polarization 
at high Galactic latitude with the {\it Planck} data is limited by sensitivity to an effective angular resolution of $\sim \,1^\circ$ in the diffuse ISM and $10\arcmin$ 
in molecular clouds where the column density is larger than $10^{22}\,$H\,cm$^{-2}$ \citep{PlanckXII2018},  while {\spicapol}  will map dust polarization with a factor $\sim \,$20--70 better resolution 
at 100--350$\, \mu$m. 

Dust polarization probes the magnetic field orientation in dust-containing regions, i.e., mostly in the cold and warm phases of the ISM, which 
account for the bulk of the gas mass, and hence of the dust mass. These ISM phases comprise the diffuse ISM and  star-forming molecular clouds. They 
account for most of the gas turbulent kinetic energy in galaxies \citep{HennebelleFalgarone2012}. 
Thus, among the various means available to map the structure of interstellar magnetic fields, dust polarization is best suited to 
trace the dynamical coupling between magnetic fields, turbulence, and gravity in the ISM. This interplay is pivotal to ISM physics and star formation. 
It is also central to cosmic magnetism because it underlies dynamo processes  \citep{Subramanian08}.

Observations have so far taught us that magnetic fields are correlated with the structure of matter in both the diffuse ISM and in molecular clouds \citep{Clark+2014,PlanckXXXII2016,PlanckXXXV2016}
but this correlation does not fully describe interstellar magnetism. Data must also be used to characterize the intermittent nature of interstellar magnetic fields. 

The magnetic structures identified in MHD simulations (Fig.~\ref{fig:simus_intermittency})
may be described as filaments, ribbons or sheets with at least 
one dimension commensurate with dissipation scales of turbulent and magnetic energy in shocks, current sheets or through ambipolar diffusion \citep{Momferratos14,Falgarone15}.  
While the viscous and Ohmic dissipation scales of turbulence are too small to be resolved by {\spicapol}, 
turbulence dissipation due to ion-neutral friction is expected to occur on typical scales  between $\sim 0.03\,$pc and $\sim 0.3\,$pc \citep[cf.][]{Momferratos14}, 
which is well within the reach of {\spicapol} for matter in the local ISM. 

Although dust polarization does not measure the field strength, the polarization angle  may be used to map these magnetic structures, as illustrated in Fig.~\ref{fig:simus_intermittency}. 
The figure shows that the largest values of the increment of the polarization angle, $\Delta \Phi$, 
delineate structures that tend to follow those of intense dissipation of turbulent energy.
{\spicapol} will allow us to identify magnetic structures such as those in Fig.~\ref{fig:simus_intermittency} 
even if their transverse size is unresolved because i) they are highly elongated  and ii) their spatial distribution in the ISM is fractal.

Regions of intermittency in interstellar turbulence correspond to rare events. Their finding requires obtaining large data sets 
combining brightness sensitivity and angular resolution, as illustrated by the CO observations with the Institut de Radioastronomie Millim\'etrique (IRAM)  
30m telescope analyzed by \citet{Hily-Blant08} and \citet{Hily-Blant09}. {\spicapol} has the unique capability to extend these pioneering studies of the 
intermittency of gas kinematics to dust polarization observations, tracing the structure of 
magnetic fields (Sect.~\ref{subsubsec:polarimetry}), with a comparable angular resolution. {\spicapol} will also greatly strengthen their statistical significance  by covering a total sky area more than two orders of magnitude larger.  

\begin{figure}
\begin{center}
\includegraphics[width=8cm]{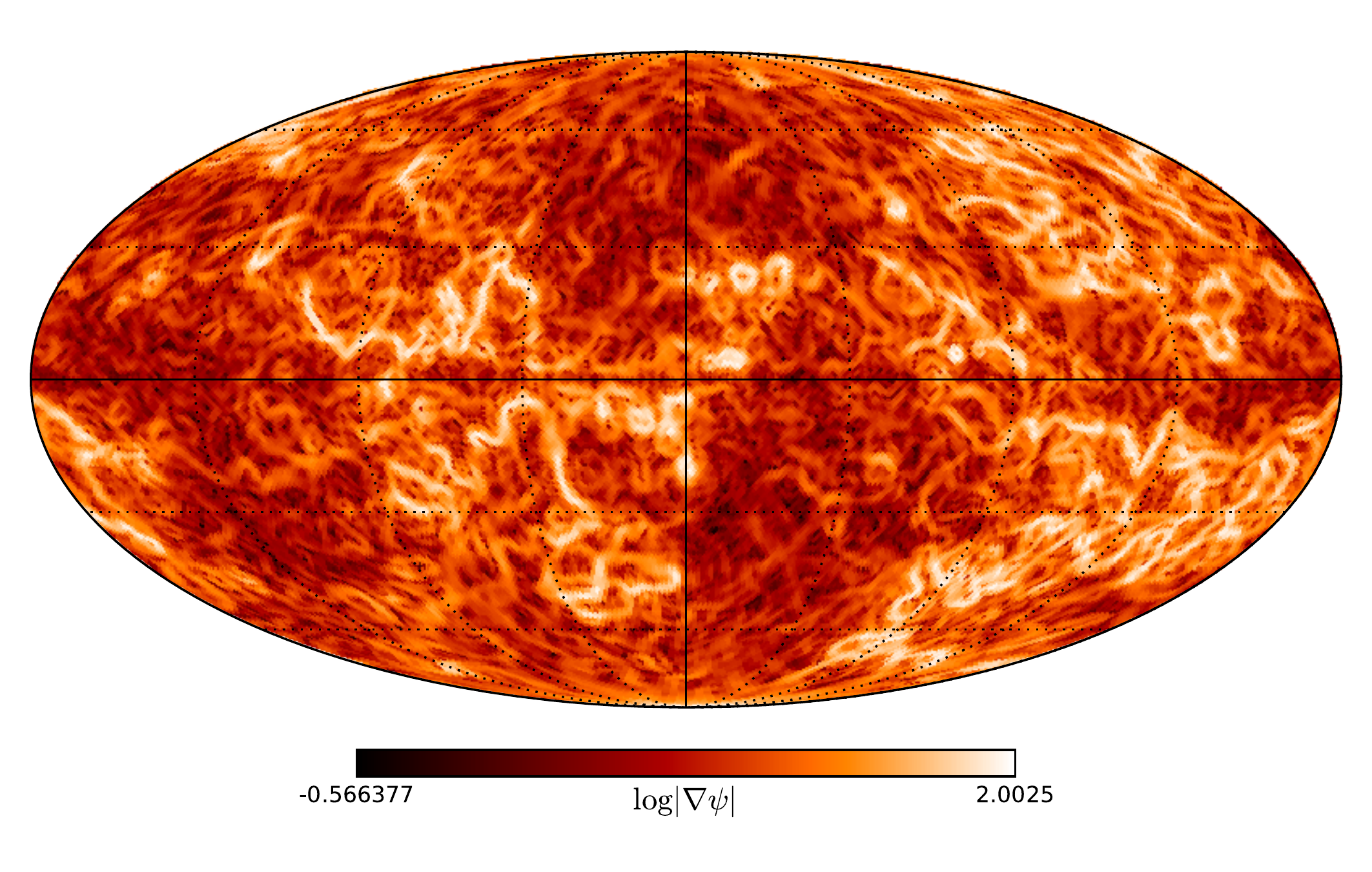}
 \caption{Non-Gaussianity of the magnetic field structure in the {\it Planck} dust polarization data.
This all-sky image, in Galactic coordinates centered on the Galactic center, presents the modulus of the angular polarization gradient, $|\nabla{\psi}|$, 
built from the {\it Planck} data at 353\,GHz smoothed to 160\arcmin\ resolution. 
Figure adapted from Appendix D of \citet{PlanckXII2018}. 
}
\label{fig:planck_intermittency}
\end{center}
\end{figure}

{\spicapol}  holds promises to reveal 
a rich array of magnetic structures, characterizing  the intermittency of interstellar magnetic fields. {\it Planck} data, at a much coarser scale, gives a first insight at the expected outcome of the observations illustrated in Fig.~\ref{fig:planck_intermittency}. 
Magnetic structures will be identified in the data as locations 
where the probability distributions of the increments of the polarization angle, and the Stokes Q/I and U/I ratios, depart from Gaussian distributions.
Compared to {\it Planck}, {\spicapol} will only map a small fraction of the sky ($\sim 1 \%$ for nearby molecular clouds and diffuse ISM observed away from the Galactic plane) but it will probe 
the field structure on much smaller scales (by a factor $30$ or more)
where the surface density of magnetic structures 
is expected to be much larger.

\subsection{Observing strategy}
\label{subsec:observations}

{\spicapol} will considerably expand our ability to map the structure of interstellar magnetic fields.
These  data will be complementary to a diverse array of polarization observations of the Galaxy. 

Stellar polarization surveys will be 
combined with {\it Gaia} astrometry \citep[e.g.][]{Tassis+2018} to build 3D maps of the magnetic fields in the Galaxy but with a rather coarse resolution, comparable to that of the density structure of the local ISM in \citet{Lallement18}. 

Synchrotron observations at radio wavelengths 
with the Square Kilometer Array (SKA) and its precursors will probe the structure of magnetic fields \citep[][]{Dickinson15,Haverkorn15}, in particular in ionized phases through Faraday rotation \citep[][]{Gaensler11,Zaroubi15}.  
SKA will also provide Faraday rotation measurements toward $\sim 10^7$ extragalactic sources \citep{Johnston-Hollitt15}, which will be available for comparison with {\spicapol} dust polarization data as illustrated in the pioneering study of \citet{Tahani+2018}. 

{\spicapol} will allow us to study interstellar turbulence over an impressive range of physical scales and astrophysical environments from the warm ISM phases to molecular clouds. 
Observations of nearby galaxies are best suited to probe the driving of turbulence in relation to galaxy dynamics (spiral structure, bars, galaxy interaction, outflows) 
and stellar feedback as discussed in Sect.~\ref{sec:galaxies}. Galactic observations will probe the inertial range of turbulence over 4 to 5 orders of magnitude 
from the injection scales ($\sim 100\,$pc - $1\,$kpc) down to $0.01\,$pc. The smallest physical scales will be reached by observing interstellar matter nearest the Sun, away from the 
Galactic plane: the diffuse ISM at high Galactic latitudes and star-forming molecular clouds in the Gould Belt. These sky regions are best suited for the study of turbulence because
the overlap of structures along the line of sight is minimized. The Gould Belt clouds are already part of the filament science case in Sect.~\ref{sec:filaments}. 
This survey will include star-forming clouds and diffuse clouds representative of the cold neutral medium.
Deeper polarimetric imaging of high Galactic latitude fields (10 hr per deg$^2$),  sampling regions of low
gas column density ($\rm A_V \sim 0.1$ to 0.3), will allow us to probe turbulence in the warm ISM phases. 
These deep imaging observations could potentially share the same 
fields 
as those used to carry out a SPICA-SMI cosmological survey.  
The size of the area that may be mapped to that depth (of order $\sim 100\,$deg$^2$) will be
optimized with the needs of this survey.  
Altogether, we estimate that {\spicapol} will cover a total area of about 500 deg$^2$ away from the Galactic Plane 
including diffuse ISM fields at high Galactic latitudes, 
which will be available to study turbulence in diverse interstellar environments.  
At the angular resolution of {\spicapol}  at $200\,\mu$m, 
these data correspond to a total of $2 \times 10^7$ polarization measurements. 
This number is 20 times larger than the statistics offered by the {\it Planck} 
polarization data. The gain in angular resolution and sensitivity is so large, that {\spicapol} will supersede {\it Planck} in terms of data statistics, 
even if the maps used cover only $\sim 1\%$ of the sky. 

A wealth of spectroscopic observations of HI and molecular gas species, tracing the gas density, column density and kinematics, 
will become available before the launch of {\spicapol} with SKA and its precursors \citep{McClure-Griffiths15}, and the advent of powerful 
heterodyne arrays on millimeter ground-based telescopes, e.g. the Large Millimeter Telescope and the IRAM 30m telescope.
Furthermore, we will be able to investigate the link between coherent magnetic structures and turbulent energy dissipation observing 
main ISM cooling lines from H$_2$, CII, and OI with the SPICA mid and far-IR spectrometers SMI and SAFARI. 
These complementary data from SPICA and ground-based observatories
will be combined  to characterize the turbulent magnetized ISM statistically. 
The data analysis will rely on on-going progress in the development of statistical methods \citep[e.g.][]{Makarenko18}, which we will use to characterize the structure of interstellar magnetic fields and their correlation with gas density and velocity. 
This process will converge toward an empirical model of interstellar turbulence, which will be related to ISM physics comparing data and MHD simulations.

\section{Magnetic fields in protostellar dense cores}
\label{sec:protostars} 

\subsection{Current state of the art}

In molecular clouds, protostellar dense cores are the ``seeds'' where the gravitational force proceeds to form stars. 
Class~0 objects are the youngest known accreting protostars: most of their mass is still in the form of a dense core/envelope ($M_{\rm env}  \gg M_\star $) 
and this phase is characterized by high accretion rates of gas from the dense core onto the central stellar object, accompanied by ejection of powerful highly 
collimated flows \citep{Andre00, Dunham+2014}. 

How many stars can be formed out of a typical molecular cloud depends not only on the physical conditions in filamentary structures, 
but also on the detailed manner the gravitational collapse proceeds within individual protostellar cores (i.e. would it form a single/binary stellar system, one or several low-mass stars or high-mass stars?).  

By the end of the protostellar phase, the star has gained most of its final mass: understanding the role of magnetic fields during the protostellar stage 
is therefore crucial to clarify 
how they affect some of  
the most remarkable features of the star formation process, 
such as the distribution of stellar masses, the stellar multiplicity, or the ability to host planet-forming disks \citep{McKee07, LiPPVI}. 

The development of numerical magneto-hydrodynamical models describing the collapse of protostellar cores and the formation of low-mass stars, 
has opened new ways to explore in more details the physical processes responsible for 
the formation of solar-type stars.  
MHD models suggest that protostellar collapse proceeding with initially strong and well aligned magnetic field produces significantly different outcomes than hydrodynamical or weakly magnetized models \citep{Fiedler93,Hennebelle08a,Masson16}.
For example, if the field is strong enough and well coupled to the core material, magnetic braking will regulate the formation of disks and multiple systems during the Class~0 phase. 
This has been the focus of recent studies \citep[][]{Hennebelle16,Krasnopolsky11,Machida11} because it could potentially explain the low-end of the size distribution 
of protostellar disks \citep[e.g.][]{Maury10, SeguraCox18, Maury18b}.

\begin{figure}
\begin{center}
\includegraphics[width=\linewidth]{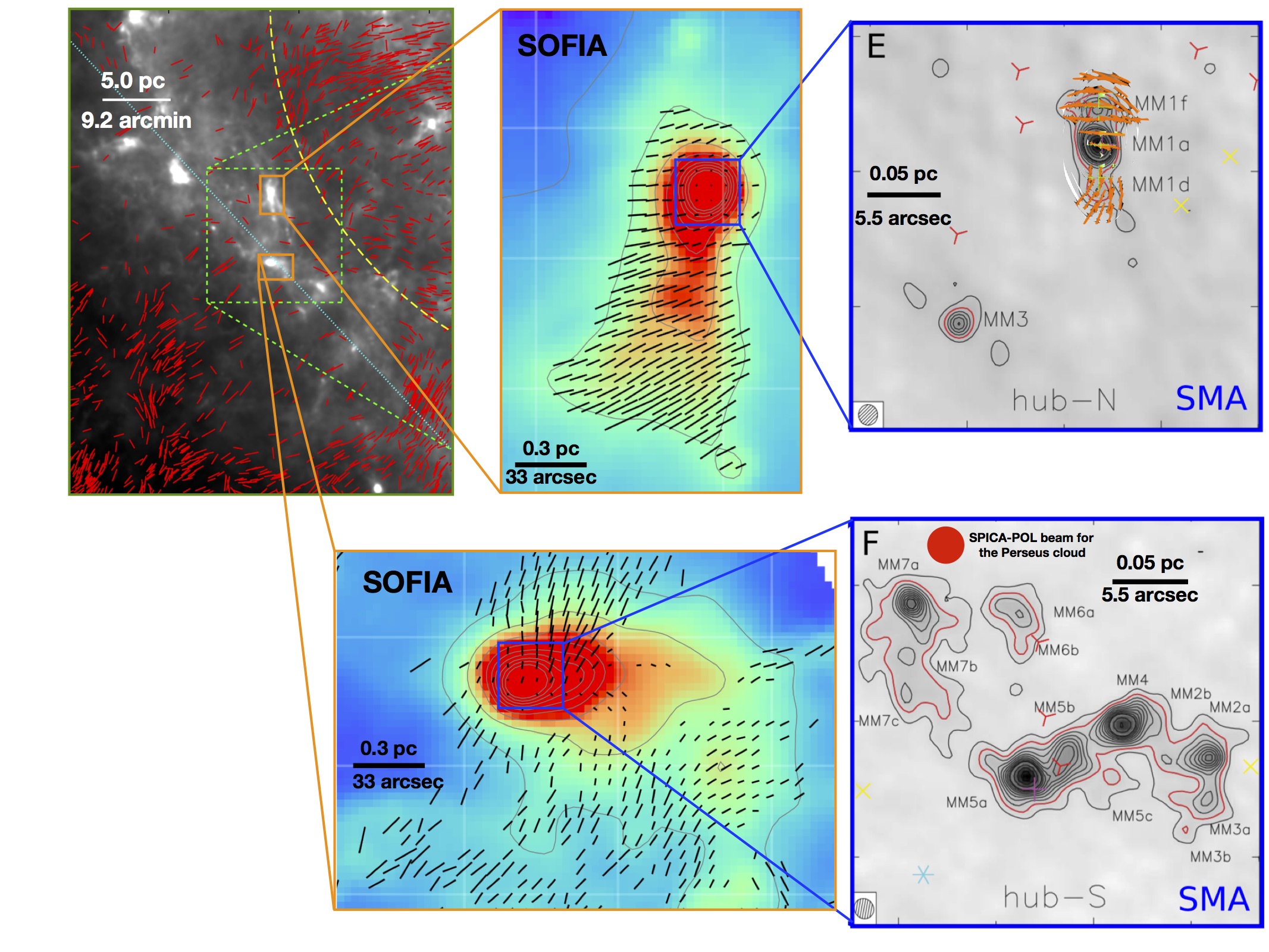}
\caption{Composite images of the G14.225-0.506 massive star forming region \citep{Busquet13, Busquet16, Santos16}. 
{\it Top left panel:} R band optical polarization vectors (red segments) overlaid on {\it Herschel}  250~$\mu$m image overlapped  (from Santos et al. 2016). 
{\it Central panels:} SOFIA/HAWC+ 200~$\mu$m  images (beam $14\arcsec$) of the Northern (top) and Southern (bottom) hubs, with black segments 
showing the magnetic field direction (F. Santos, private communication).
{\it Right panels:} Submillimeter Array (SMA) images of the 1.2~mm emission toward the  center of the Northern (top) and Southern (bottom) hubs  \citep{Busquet16}, 
with orange segments showing the magnetic field direction (A\~nez et al. in preparation). 
}
 \label{Fig_G14}
\end{center}
\end{figure}

All protostellar cores are magnetized to some level and current observations  suggest that at least in some cases the magnetic field at core scale is remarkably well organized, pointing toward scenarii with strong field even at the high column densities typical of protostellar cores \citep[e.g., IRAS 4A, G31.41, G240.31, NGC 6334, L1157, B335:][]{Girart06, Girart09, Qiu14, Li-HB15, Stephens13, Galametz18, Maury18a}, while in other cases (e.g., \citealt{Girart13, Hull17b, Ching17}, and Fig.~\ref{Fig_G14} for an example in the G14 massive star forming region) the core-scale magnetic field shows very complex morphology. 
In Fig.~\ref{Fig_G14}, for instance, it is noteworthy that 
the northern hub of the G14 region, with a more uniform magnetic field, has a lower level of fragmentation than the southern hub (that shows a more perturbed magnetic field). 
These observations suggest the field may remain organized at scales where collapse occurs in most solar-type progenitors, and also at least some of the massive cores \citep{Zhang14, Beuther+2018}. 
Current results may be biased, however, because present single-dish facilities selectively trace magnetic fields from the brightest regions within star-forming cores 
(dust polarization is only detected at the highest column densities, see Fig.~\ref{Fig_G14}).


\subsection{Role of magnetic fields in controlling the typical outcome of protostellar collapse}


{\spicapol} can perform 
statistical studies in unprecedentedly large samples of protostellar cores, testing for example whether the magnetic field in cores is directly inherited from their environment (if the magnetic field in low-density filamentary structures, see \S ~\ref{subsec:filaments}, connects to the magnetic field in high-density cores, which is expected in the strong field case), or if the field in cores is disconnected from the local field in the progenitor cloud (weak field case). 
While the ALMA interferometer can only provide constraints on the magnetic field topology at the smallest scales ($\sim \,$0.02\arcsec $\,$ to $\sim \,$20\arcsec, i.e. $<5000$ au in Gould Belt clouds), the polarization capabilities of other facilities probing larger spatial scales (SMA, NIKA2, and HAWC+) are severely sensitivity limited. 
Accordingly,  studies linking cores and filaments can currently be carried out in bright, massive star-forming regions mostly (see Fig.~\ref{Fig_G14}), 
and only in a handful of nearby solar-type protostellar cores. 

\begin{figure}[ht]
\begin{center}
\includegraphics[width=0.9\linewidth]{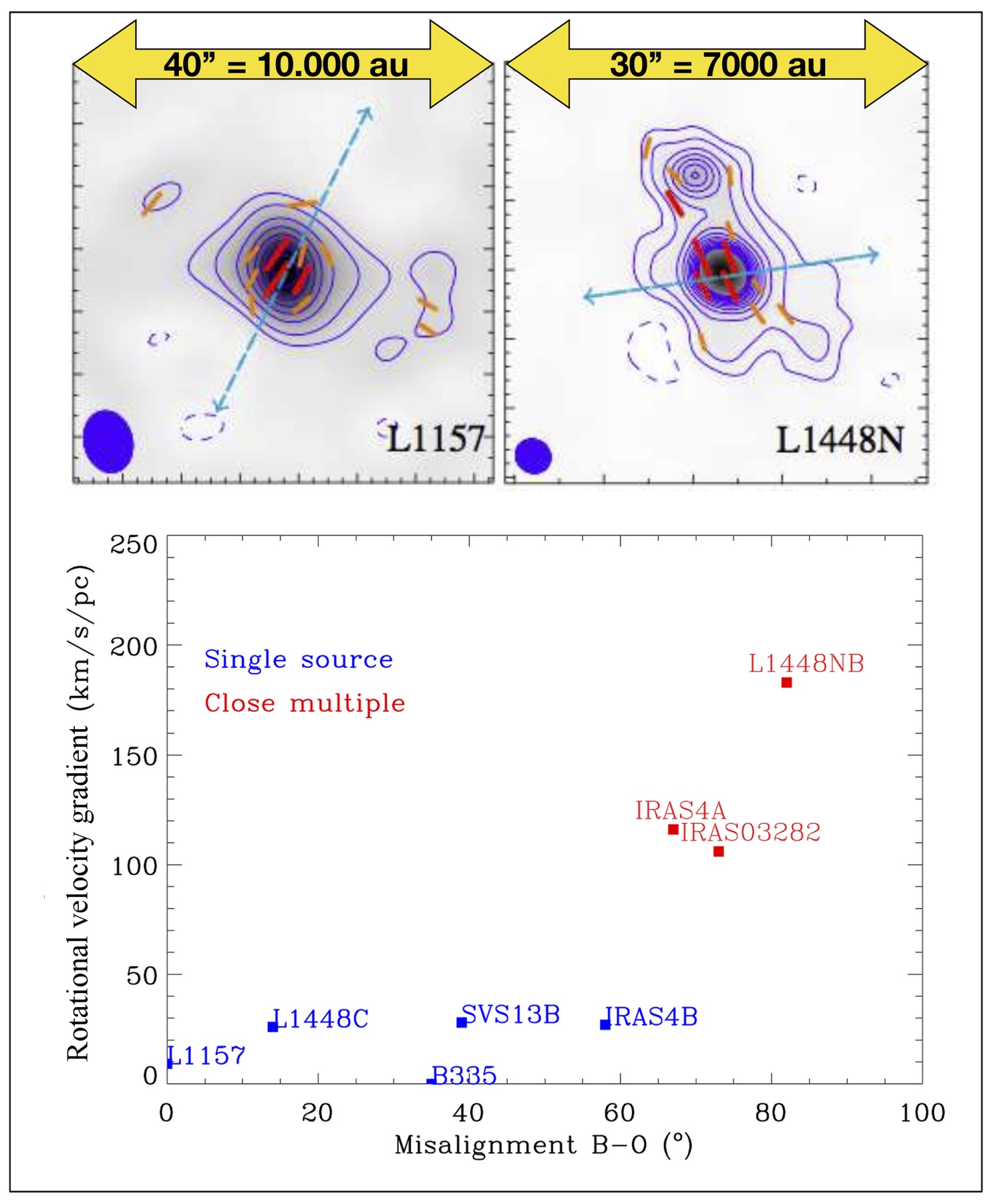}
\caption{Potential role of the magnetic field topology at core scales in the formation of disks and multiple systems. 
{\it Top:} 
Magnetic field (red/orange line segments, from dust polarization observations with the SMA at 850$\,\mu$m) in two solar-type Class~0 
protostellar cores \citep{Galametz18}. 
The blue arrows indicate the jet/rotation axis of these cores,  aligned with the core-scale magnetic field in L1157 (left), 
and mostly orthogonal to it in L1448N (right). 
{\it Bottom:} Level of core rotation 
(from kinematic observations at core scales, \citealt{Yen15a} and Gaudel et al. in prep) 
as a function of the misalignment between the rotation axis and the magnetic field (observed 
at core scale with the SMA -- \citealt{Galametz18}) in a sample of nearby Class~0 sources. 
There is a hint that large 
misalignments of the magnetic field at core scales  
lead to sources with large rotational gradients and multiple systems at smaller scales (red symbols).
}
 \label{Fig_SMA}
\end{center}
\end{figure}

Some indications have been found, in small samples ($<20$ objects), that the topology of the magnetic field at core scales 
may be linked to the distribution of angular momentum in solar-type cores, and hence that the magnetic field may be of paramount importance 
to set the initial conditions for the formation 
of protoplanetary disks and multiple systems 
(see Fig. \ref{Fig_SMA} and \citealt{Galametz18, SeguraCox18}).
An example of the type of studies that {\spicapol} could extend to the full mass function of protostellar cores in a statistical fashion is shown in Fig.~\ref{Fig_G14} and Fig.~\ref{Fig_SMA}: 
these two figures illustrate the tentative link between the magnetic field topology at core scale and the disks and multiplicity fraction found {\it within} protostellar cores at smaller scale.
The magnetic field properties found with {\spicapol} at dense core scales can be compared with 
the protostellar properties observed with interferometers at smaller scales
to build correlation diagrams similar to the one shown in Fig.~\ref{Fig_SMA}.
In this way, {\spicapol} observations 
can test the hypothesis, tentatively suggested by current studies of the brightest protostars, that magnetic fields 
regulate the formation of disks and multiple systems during the main accretion phase. 
Observations of 
large samples of protostars 
could be carried out thanks to the sensitivity and spatial resolution of  {\spicapol}, 
which is crucially needed not only to populate diagrams such as Fig.~\ref{Fig_SMA}, but also because only statistics will allow us to solve the degeneracy 
induced by projection effects 
intrinsically linked to dust polarization (tracing only the magnetic field component in the plane of the sky).
Moreover, {\spicapol} will provide information on the geometry of magnetic field lines 
across the full protostellar core mass distribution, 
probing different behaviors in different mass regimes, and potentially as a function of environment in different star-forming regions. 


The angular resolution and surface brightness dynamic range of {\spicapol} will make it possible to resolve most $\sim \,$2000--20000~au
protostellar cores in nearby star-forming regions out to 250 pc.
A wide-field {\spicapol} survey of all nearby clouds as envisaged in \S ~\ref{subsec:filaments} ($\sim 2\,$hr per square degree) will map dust polarization (fraction $>1\%$) 
at core scales with signal-to-noise ratio $>7$, in complete populations of $\simgt$1000 protostars (Class~0 and Class~I) and their parent cloud/environment,  
from massive protostellar cores down to the low-mass progenitors of solar-type stars. 
In contrast, current millimeter/submillimeter polarimetric instruments, such as SCUBA2-POL, NIKA2-POL, SOFIA/HAWC$+$, or BLAST-TNG (cf. \S ~\ref{subsec:spica-adv})
are limited to the subset of the $\sim$50-100 brightest cores, 
and without the important context provided by the magnetic field information in the parental clouds. 

%

\section{Role of magnetic fields in high-mass star and cluster formation}
\label{sec:massive-sf} 

As mentioned in \S ~\ref{subsec:fil-paradigm}, supercritical molecular filaments are believed to be the preferred birthplaces of solar-type stars.
It is however unclear 
whether the filamentary paradigm -- or an extension of it,  based on unusually high line masses or levels of turbulence \citep[e.g.][]{Roy+2015}  -- also applies to high-mass star formation 
and can lead to the quasi-static formation and monolithic collapse of massive prestellar 
cores,  then forming high-mass protostars. 
The most recent observational results, partly obtained with {\it Herschel}, suggest that high-mass stars and stellar 
clusters form in denser, more dynamical filamentary structures called ridges\footnote{By ridge, we do not mean the crest of a given filament 
but a massive elongated structure ($> 1000\, M_\odot $ of dense molecular gas with $n_{H_2} > 10^5\, {\rm cm}^{-3}$), 
consisting of a dominant  highly supercritical filament and an accompanying network of sub-filaments, often themselves supercritical.} 
which exceed the critical line mass of an isothermal filament by up to  $\sim $ two orders of magnitude (see, e.g., the review by \citealp{Motte+2018a}). 
Such 
massive structures may originate from highly dynamical events at large scales like converging flows and cloud-cloud collisions, continuing on median scales 
through the global collapse and filament feeding of ridges. 
The role of magnetic fields in this scheme is poorly known and 
may be as crucial as for low-mass star formation.

In the hypothesis 
of large-scale cloud collapse, 
gravity overcomes the magnetic field support and the magnetic field follows the infall gas streams from cloud scales ($\sim$100~pc) to accumulation points 
at scales between $\sim$1~pc to $\sim$0.1~pc, with a typical hourglass geometry toward these accumulation points \citep[][]{Girart09,Cortes+2016}.
Large-scale collapse leads to very dense, massive structures at pc scales, which are either spherical (hubs) or 
elongated (ridges) \citep[][]{HartmannBurkert2007,Schneider+2010,Hill+2011,Peretto+2013}.
Pilot works with ground-based facilities (SMA) toward the DR21 ridge in Cygnus~X \citep[][]{Zhang14,Ching17} 
show that the magnetic field is ordered at the scale of the ridge and mostly perpendicular to its main axis, as for 
low-mass supercritical filaments (cf. \S ~\ref{subsec:fil-paradigm} and Fig.~\ref{taurus_planck}), 
suggesting mass accumulation along field lines.
However, while large-scale collapse and strong ordered magnetic fields are probably a key ingredient, 
the detailed physical processes at the origin of ridges remain, for now, a mystery. 

At some point, ridges fragment into hundreds of protostellar cores in local, short, but violent bursts of star formation, 
leading to exceptionally large instantaneous star formation rates \citep[][]{NguyenLuong+2011, Louvet+2014}.
This clustered mode of fragmentation in ridges may differ in nature 
from the filamentary mode of fragmentation leading to low-mass star formation at significantly lower average densities. 
As a matter of fact, top-heavy core mass functions, overpopulated with high-mass protostellar cores, 
begin to be found with ALMA in the massive, young ridges of the Galactic plane \citep[][]{Csengeri+2017,Motte+2018b,YCheng+2018}.
The magnetic-field configuration (field topology and field strength) inside the hubs 
and ridges, at scales of a few 0.1~pc, 
may limit 
fragmentation (see \citealp{Commercon+2011} for MHD simulations and Fig.~\ref{Fig_G14} for recent observations)
and favor the formation of massive protostellar cores against their low-mass counterparts. 
Dynamical processes associated with local accretion streams 
and global collapse may also favor the growth of high-mass protostellar cores due to competitive accretion \citep[][]{Smith+2009}. 
Elucidating the relative roles of -- and coupling between -- magnetic fields and dynamics, 
is therefore of crucial importance for understanding the origin of high-mass stars and their associated clusters.
For example, if cluster-scale global collapse is required to form massive stars near the bottom of the gravitational potential well, 
the collapsing flow should drag the magnetic field on the cluster scale into a more or less radial configuration. If, on the other hand, 
only localized collapse of a pre-existing massive core is required to produce a massive star \citep{McKee03},  
the collapse-induced field distortion is expected to be limited to the smaller, core region. 

Observationally, this requires 
probing the magnetic field configuration from cloud scales ($\sim$100 pc) to 
protostellar scales ($\sim$0.01 pc) within massive dense ridges/hubs 
(down to so-called ``massive dense cores'' or MDCs; 0.1--0.3pc; \citealp{Motte07,Bontemps+2010b}).
%
%
Ultimately, a spatial dynamic range\footnote{The spatial dynamic range is defined as the ratio of the largest to the 
smallest spatial scale accessible to an instrument or an observation (see also Sect.~\ref{subsec:spica-adv}).} as high as $10^4$ 
(from 100~pc to 0.01~pc) is thus needed. 
In nearby high-mass star-forming regions, located at $\sim 1$ to 3~kpc, this translates to angular scales from a few ($\sim \,$2--6) degrees 
down to $\sim 0.1\arcsec$. 
While the $0.1\arcsec$ scale of individual pre-/protostellar cores in these regions is only reachable with SMA or ALMA, 
the inner scale of massive dense ridges/hubs or MDCs ($\sim 0.2\,$pc, or $\sim14\arcsec-40\arcsec$ at $\sim$1--3~kpc distance) 
can be reached with the angular resolution of {\spicapol}.
This MDC scale is of particular importance for high-mass star and cluster formation since 
high-mass protostellar cores appear to form in only a subset of MDCs, possibly those with a high level of magnetization 
(see, e.g., \citealp{Motte+2018a,TCChing+2018} and Fig.~\ref{Fig_G14}).
%
%
With a typical spatial resolution of $10\arcmin $, {\it Planck} polarization data already provide some indications 
on the magnetic-field geometry at scales between 100~pc and  $\sim \, $3--9$\,$pc, 
but  {\it Planck} maps are strongly limited by the confusion arising from several layers 
of dust emission along the line of sight within the large beam. 
The spatial resolution of {\spicapol} is required to separate the contributions of these layers 
and focus on the polarization signal from high-mass star-forming ridges, hubs, and MDCs. 
The high sensitivity of {\spicapol}  will also be crucial to trace the magnetic 
field topology all the way to the outer environment of 
ridges and hubs, where the column density of dust reaches values below $A_V \sim 1$--2.

\section{Magnetic fields in galaxies}
\label{sec:galaxies} 

\noindent Magnetic fields are an important agent that influences the
structure and evolution of galaxies \citep[e.g.][]{Tabatabaei+16}. The magnetic pressure in the ISM  
is comparable in magnitude to the thermal,
turbulent, and cosmic ray pressures \citep[e.g.][]{Ferriere01,BoularesCox1990,Beck2007}, so
the magnetic field contributes significantly to the total pressure
which supports a galactic gas disk against gravity. 
The interplay between the magnetic field, gravity, and turbulence is central to the process of star formation 
\citep[see, e.g.,][]{McKee07, HennebelleFalgarone2012, Crutcher2012}, 
both on the scale of individual stars and filaments (cf. \S ~\ref{sec:filaments} and \S ~\ref{sec:protostars}), and for the formation of molecular clouds out of the magnetized diffuse interstellar gas \citep[e.g.][]{Kortgen+2018}. On even larger scales in galaxies, magnetic fields control the density and distribution of cosmic rays \citep[e.g.][]{KoteraOlinto2011}, mediate the spiral arm shock strength \citep[e.g.][]{ShettyOstriker2006}, and may even modulate rotation of galaxy gas disks \citep[e.g.][]{Elstner+2014} and quench high-mass star formation \citep[e.g.][]{Tabatabaei+18}.  Magnetic fields play an important role in launching galactic
winds and outflows \citep[][]{Heesen+2011}, regulate gas kinematics
at the disk-halo interface \citep{HenriksenIrwin2016}, and
ultimately connect galaxies to the intergalactic medium \citep{Bernet+2013}.

\subsection{Current observational status}

\noindent Significant progress has been made in recent decades to characterize the interstellar magnetic fields of external galaxies, with measurements of the magnetic field strength and orientation obtained for about one hundred nearby galaxies \citep[see the appendix of ][]{BeckWielebinski2013}\footnote{The list is continually updated in the arXiv version at {\tt https://arxiv.org/abs/1302.5663} }. This effort has established a broadly consistent picture of the large-scale ($>1$\,kpc) properties of galactic magnetic fields. As in the Milky Way, the interstellar magnetic field in external galaxies can be described as a combination of large-scale regular fields and small-scale turbulent fields. Observations of face-on spiral galaxies demonstrate that galaxies typically host spiral fields in the disk, with the observed large-scale magnetic field orientations appearing similar to the material spiral arms. The vertical structure of the field is more easily probed via observations of edge-on systems, which typically show an X-shaped structure such that the field tends to become more inclined (and eventually perpendicular) with increasing distance from the midplane. Field strengths vary, but are usually in the range of several to tens of $\mu$G, with roughly similar contributions from ordered and random field components \citep{BeckWielebinski2013}. \\

\noindent To date, magnetic field properties in external galaxies have mostly been investigated via observations of synchrotron emission at GHz frequencies \citep[for a review, see][]{BeckWielebinski2013,Beck2015}. 
At the same frequencies, Faraday rotation of background radio sources provides an alternative, direct determination of the direction and strength of magnetic field within galaxies. This technique has been used for detailed studies of the Milky Way's magnetic field \citep[e.g.][]{TerralFerriere17,Mao+10}, but its application to external galaxies has thus far been limited to the most nearby galaxies \citep[which have a large angular size and thus a sufficient number of bright background sources, e.g. M31, LMC,][]{Han+98,Gaensler+05}. Faraday rotation measurements for a much larger sample of nearby galaxies is an important science driver for SKA \citep[see e.g.][]{Beck+15}.\\

At other wavelengths, the magnetic field structure of a much smaller number of external galaxies has been surveyed using optical polarization e.g. the Magellanic Clouds (LMC and SMC; Mathewson \& Ford 1970), NGC1068, and M51 \citep{Scarrott+1987}. Wide-field imaging of polarized extinction at infrared (IR) wavelengths is a newer capability that has been used to probe the magnetic field structure in nearby edge-on galaxies, with results that are generally in good agreement with radio observations \citep[e.g.][]{Clemens+2013,MontgomeryClemens2014}. The IR extinction technique is less suited to observations of face-on galaxies, due to the relatively short path length that can produce internal extinction \citep[see e.g.][for the case of M51]{PavelClemens2012}. \\

\noindent {\spicapol} will probe the magnetic field structure via observations of dust polarization in emission. An important advantage of this technique compared to radio synchrotron observations is that it traces the magnetic field structure in the cold gas where star formation occurs, with minimal contamination from the warm ionized gas in the halo \citep[e.g.][]{Mao+2015}. Studies in nearby galaxies further suggest that the magnetic field is coupled to the interstellar gas independently of the star formation rate \citep[e.g.][]{Schinnerer+13,Tabatabaei+18}, highlighting the importance of tracing the field in the neutral ISM. The SCUBA-POL camera on the James Clerk Maxwell Telescope (JCMT), 
operating at 850\,$\mu$m, obtained the first such dust polarization observations of external galaxies \citep[e.g.][]{Matthews+2009}, 
but was only able to access extremely bright extragalactic regions, such as the centre of the nearby starburst system M82 \citep[e.g.][]{Greaves+2000}.
The {\it Planck} mission has recently provided all-sky measurements of the polarized submillimeter dust emission, but with an angular resolution 
($\sim\,$10\arcmin\ at 353\,GHz) 
sufficient to resolve sub-kiloparsec scales only in the closest Local Group galaxies ($<1$\,Mpc). 
With $\sim$arcsecond resolution, a key opportunity for ALMA will be targeted imaging of the detailed magnetic field structure 
in extragalactic molecular clouds, but wide-field polarization surveys of nearby galaxies will remain impractical, 
due to prohibitive integration times for fields larger than a few square arcminutes. 

\begin{figure*}
\begin{center}
\includegraphics[width=15.75cm,angle=0]{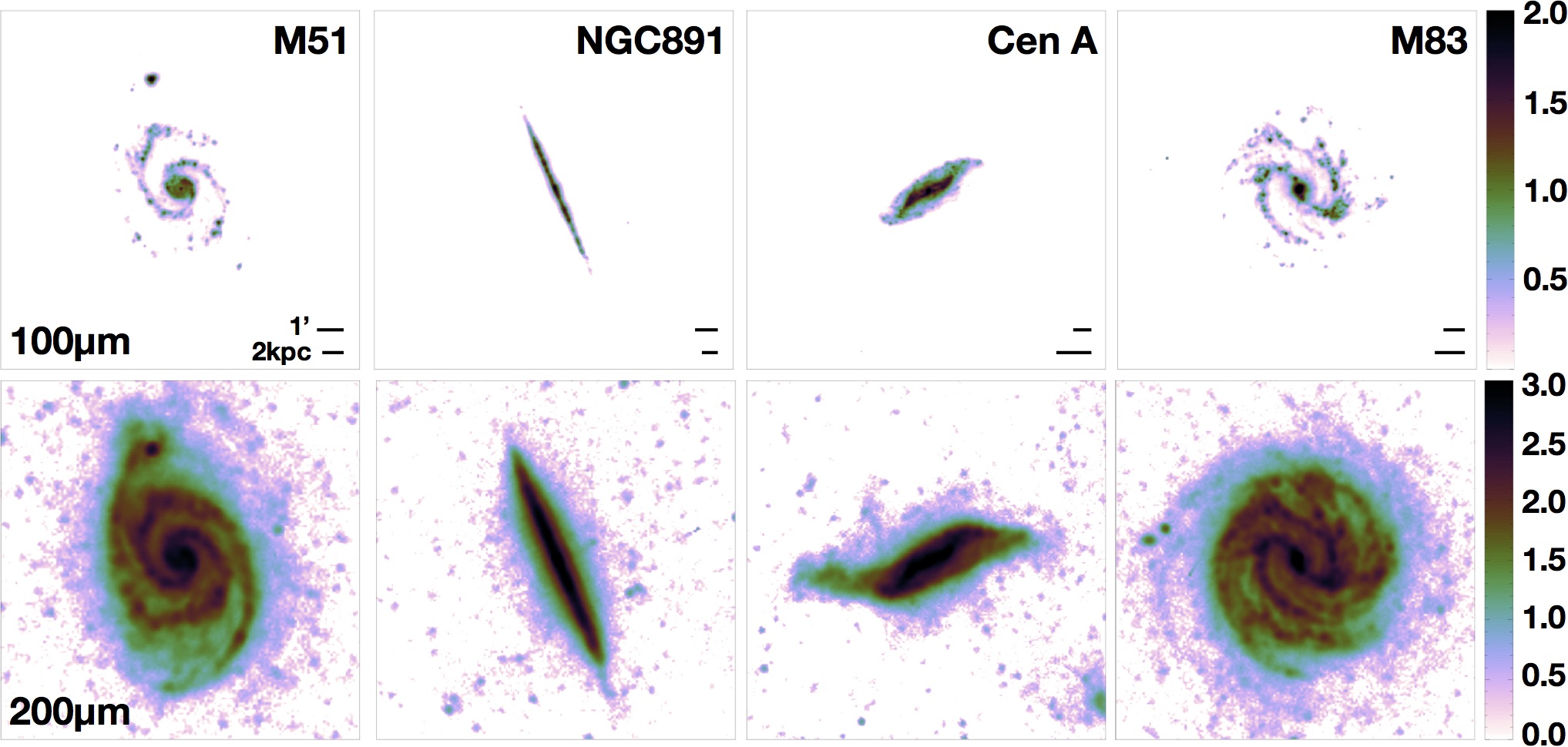}\hfill
    \caption{Expected signal-to-noise in 100\,$\mu$m (top row) and 200\,$\mu$m (bottom row) polarized intensity after 2 hr on-source integration with {\spicapol} for four nearby ($d<10$\,Mpc) galaxies.  The maps are constructed using {\it Herschel} data at 70$\mu$m and 250$\mu$m as input. We assume the {\spicapol} performance parameters given in Table~\ref{tab:POL}, 
a dust spectral index of $\beta=1.9$, and a conservative polarization fraction of 1\%.  The color scale, which runs between a signal-to-noise ratio 
of 1 and 100 (top row) or 1 and 1000 (bottom row), uses a logarithmic stretch, with green indicating a signal-to-noise of 10.}
\label{fig:ng_polsnr}
\end{center}
\end{figure*}

\subsection{Key opportunities for {\spicapol} on nearby galaxies}

\noindent Mapping the structure of interstellar magnetic fields in the cold ISM of nearby galaxies is crucial to understand how magnetic fields influence gas dynamics in galaxies, and in particular the role of the field in regulating star formation, driving galactic outflows, and fuelling galactic nuclei. Observational studies of these processes in nearby galaxies complement Milky Way studies, which typically have superior spatial resolution, but may be limited by distance ambiguity and line-of-sight confusion.  Among current and near-future facilities, only {\spicapol}  
will be able to make these measurements
across a representative sample of external galaxies, probing a much wider range of ISM conditions than those encountered in the Milky Way, and to conduct spatially complete mapping of the field structure in Local Group targets. \\

\noindent To highlight {\spicapol}'s unique capabilities for such an effort, Fig.~\ref{fig:ng_polsnr} shows the estimated signal-to-noise ratio in polarized intensity for several iconic nearby galaxies 
at $100\,\mu$m (top row) and $200\,\mu$m (bottom row) after 2~hr on-source integration with {\spicapol}. These maps are constructed assuming the performance
parameters given in Table~\ref{tab:POL}
and a conservative polarization fraction of 1\%. At the intrinsic resolution of the 100\,$\mu$m band, the sensitivity is sufficient for tracing the detailed polarization structure of bright features such as spiral arms, bar dust lanes, and galaxy centers. For the galaxies in Fig.~\ref{fig:ng_polsnr}, several hundred independent 100\,$\mu$m polarization vectors would be obtained. Measurements in the more sensitive 200\,$\mu$m and 350\,$\mu$m bands would essentially cover the entire galactic disk within $0.6\,R_{25}$ (where $R_{25}$ is the optical radius).
The excellent signal-to-noise ratio that can be achieved with a modest integration time per galaxy means that {\spicapol} could 
conduct the first systematic survey of the polarized far-IR dust emission -- and hence magnetic field structure -- in $\sim100$ nearby galaxies. 
This is slightly larger than the combined sample of galaxies targeted by the VNGS and KINGFISH {\it Herschel} nearby galaxy projects, 
and would require only $200$\,hr of on-source observing time. 
In the remainder of this section, we highlight some of the potential science drivers for such a {\spicapol} nearby galaxy survey.

\subsubsection{Testing and refining Galactic dynamo models}

\noindent The currently favored paradigm for interstellar magnetic fields is that they are amplified by dynamo action. In this scenario, weak primordial fields in young galaxies are quickly amplified by a small-scale turbulent dynamo, which continuously supplies turbulent fields to the ISM after the formation of a galactic disk in $\lesssim10^{9}$\,yr \citep{Schleicher+2010}.  The large-scale field is then amplified by the mean-field $\alpha - \Omega$ dynamo effect \citep[e.g.][]{Ruzmaikin+88}, whereby the combination of differential rotation of the galactic disk ($\Omega$ - effect) and helical turbulence
($\alpha$ - effect) presumably driven by supernova explosions \citep{FerriereSchmitt2000}, produce small-scale turbulent and organize some fraction of them into regular large-scale patterns.\\

\noindent The mean field dynamo is expected to generate a regular magnetic field with both poloidal and azimuthal components, and nearly all polarized synchrotron observations of face-on disk galaxies show a large-scale spiral pattern. To date, the magnetic field pitch angles $p_{B}$ and azimuthal structure that have been observed in nearby disk galaxies via observations of polarized radio synchrotron emission are broadly compatible with the predictions of mean field dynamo theory \citep{Fletcher2010, vanEck+15}. Yet the precise nature of the magnetic field generated via the mean-field dynamo depends on properties of the host galaxy. For example, the rotation curve determines the shear strength in a differentially rotating galaxy disk, and hence how the azimuthal field component is generated from the poloidal field. The $\alpha$-effect -- by which a poloidal field component is generated from the azimuthal field -- is thought to be powered by supernova explosions, which depend on a galaxy's star formation rate. As our knowledge of external galaxies grows, the logical next step is to refine dynamo models for specific galaxies to include all relevant observed galaxy properties -- e.g. the ionized and molecular gas density distributions, rotation curve, star formation rate, gas inflow and outflow rates -- and test the model predictions for individual galaxies against the observed properties of the magnetic field. The first attempt to do this systematically for a sample of galaxies \citep{vanEck+15}  was hampered by inconsistencies in the available radio observations. A sample of galaxies observed with the same instrument at the same resolution and sensitivity is necessary to allow the details of dynamo theory, such as how the dynamo saturates, to be tested against data. {\spicapol}'s moderate resolution, full-disk sampling of the magnetic field structure across a statistically significant sample of nearby galaxies would provide precisely this test.

\subsubsection{Magnetic fields and gas flows in barred and spiral galaxies}

\noindent Mapping the structure of the magnetic field across a sample of nearby galaxies is needed to understand the typical dynamical importance of the field on galactic scales. Of particular interest is how gas flows in galaxies -- e.g. gas streaming along spiral arms, inflow along bar dust lanes, and starburst-driven outflows -- interact with the field. While independent estimates of the field strength will still be required, {\spicapol} observations at sub-kiloparsec resolution of the magnetic field structure and complementary spectral line data for tracing interstellar gas kinematics will be extremely valuable for studying the interplay between the field and motions within the cold gas reservoir across the local galaxy population.\\

\noindent In face-on disk galaxies, the large-scale field traced by radio polarization observations tends to follow a spiral pattern. This pattern is expected from mean field dynamo theory, and is not directly connected to a galaxy's baryonic (i.e. gas/stellar) spiral structure. Observations of spiral galaxies indeed show that the field pattern is not always spatially coincident with the spiral arms, and in several cases \citep[most famously NGC6496][]{Beck2007} the ordered field pattern is most pronounced in the interarm region. Some of the large-scale field patterns in galaxies may be due to the combined action of shear and compression in the interstellar gas, which renders the turbulent field anisotropic (and hence ordered) over large scales \citep[e.g. in M51,][]{Fletcher+2011,Mulcahy+14,Mulcahy+16}. Current observations also suggest that the average pitch angle of the regular spiral field pattern is often similar to the pitch angle of the local spiral arm $p_{M}$. This is not a direct prediction of mean-field dynamo theory, but would be expected if spiral shocks amplify the magnetic field component parallel to the shock. Significant discrepancies between $p_{B}$ and $p_{M}$ in the inter-arm region, as well as large azimuthal and radial variations in $p_{B}$, are also observed, the origin of which are not yet well understood.\\

\begin{figure*}
\begin{center}
\includegraphics[width=15.0cm,angle=0]{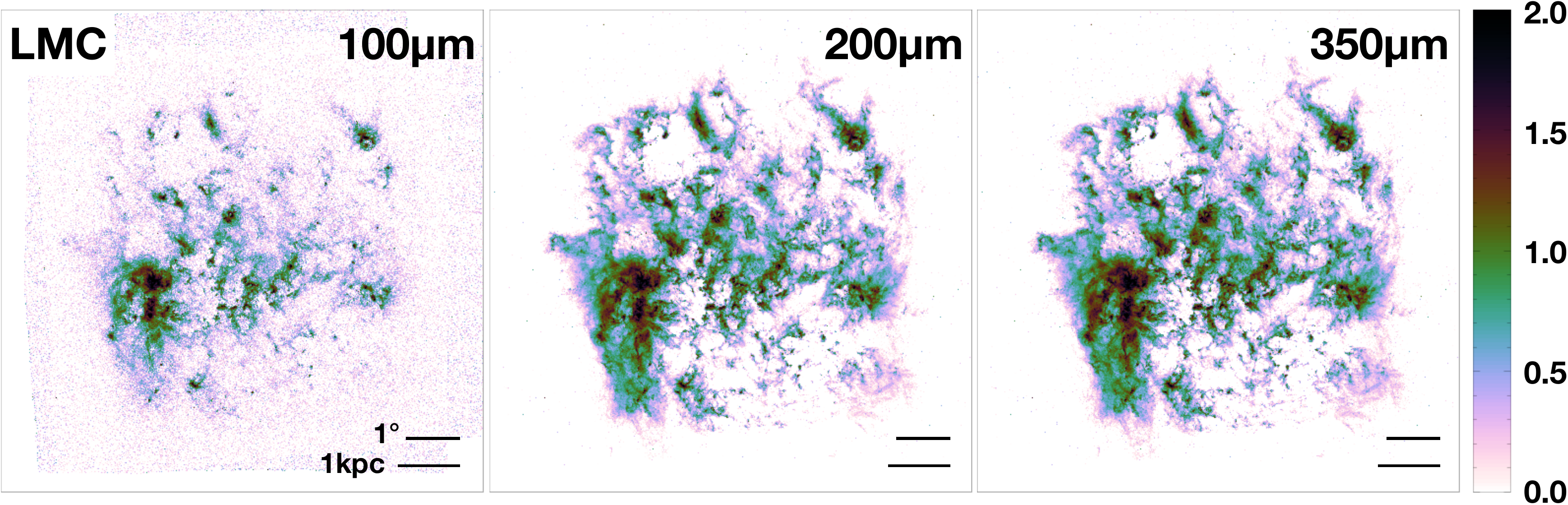}\hfill
 \caption{Expected signal-to-noise in polarized intensity for {\spicapol} observations of the LMC after 50 hr on-source integration.  
 The maps are constructed using {\it Herschel} data at 100$\mu$m and 250$\mu$m  as input 
 \citep{Meixner+2013}. 
 We assume the {\spicapol} performance parameters given in Table~\ref{tab:POL},  
a dust spectral index of $\beta=1.9$, and a conservative polarization fraction of 1\%.  
The color scale, which runs between a signal-to-noise ratio of 1 and 100, uses a logarithmic stretch, with green indicating a signal-to-noise of 10.}
\label{fig:mc_polsnr}
\end{center}
\end{figure*}

The central regions of barred galaxies are the site of fast radial gas inflow, strong shocks, and intense star formation. Barred galaxies often show strong gas streaming along the shock fronts at the edge of bars, which develop because the gas is rotating faster than the bar pattern. Radio polarization observations of the prototypical barred galaxy NGC1097 \citep{Beck+2005} reveal strongly polarized emission along the bar with field orientations parallel to the gas streamlines. The observed polarization pattern suggests that the field is amplified and stretched by shear in the compression region, and that the field is frozen into the gas and aligned with the gas flow over a large part of the bar. If this result holds generally in barred galaxies, the polarization pattern in bars -- especially using a tracer that preferentially probes the dense interstellar gas -- would provide important complementary information on the plane-of-sky gas flows to the line-of-sight kinematic information obtained from molecular emission lines. In combination with estimates for the magnetic field strength, information about the magnetic field structure in the central regions of barred galaxies would also provide useful constraints for models of AGN fuelling. One of the main problems in this area is to generate mass inflow rates that are compatible with the observed nuclear activity. Magnetic stress in circumnuclear rings \citep[e.g.][]{Beck+1999} and fast MHD density waves \citep[e.g.][]{Lou+2001} have been proposed as potential mechanisms, but current observational data for the field strength and structure in the inner regions of barred galaxies is not sufficient for a rigorous test of these models.

\subsubsection{Magnetism in dwarf galaxies}

\noindent Due to their slow rotation, the amplification of magnetic fields should be less efficient in dwarf galaxies. Yet observations of radio polarized intensity show that several nearby low-mass galaxies host large-scale ordered fields, e.g. the Magellanic Clouds, NGC4449 and IC10 \citep{Chyzy+2003, Gaensler+05,Mao+12,Chyzy+16,Heesen+18}. 
Dwarf galaxies are also more likely to exhibit star formation powered outflows and galactic winds, due to their shallow gravitational potential. The magnetized nature of these outflows has been observed in some dwarf systems \citep{Chyzy+2000,Kepley+2010}, consistent with some models of a cosmic ray driven dynamo \citep{Siejkowski+2014,DuboisTeyssier2010}. To date, all dwarf galaxies with detected ordered magnetic fields are star-bursting, participating in a galaxy-galaxy interaction, and/or experiencing significant gas infall, suggesting the importance of enhanced turbulence for the magnetic field properties and evolution of these systems. {\spicapol} observations of a sample of local dwarf galaxies with a range of masses, interaction properties and star formation histories, would provide valuable input for theories for the amplification of magnetic fields in such systems, and their role in magnetizing the intergalactic medium (IGM). 

\subsubsection{The Magellanic Clouds}

\noindent The Large and the Small Magellanic Cloud (LMC, SMC) are the closest gas-rich galaxies to the Milky Way. A {\spicapol} survey of the Magellanic Clouds
would for the first time probe the magnetic field structure in the cold ISM across all spatial scales between the clumpy sub-structure within GMCs ($\sim2$\,pc) and the galactic disk (several kpc). Observations across such a large range of spatial scales are needed to decipher the dynamical importance of the magnetic field for the inherently hierarchical process of star formation, i.e. from the formation of GMCs out of the diffuse ISM, down to the formation of individual stars. Spatially complete surveys of dust emission in the Magellanic Clouds with ALMA are unfeasible due to their large angular size ($\sim50$ and $\sim10$~deg$^2$ for the LMC and SMC respectively).\\

\noindent As an example of what could be achieved with {\spicapol}, Fig.~\ref{fig:mc_polsnr} shows the estimated signal-to-noise ratio 
for a 50\,hr polarimetric imaging survey of the LMC at $100\, \mu$m, $200\,\mu$m, and $350\, \mu$m. This hypothetical survey 
would achieve a signal-to-noise ratio of $3$ for the polarized intensity at 100$\,\mu$m for interstellar  gas with column densities above $\sim2.5 \times 10^{21}$\,cm$^{-2}$ \citep[equivalent to 
$A_{V}\sim0.4$ in the LMC--][]{WeingartnerDraine2001}. This sensitivity would yield $\sim$0.5 million independent measurements of the magnetic field orientation in the interstellar gas on spatial scales of 2\,pc, including in the column density regime of the atomic-to-molecular phase transition. At 200\,$\mu$m and 350\,$\mu$m, a similar number of significant detections of the magnetic field orientation would be achieved in even more diffuse gas ($\sim1 \times 10^{21}$\,cm$^{-2} \approx 0.15$\,mag). This represents $\sim$ two orders of magnitude increase in detail over measurements 
with {\it Planck}'s 353\,GHz channel in the Magellanic Clouds, and would provide the first spatially complete view 
of the parsec-scale magnetic field structure in the molecular gas reservoir of any galaxy.

\subsubsection{Wavelength dependence of polarization in U/LIRGs and AGNs}

The polarization of luminous external galaxies such as Luminous Infrared Galaxies (LIRGs), Ultraluminous Infrared Galaxies (ULIRGs) and active galactic nuclei (AGNs), in the far-IR and submillimeter can arise from synchrotron emission but also from emission or absorption by aligned dust grains in the optically thick clouds that surround young stars and AGN tori \citep[e.g.][]{Efstathiou+97, Aitken+02}. Information in the far-IR and submillimeter can be combined with information at 10\,$\mu$m and 18\,$\mu$m as well as near-IR data 
from the ground to study the switch in position angle by about 90 deg 
 that is predicted as polarization changes from dichroic absorption 
at shorter wavelengths to dichroic emission at longer ones. Several highly polarized galaxies in the mid-IR were found 
by \citet{SiebenmorgenEfstathiou01} with Infrared Space Observatory (ISO) 
and more recently by \citet{LopezRodriguez+18a} with CanariCam on the 10.4-m Gran Telescopio Canarias (GTC). \\

\noindent This is a science area where significant progress can be achieved with {\spicapol}, which will provide sensitive polarization measurements at 100--350\,$\mu$m for a large sample of luminous external galaxies. Such information is currently available for very few objects.  In a recent study of the nearby radio galaxy Cygnus A using data from HAWC+ onboard SOFIA,  \citet{LopezRodriguez+18b} showed that this approach can be very useful for unravelling the polarization mechanisms in the infrared and submillimeter and providing an independent method of estimating the contributions of AGN tori and starbursts to the SEDs. 
Exploring the role of AGNs and star formation in galaxies is a scientific objective of wide interest. 
The opportunity to study 
multi-wavelength polarization with {\spicapol}  will be complementary to other methods 
such as spectroscopy \citep[e.g.][]{GonzalezAlfonso+17} and traditional SED fitting of the total emission \citep[e.g.][]{Gruppioni+17}.

\subsection{Distant galaxies and the potential detection of the Cosmic Infrared Background polarization}
\label{subsec:cib}

The build-up of coherent magnetic fields in galaxies and their persistence along cosmic evolution is being
investigated with analytical models of the galactic dynamo \citep[e.g.][]{Rodrigues19} and numerical simulations 
of galaxy formation \citep[e.g.][]{Martin-Alvarez18}. These studies suggest that the mean-field dynamo is effective early in the evolution of galaxies 
but, today, polarization data available to trace 
the redshift evolution of galactic magnetic fields are very scarce \citep{Mao17}. 
While SKA holds exciting promises to extend observations of cosmic magnetism to the distant universe  \citep[e.g.][]{Basu+2018, Mao18},
we argue here that \spicapol\ can also uniquely contribute providing the first polarimetric extragalactic survey at far-IR wavelengths. 

To quantify what could be achieved with {\spicapol}, we consider the point source sensitivity of a polarimetric extragalactic survey for an integration time of 
10 hr per deg$^2$ (Table~\ref{tab:POL}). At the detection limit of {\it Herschel} imaging surveys in total intensity, the signal-to-noise ratio of {\spicapol} in Stokes Q and U 
is $\sim 200$ at $200\, \mu$m, and $\sim 100$ at $100\, \mu$m and $350\, \mu$m.  This sensitivity needs to be compared to the few existing values of 
the far-IR polarization fraction for galaxies as a whole. 

The net polarization fraction resulting from the integrated emission of galaxies depends on the existence of a coherent mean magnetic field 
and on viewing angle. 
In disk galaxies, the polarization angle is aligned with the projection of the galaxy angular momentum vector on the plane-of-the-sky, and the polarization 
fraction increases from a face-on to an edge-on view. Integrating the {\it Planck} dust polarization maps at 353\,GHz over a $20^\circ$ wide band centered 
on the Galactic plane, \citet{deZotti18} found a polarization fraction $p= 2.7$\%. Within a simple model, $p$ is expected to scale as sin$\,i$, 
where $i$ is the inclination angle of the galaxy axis to the line of sight. For this scaling, the mean $p$ fraction averaged over inclinations is $1.4\%$. 
This may be taken as a reference value for spiral galaxies like the Milky Way, but $p$ is likely to be on average lower for distant infrared galaxies. 
Indeed, SOFIA polarization imaging of the two template starburst galaxies, M82 and NGC253, revealed regions with different polarization orientations, 
which tend to average out when computing the integrated polarized emission yielding an overall mean $p \sim 0.1$\% \citep{Jones19}. 

Even if {\spicapol} only detects polarized emission from only a small fraction of {\it Herschel}  galaxies, 
the number of detections will be significant given the present dearth of such measurements.  
If the detections are numerous, the emission from galaxies could even limit
the  polarization sensitivity of {\spicapol} deep surveys.  
This is a possibility that needs to be assessed. 
Beyond the study of individual galaxies, we anticipate that the main outcome of a deep polarimetric extragalactic survey with \spicapol\ 
could follow from a statistical analysis of the data. 

Statistical analysis is the reference in cosmology and much can be learned  without detecting sources individually. 
In particular, the cross-correlation of surveys across the electromagnetic spectrum is a powerful means commonly used. 
In the far-IR, this is illustrated by the results obtained stacking {\it Herschel} data on positions of extragalactic sources in the near- and mid-IR. 
This approach has been successfully used to statistically identify sources accounting for the the bulk of Cosmic Infrared Background (CIB), although they were 
too faint to be detected individually \citep{Bethermin12,Viero13_stacking}.
{\spicapol}, which will extend these studies to polarization, is uniquely suited to detect  
 -- or set tight constraints on --  the CIB polarization. 
We note that the analysis of point sources circumvents the difficulty of separating the CIB 
from the foreground polarized emission of the diffuse Galactic ISM. 

For polarization, data stacking needs to
be oriented to align polarization vectors.
This can be achieved by using, e.g., galaxy shapes measured from near-IR surveys at the appropriate angular resolution. 
Another interesting path will be the study of correlations between {\spicapol} extragalactic survey data and maps of the cosmic web 
inferred from weak-lensing surveys. 
The angular momenta of galaxies are not randomly oriented on the sky. The cosmic web environment has a strong influence on galaxy formation and evolution, 
and tidal gravitational fields tend to locally align the spins of dark-matter halos and galaxies. 
Such alignments, observed in dark-matter simulations, bear information on galaxy formation and evolution, as well 
as on the growth of structure in the Universe  \citep{Kirk15}. 
In this picture, low-mass haloes tend to acquire a spin parallel to cosmic web filaments, while the most massive haloes, which are typically the products of later mergers, have 
a spin perpendicular to filaments \citep[e.g.][]{Codis12,Dubois14}. 
Quasar observations provide observational evidence of a correlation between the polarization orientation of galaxies 
and the large-scale structure of the Universe \citep{Hutsemekers14}, but only for a small number of sources. 
{\spicapol} can uniquely contribute  to characterizing this correlation for infrared-luminous galaxies.

The scientific goals outlined here are new and promising but still qualitative. Modelling is required to assess the scientific outcome of 
a deep polarimetric extragalactic survey with {\spicapol}  and decide on the best observing strategy in terms of survey depth 
and sky coverage. 



\section{Constraining Dust Physics}
\label{sec:dust-physics} 

The polarization of thermal dust emission depends on 
the shape, size, composition, and alignment efficiency of dust grains, and also on the 3D structure of the Galactic magnetic field on the line of sight and within the instrument beam. 
Dust polarized emission can therefore bring specific constraints on the alignment mechanism of dust grains and possibly on the grain shape.
Despite this complex nature, it can also be used to constrain the optical properties of aligned dust grains (emissivity, spectral index), which are large grains at thermal equilibrium \citep{PlanckXXI2015,PlanckXXII2015}. 


As described below, {\spicapol}, with its high angular resolution and good wavelength coverage of the polarized dust SED, will be a key instrument to provide new constraints 
on grain alignment theories and inform the evolution of aligned grain properties  from the diffuse ISM to the densest cloud cores.



\subsection{Probing the grain alignment mechanism}

Grain alignment is subordinate to various processes. First, grains must rotate 
supra-thermally\footnote{The grain thermal energy, which is radiatively balanced, is not in equipartition with the grain rotational energy, 
allowing for the suprathermal rotation of large grains under specific torques \citep{Purcell1979}.} 
to be well aligned. 
Radiative torques, or chemical torques resulting from the formation of H$_2$ molecules 
on the surface of grains are good candidates for grain spin-up. Second, alignment torques of magnetic, radiative, or mechanical origin \citep[][]{Hoang_Lazarian2016} 
are needed to align the supra-thermally rotating dust grains. 
Compared to the time evolution of molecular clouds and cores, the alignment of grains along magnetic field lines by radiative torques is a fast process, 
with a timescale on the order of $10^3$--$10^4\,$~yr \citep{HL14}\footnote{More precisely, the grain alignment timescale 
by the RATs mechanism is about three orders of magnitude shorter than the 
local free-fall time.}, so that situations where grain alignment would be out-of-equilibrium can be safely ignored. 

While the mechanisms of dust grain alignment are still debated (see, e.g., Sect.~\ref{subsubsec:alignment}), polarization measurements 
at UV to optical wavelengths imply that the alignment efficiency of dust grains is sensitive 
to grain size. 
Such a behavior, well observed in the Mie regime where absorption and scattering are size-dependent \citep{BH83}, is however more difficult to extract from the polarized thermal SED because 
the dust temperature and dust spectral index depend more on the grain shape, internal structure, and composition (through its emissivity) than on the exact grain size.




\begin{figure}
\begin{center}
\includegraphics[width=\columnwidth]{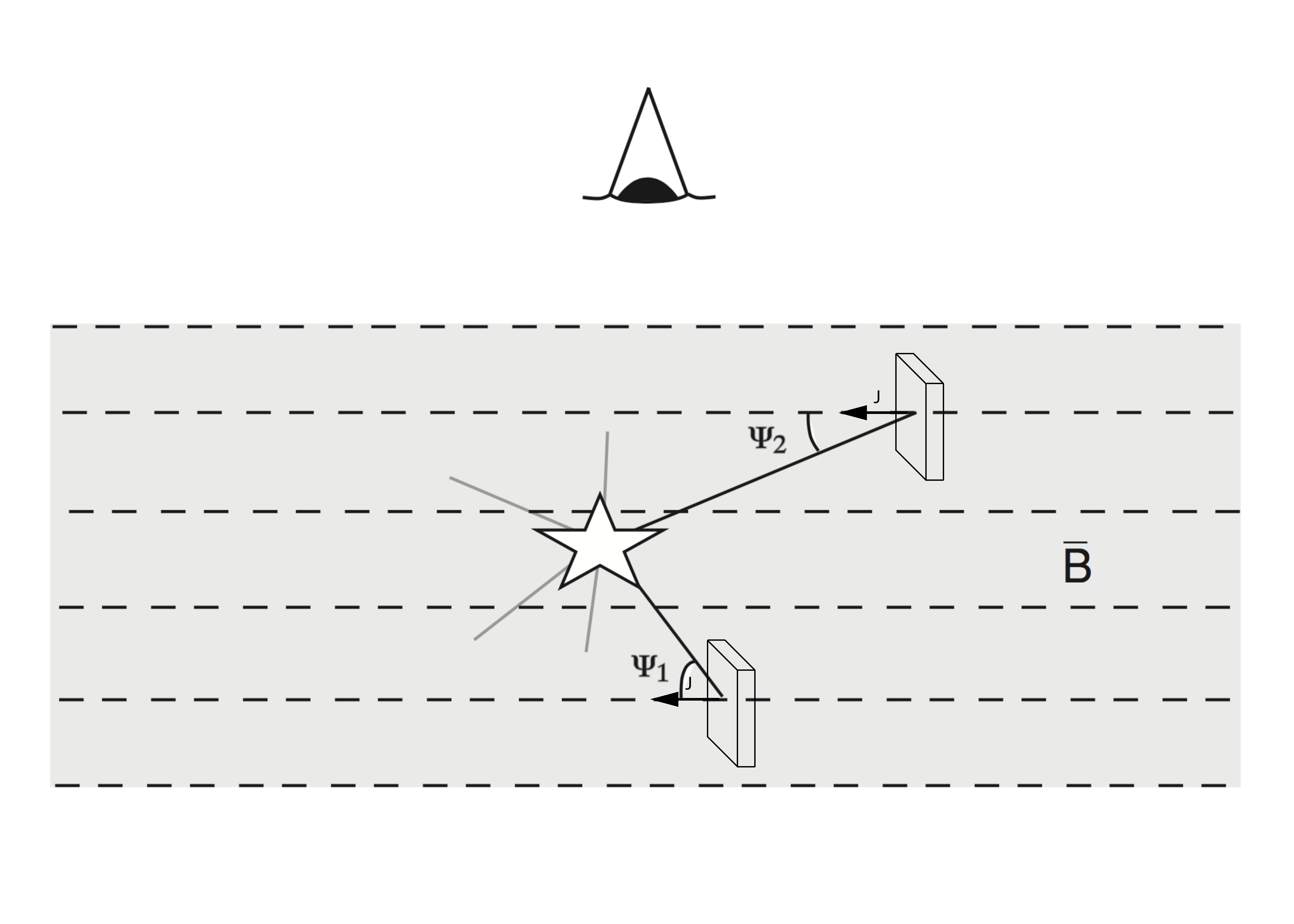}
\caption{\label{Andersson_geometry} Sketch of the geometry around a single star dominating the heating of the local ISM. The magnetic field direction is represented by the horizontal dashed lines. The aligned dust grains are sketched as prolate rotating parallelograms. If RATs dominate, dust alignment will be more efficient in regions with low values of $\psi$, the angle between the stellar radiation and magnetic field directions. 
Figure adapted from \cite{Andersson_Potter_2010} and \cite{Andersson+2011}.}
\end{center}
\end{figure}

As already mentioned in Sect.~\ref{subsubsec:alignment}, the leading grain alignment theory is Radiative Alignment Torques (RATs) \citep[e.g.][and references
therein]{Lazarian_Hoang2007}. 
The RATs alignment mechanism, if present, will lead to characteristic signatures in observations. For instance, in this theory, the alignment efficiency is directly dependent on the angle between the incident radiation field and the magnetic field direction ($\psi$).  
When dust is heated by a single nearby star or in starless dense cores where the field is strongly attenuated and  anisotropic, the incident radiation field direction is well characterized. Since the magnetic field orientation projected on the plane of the sky can be determined
from the polarized signal, mapping dust polarization in such regions can in principle be used to test alignment by RATs. 
This is illustrated in Fig.\,\ref{Andersson_geometry} which sketches the relative geometry of the radiation field and magnetic field in a region of the ISM where the radiation field is dominated by a single star. In such regions, the RATs alignment theory predicts 
a stronger alignment and therefore a higher polarization fraction where $\psi$ is close to 0 deg. 

Despite an expected clear signature, direct observational evidence of this angular effect has been scarce. The only positive detection reported in emission is by \cite{Vaillancourt_Andersson2015} who detected a periodic modulation of the dust polarization fraction around the Becklin-Neugebauer Kleinmann-Low (BNKL) object in Orion OMC-1. 
Some authors also claim to have evidenced a correlation between dust temperature and polarization fraction, as expected for dust grains aligned through the RATs mechanism 
\citep{Andersson+2011,Mats2011}.
There are also a few reports that indicate a possible influence of  H$_2$ formation \citep[e.g.][in IC63]{Andersson+2013}. 
In any case, only a handful of cases have been investigated so far and there is certainly a bias in the literature for publishing detections rather than non-detections. Given the intrinsically tangled 
nature of magnetic field geometry, chance coincidences are very difficult or even impossible to exclude 
for these few isolated studies and a statistically representative study is clearly needed. 

Such a study has not been carried out so far 
using  {\it Planck} all-sky data, essentially because the number of interstellar regions where dust is directly and predominantly heated by a single star is very low at the {\it Planck} angular resolution. 
Attempts to unambiguously detect a statistical increase of the polarization fraction with dust temperature in the {\it Planck} data, which would also be attributable to radiation-enhanced spin up and alignment of dust grains, have not led to a strong conclusion. 
\cite{PlanckXII2018} showed that it is possible to disentangle, statistically,  between what can be attributed to variations in grain alignment efficiency or grain properties, 
and what is due to line-of-sight and beam averaging of magnetic field structures. 
This study demonstrated that there is no strong variations in grain alignment efficiency in the diffuse ISM (up to a column density $N_{\rm H} \sim 2\times10^{22}$ cm$^{-2}$), 
but, due to the low angular resolution of the {\it Planck} data, could not conclude in the case of the high-density 
ISM. 
\cite{PlanckXII2018} did not find any correlation either between dust temperature and polarization fraction in the diffuse ISM. 
In conclusion, the analysis of the {\it Planck} all-sky data have so far not allowed to strongly confirm or rule out any specific grain alignment theory, 
but have provided an upper limit to the drop of alignement in the diffuse ISM. 

Owing to its much higher angular resolution and  sensitivity, {\spicapol} will allow us to systematically map the polarization of dust emission around thousands of individual stars heating the nearby ISM locally. 
The good coverage of the polarized SED will allow us to measure the temperature of aligned dust grains responsible for the polarized emission.
Analysis of the polarization fraction as a function of the angle between the known radiation field and the magnetic field direction derived from polarization will allow us to test,  for the first time, the RATs alignment theory with statistical significance, in a way that is not affected by local variations and the complexity of individual objects. At the same time, we will be able to detect, if present, the polarized emission resulting from the small temperature difference between grains heated face-on and edge-on immersed in an anisotropic radiation field \citep{Onaka+1995}. This process, which is only efficient 
at short wavelength \citep[$\lambda \le 100\,\mu$m,][]{Onaka2000}, would appear in {\spicapol} data as a characteristic difference between the polarization fraction and 
angle measured at $100\,\mu$m and the ones measured by unaffected channels at longer wavelengths.  

Altogether, the high resolution, sensitivity, and spectral coverage of {\spicapol} will set unprecedented, probably unexpected constraints on 
the physics of grain alignment in star-forming regions, a topic which {\it Planck} observations could hardly address.



\subsection{Dust polarization as a proxy for dust evolution}

\begin{figure}
\begin{center}
\includegraphics[width=\columnwidth]{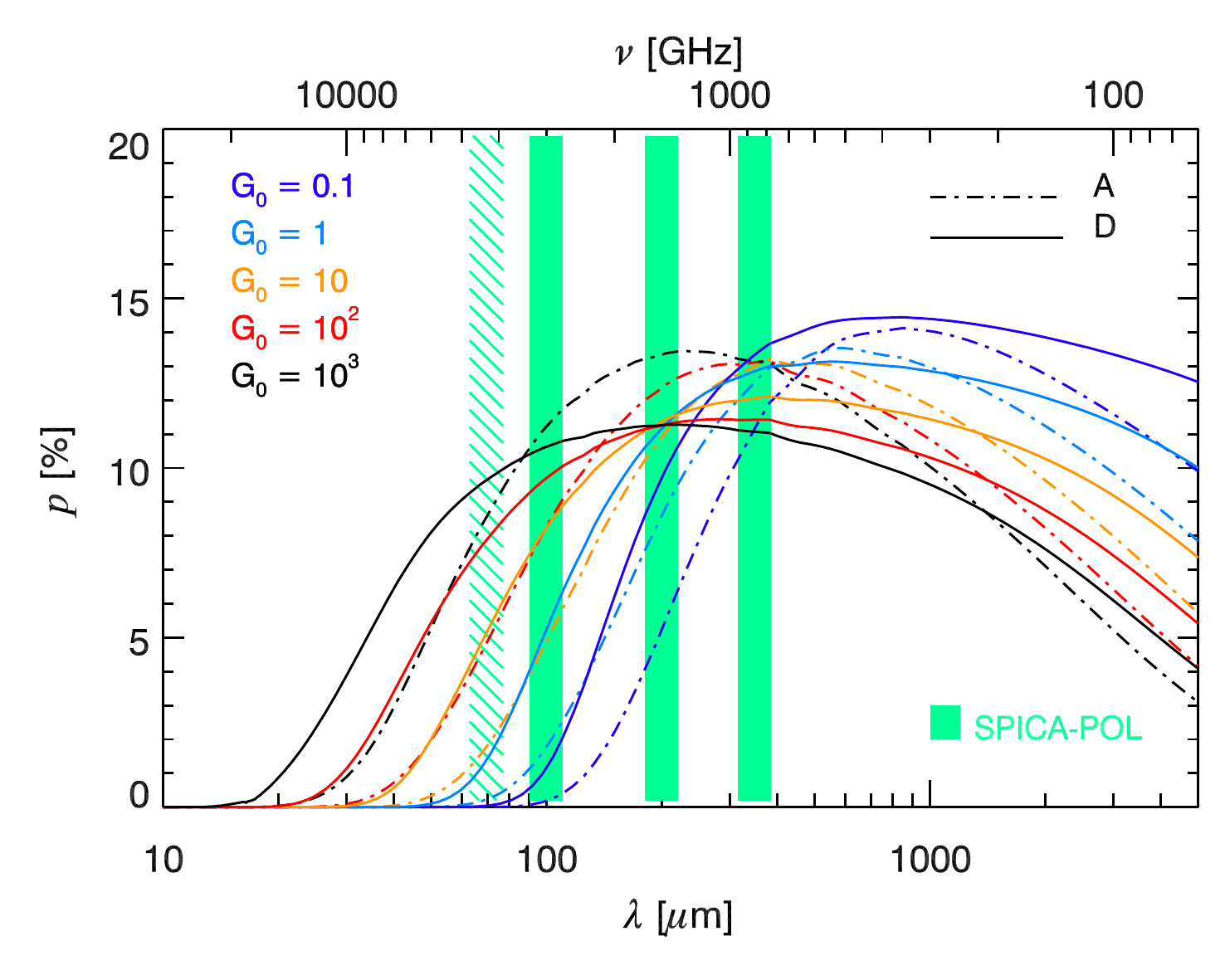}
\caption{Polarization fraction as a function of wavelength predicted using the \texttt{DustEM} (\url{http://www.ias.u-psud.fr/DUSTEM}) numerical tool \citep{Compiegne2011,Guillet+2018}. 
The vertical bands show the {\spicapol} photometric channels. 
The dashed band shows a suggested shifted location for the short-wavelength band of {\spicapol} at $70\,\mu$m, which would better cover the Wien part of the polarized dust SED.
In model A, only silicate grains are aligned, while carbon grains are randomly aligned. In model D, both silicate and carbon are aligned, with carbon inclusions incorporated in the silicate matrix (6\% in volume). Figure adapted from \cite{Guillet+2018}.}
\label{model_dustem} 
\end{center}
\end{figure}

The wavelength range covered by {\spicapol} will allow us  to disentangle between various dust models. The Wien part of the polarized dust emission is currently not constrained. 
It is in this wavelength range that dust models present the strongest differences in spectral variations of $P/I$ (Fig.~\ref{model_dustem}), 
in particular between those where carbon grains are aligned and those where they are not \citep{Draine_Hensley2013,Guillet+2018}. 
Moreover, as dust models now predict both emission and absorption properties of dust grain populations in polarization \citep[][]{Siebenmorgen+2014,Draine_Hensley2017,Guillet+2018}, 
joint observations of common targets with {\spicapol} and survey experiments targeting extinction polarization of background stars, such as PASIPHAE \citep[][]{Tassis+2018}, 
can be used to further test such models. 

High-resolution polarization observations with {\spicapol} will allow us to probe dust properties in dense environments and to further characterize 
dust evolution between diffuse and dense media \citep[e.g.][]{Kohler+2015}.
Observationally, a number of studies based on emission and extinction data have provided  evidence of grain growth within dense clouds. 
One of the main results is an increase of the far-IR/submillimeter emissivity by a typical factor of 2--3 compared to standard grains in the diffuse medium
\citep[cf.][]{Stepnik+2003, PlanckCollaboration25_2011, Ysard+2013, Roy+2013, Juvela+2015}, 
as predicted by calculations of aggregate optical properties \citep[][]{Ossenkopf+1994,Kohler+2012,Kohler+2015}. 
Another striking result is the so-called {\it coreshine} effect, i.e., enhanced mid-IR light scattering detected with {\it Spitzer} toward a 
number of dense cores \citep[][]{Pagani+2010, Steinacker+2010}, implying the presence of larger \citep{Steinacker+2015} or, taking 
into account the change in dust optical properties, only moderately larger \citep{Ysard+2016}, dust grains.

Stochastic emission by very small grains ($a\ll 10$\,nm) is known to contribute significantly to the 60\,$\mu$m and 100\,$\mu$m emission bands.
This contribution is estimated to be on the order of 13\% at 100\,$\mu$m, and 45\% at 60\,$\mu$m \citep[e.g.][]{Jones+2013}. 
When the 100\,$\mu$m band is used to derive the dust temperature\footnote{This is the case for {\it Planck} studies using 100\,$\mu$m IRAS data, 
but not for {\it Herschel} results based on SED fitting between 160\,$\mu$m and 500\,$\mu$m.}, this contamination 
affects the accuracy of mass determinations using dust continuum measurements. 
The formation of dust aggregates first removes the very small grains from the gas phase, as suggested by observations 
showing a significant decrease in the 60\,$\mu$m emission \citep[][]{Laureijs+1991,Bernard+1999, Stepnik+2003, Ysard+2013} 
and as predicted by dust evolution models \citep[][]{Ossenkopf+1994, Kohler+2015}.
Because small grains are not aligned with the magnetic field,
such contamination is absent from the polarized thermal 
emission SED\footnote{This is probably also valid for the zodiacal light emission, which severely contaminates the 60\,$\mu$m band near the Ecliptic plane in total intensity, 
but should not in polarization because the large warm grains responsible for this emission are not known to be aligned. This will however have to be checked.}. 
As a consequence, the dust temperature 
derived\footnote{In the Rayleigh regime ($a\ll \lambda$) that characterizes thermal dust emission, the influence of the magnetic field and alignment efficiency on polarization observables 
is achromatic, and therefore does not affect the spectral dependence of the SED, but only its amplitude. The spectral index of the polarized SED will therefore characterize 
the optical properties of aligned grains.}  
from the polarized dust  SED that {\spicapol} can observe will only reflect the temperature of large aligned dust grains in the transition from the diffuse to the dense ISM, 
a constraint that 
will be used in addition 
to that inferred from the total intensity SED to study dust evolution processes.

Just like unpolarized emission, polarized  dust emission will also probe variations in dust emissivity as expected from dust evolution in dense clouds. 
Dense environments could not be properly characterized in polarization at the low resolution of the {\it Planck} data. Unlike unpolarized emission, 
the anisotropic nature of polarized emission makes it sensitive to the grain shape. The formation of dust aggregates by grain-grain coagulation 
must have its counterpart in polarization, and {\spicapol} will detect signatures which will have to be analyzed through detailed modeling of the coagulation process. 
Here again, the combination of {\spicapol} data with the increasing amount of high-resolution polarization observations in the optical and the near-IR 
will provide strong constraints on dust evolution \citep{PlanckXXI2015,PlanckXII2018}.

Observations of  total dust emission intensity at far-infrared and submillimeter wavelengths with {\it Planck} and {\it Herschel} have also brought surprises. 
One of them is evidence that the logarithmic slope of the dust emission SED at long wavelengths, 
often referred to as the dust emissivity index, $\beta$, exhibits significant variations 
at large scales across the Galaxy. The {\it Planck} all-sky data clearly show variations of $\beta$ along the Galactic Plane, 
from very steep SEDs toward inner regions of the Milky Way 
to much flatter SEDs ($\beta \simeq1.5$) toward the Milky Way anticenter \citep{PlanckXI+2014}. 
This has also been confirmed in the far-infrared by the analysis of Hi-GAL data \citep{Paradis+2012}. Even larger variations have been found 
in observations of external galaxies, with the SMC and LMC having $\beta\simeq1.3$ and $\beta\simeq1.0$, respectively \citep{PlanckXVII2011}. 
Such variations are observed in the {\it Herschel} data within individual nearby galaxies such as M31 \citep[][]{Smith+2012} and M33 \citep[][]{Tabatabaei+2014}. 
The origin of these variations is currently unclear and three main classes of dust models have been proposed to explain them. 
The first type involves the mixing of different materials during the dust life-cycle \citep{Kohler+2015,Ysard+2015}. 
The second type of models invokes Two-Level-System (TLS) low-energy transitions in the amorphous material composing dust grains \citep{Meny+2007} 
as the cause for the flattening of the SED. The third type of models proposes that magnetic inclusions in dust grains  \citep{Draine_Hensley2013} 
could produce the observed variations \citep{Draine_Hensley2012}. Determining the origin of these variations is critical in many respects, 
not only to understand the dust cycle in the ISM, but also for accurate mass determinations from dust continuum measurements (which require 
good knowledge of the dust emissivity, its wavelength dependence, and its spatial variations).

In this domain again, extensive polarimetric imaging at far-IR wavelengths is likely to play a critical role in the future. 
Dust models based on dust evolution have not yet presented their predictions in polarization, but the other two classes of 
models mentioned above predict significantly 
different behaviors for the polarization fraction as a function of wavelength. TLS-based models essentially predict a flat spectrum for the polarization fraction, 
a prediction compatible with {\it Planck} 
observations. In contrast, metallic-inclusion models predict variations of the polarization fraction in the submillimeter \citep{Draine_Hensley2013}, which are not observed. It will 
be possible to evidence those distortions of the polarization SED by comparing far-IR {\spicapol} observations with existing submillimeter {\it Planck} data for the Magellanic clouds 
and polarization data obtained with new ground-based polarimetric facilities such as NIKA2-POL and SCUBA2-POL for Milky Way regions/sources and nearby galaxies. Correlating 
changes in the polarized SED with variations of $\beta$ will allow us to constrain models of the submillimeter dust emissivity  in a very unique way. 

\subsection{Toward a tentative detection of polarization by dust self-scattering in the densest cores}

In the past decade, interesting constraints on grain sizes in dense clouds have come from the detection of the ``coreshine'' effect \citep{Pagani+2010}, 
which results from scattering of near-IR stellar photons by dust grains present in the cloud. 
This has been interpreted as evidence of grain growth ($a=1\, \mu$m, \citealp{Steinacker+2010}) or grain compositional and structural evolution 
with a modest size increase ($a<0.5\, \mu$m, \citealp{Ysard+2016}).
More recently, it has been demonstrated that very large ($a > 10\, \mu$m) dust grains are able to produce polarization by scattering thermal dust emission, 
a process that is called 'self-scattering'. This was first predicted to be observable in protoplanetary disks \citep{Kataoka+2015} and 
then confirmed by numerous ALMA observations \citep{Kataoka+2016, HYang+2017, Girart+2018}. 

{\spicapol}, with its $100\,\mu$m polarized channel, would be able 
to detect and characterize the spectral dependence of polarization by scattering due to $\sim 15\,\mu$m dust grains \citep{Kataoka+2015}, if present.
For the effect to be observable, the thermal emission of dust must first present a quadrupolar anisotropy: scattering grains must receive more far-IR irradiation
along one direction in the plane of the sky than along the perpendicular direction. Such a condition is naturally met in a dense protostellar core, 
or in the presence of density gradients. Second, high local densities ($>10^6\, {\rm cm}^{-3}$) must be present along the line of sight so that dust grains 
can have grown to the very large sizes ($a \sim \lambda/2\pi$) needed for scattering to occur in the far-IR. 
Observing such a high-density 
medium should be feasible at $100\,\mu$m at the resolution of SPICA, 
but simulations of grain growth and polarization by scattering are needed to confirm this idea. 

Polarization by self-scattering at $100\,\mu$m with {\spicapol} will most likely concern only a few lines of sight through the densest cores.
Because polarization due to scattering sharply declines at wavelengths larger than the grain size, 
it will not alter the polarized emission from aligned dust grains at $200\,\mu$m and $350\,\mu$m used 
to trace the local magnetic field orientation in molecular clouds (\S ~\ref{sec:filaments} to \S \ref{sec:massive-sf}).

\section{Molecular clouds and the origin of cosmic rays}
\label{sec:cosmic-rays} 

Polarimetric imaging of molecular clouds (MCs)\footnote{In this section, ``molecular clouds'' are used in a broad sense, ranging from  
small individual dark clouds $\sim \,$2--10$\,$pc in size to GMCs $\sim \,$50$\,$pc in diameter (see Sect.~\ref{sec:filaments}). There is no clear cut-off size for the physical effects discussed here.}  
with  {\spicapol} will also be very useful for the study of the origin of cosmic rays (CRs). Indeed, CRs pervade the whole galaxy, and their interaction with the dense  gas of MCs has two important consequences. First, the interactions of high energy CRs (kinetic energy larger than a few hundred MeV) with the gas make MCs bright $\gamma$-ray sources. Second, CRs of low energy ($\approx 1-100$ MeV) are the only ionizing agents able to penetrate MCs and  regulate the ionization fraction of MC dense cores. For these reasons, observations of enhanced levels of $\gamma$-ray emission or ionization rates from MCs reveal the presence of a CR accelerator in their vicinity (see review by \citealp{GabiciMontmerle2015}).


A first problem arises, that of the \textit{propagation} of cosmic rays inside molecular clouds, for which the strength, 
and above all, the topology of the magnetic field around and inside them plays a central role. 
At high energies, CRs are unaffected by the magnetic field, so that they interact with all the gas (atomic as well as molecular): 
the $\gamma$-ray emissivity is simply proportional to the product (CR flux $\times$ cloud mass). In other words, for a given cloud mass, 
determining the $\gamma$-ray luminosity of a molecular cloud allows the local CR flux to be measured, irrespective of the magnetic field. 
Over galactic scales, it is well established that the CR flux deduced in this way is essentially uniform \citep{Ackermann+2011}, which means 
that the CR diffusion away from their sources is efficient enough not to be sensitive to large-scale spatial features like the spiral arms. However, 
there may be $\gamma$-ray ``hot spots'' close to CR sources, and this is precisely what happens when a supernova remnant (SNR) collides or 
is located in the vicinity of a molecular cloud. 
Many examples of such SNR-MC associations are known \citep{GabiciMontmerle2015}. The reason for the enhanced $\gamma$-ray emission 
is that SNR shock waves accelerate CR in situ, via the so-called Diffuse Shock Acceleration, or DSA, mechanism \citep{Drury1983}, so this 
particular configuration can be considered as a ``CR laboratory'': If the shock-accelerated CRs are insensitive to the local magnetic fields when 
they reach high energies, the process by which they do, in other words the \textit{acceleration} mechanism itself depends very much on it, and 
again on its topology close to the SNR shock. For instance, it is well known from theoretical DSA models 
that the acceleration efficiency depends strongly on the angle between the shock front and the local magnetic field lines \citep{CaprioliSpitkovsky2014}. 
So, again, knowing the magnetic field topology on small spatial scales in MCs impacted by SNRs would allow a detailed study of the CR acceleration 
process in the vicinity of the shock front.

In particular, a new picture of the ISM in star-forming regions has now to be taken into account: in the standard picture summarized above, 
the gaseous medium in which the SNR shock propagates is assumed uniform. But as recent work has shown (see Sect.~\ref{subsec:fil-paradigm} 
and Fig.~\ref{taurus_planck}), the structure of MCs is not uniform, but \textit{filamentary}, down to scales of parsecs (in length) and $\sim 0.1$ pc (in width), 
i.e., precisely those that are accessible to  {\spicapol} at distances $\approx 1$ kpc. 
One of the key changes is then that the shock would cross the ambient magnetic field lines at all angles, 
and likely perturb them and enhance the turbulent component of the magnetic field, on spatial scales 
comparable to that of the filaments: for not-too-distant sources, {\spicapol}  would then act as a ``magnifying-glass'' to study 
the shock-ambient gas interactions at unprecedented small spatial scales, 
and put entirely new, perhaps even unexpected, constraints on DSA models (e.g., 
see the CR escape issues raised by \citealt{Malkov+2013}). 

At low energies, the situation is markedly different, 
because the transport properties of CRs become very sensitive to a variety of processes governed by the magnetic field properties 
which may hamper the penetration of CRs into MCs, and reduce the rate at which the gas is ionized by these particles \citep{Phan+2018}.
Like $\gamma$-ray production at high CR energies, the ionization by low-energy CR can also be measured by way of infrared and millimeter-wave observations, 
which detect lines of various molecules and radicals (like H$^+_3$, HCO$^+$, DCO$^+$, etc. -- cf. \citealp{Indriolo+2015}). 
This has been done for many MCs in the Galaxy, but more recently 
also for SNR-MC collision regions \citep{GabiciMontmerle2015}: 
here again an enhancement of MC ionization has been observed. 
The results tentatively suggest a proportionality between the SNR-accelerated high-energy and low-energy 
CR fluxes, constraining the acceleration mechanism, or a proportionality between the partially irradiated, ionized gas, 
and the fully irradiated, $\gamma$-ray emitting gas, or both.

More generally, both low- and high-energy CRs are affected in their propagation in the diffuse ISM by \textit{diffusion} effects, which 
are still poorly known --and directly influenced by magnetic fields. The spatially average diffusion coefficient of CRs in the Galaxy is constrained 
by a number of observations, and is often assumed to be isotropic on large Galactic scales ($\gg 100$ pc). On the other hand, in order to explain 
a number of $\gamma$-ray observations of SNR-MC associations (characterized by spatial scales of $\sim 10-100$ pc), a diffusion coefficient about 
two orders of magnitudes smaller (i.e., slower bulk propagation) than the average Galactic one needs to be assumed. However, such a discrepancy 
could be reconciled if CR diffusion is in fact anisotropic on such small scales \citep{NavaGabici2013}. An anisotropic diffusion is indeed expected 
for spatial scales smaller than the magnetic field coherence length \citep{Malkov+2013}. Knowing the topology of the magnetic field in such regions is thus 
of paramount importance in order to interpret  $\gamma$-ray observations correctly.



\begin{figure}[!ht]
\begin{center}
\includegraphics[width=\columnwidth]{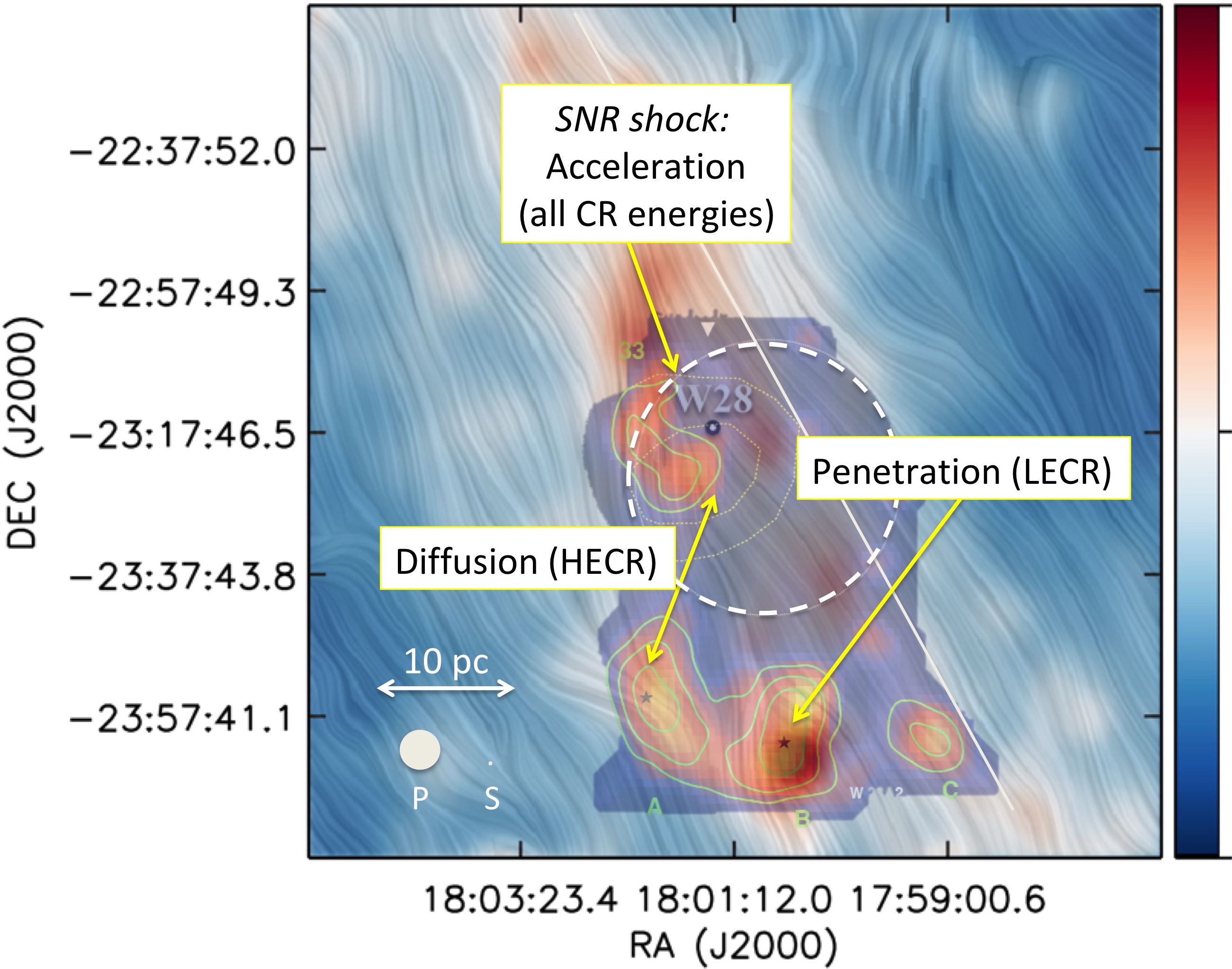}
\caption{The region surrounding the W28 SNR ($d \sim 2$ kpc; shock approximated by the dashed white circle), 
as seen in cold dust emission at $353\, $GHz 
by \emph{Planck} (color image with background $B$-field ``drapery'' 
from polarization data), 
in TeV-GeV $\gamma$-rays (white contours), and CO (green areas, well correlated with the $\gamma$-ray sources -- \citealp{Aharonian+2008},). 
The labels highlight the various CR processes  discussed in the text, at high energies (HECR) and low energies (LECR). 
The \emph{Planck} and {\spicapol}  beams are indicated by a light green circle 
(label ``P'') and a dot (label ``S''), respectively. }
 \label{w28}
\end{center}
\end{figure}

All of these issues can be illustrated by a recent study of the W28 SNR (cf. Fig.~\ref{w28}), a well-known example of an SNR-MC collision \citep{Vaupre+2014}. 
This SNR is located in the Galactic plane, at $d \approx 2$ kpc from the Sun, with an estimated (very uncertain) age $\approx 10^4$ yr. 
At this distance, the SNR apparent diameter ($\sim 30'$) gives a linear diameter $D \approx 20$ pc.
An observation by the High Energy Stereoscopic System (\emph{HESS}) \v{C}erenkov telescope, 
in the TeV $\gamma$-ray range \citep{Aharonian+2008}, covering 
a large field-of-view of $\sim 1.5^\circ \times 1.5^\circ$ (with a resolution of $\sim 0.1^\circ$), has revealed a complex of several 
resolved $\gamma$-ray sources. One of the sources, which is spatially correlated with a part of the SNR shock outline, was also 
detected as a bright GeV source by the \emph{Fermi} satellite, contrary to the other sources, which are either dimmer or undetected \citep{Abdo+2010}. 
This multiple source was soon correlated with a complex of molecular clouds mapped 
in CO by the \emph{NANTEN}\footnote{NANTEN means ``Southern Sky'' in Japanese.} telescope, 
showing that the SNR was in fact colliding with the molecular cloud associated with the GeV-TeV source 
(a physical contact being confirmed by the existence of several OH masers), the other sources being away, far upstream of the SNR shock. 

Calculations indicated a factor $\approx 100$ enhancement of the local high-energy CR flux, qualitatively consistent with a local CR acceleration 
by the SNR shock. Using the IRAM 30-m telescope 
to observe various molecular and radical tracers (H$^{13}$CO$^+$, C$^{18}$O, etc.) 
in the millimeter range, \citet{Vaupre+2014} were also able to calculate the MC ionization rate $\zeta$ at several locations. 
They found $\sim 2-3$ order of magnitude enhancements of $\zeta$ (or lower limits) over its average Galactic value ($\zeta_0 \approx 4-5 \times 10^{-17} \rm{erg~s}^{-1}$ --
e.g. \citealp{Indriolo+2015}), 
for the cloud correlated with GeV-TeV emission, i.e., indirect evidence for a similar enhancement of the low-energy CR flux, but no such enhancements 
for the clouds far upstream of the SNR shock. Within the ``GeV-TeV bright'' cloud, the measurements were separated by the IRAM telescope resolution, 
$\sim 12 ''$, i.e., comparable to (only 1.5 times better than) the  {\spicapol}  resolution at $200\, \mu$m (or a linear scale $\sim 0.15$ pc). 
The (projected) distance to the other clouds is $\sim 10$ pc, and this is seen as the diffusion length for high-energy CRs (TeV CRs reaching 
the distant clouds before the GeV CRs).

Thus, the W28 SNR and its environment provide us with a case study with all the ingredients needed to improve our understanding of the origin of CRs, 
and their relation with magnetic fields down to scales $\sim 0.15$ pc, i.e., {\it much smaller than observable before}: $(i)$ CR \textit{acceleration} by the SNR shock;
 $(ii)$ \textit{diffusion} of CRs between clouds as a function of energy; 
 $(iii)$ \textit{penetration} of low-energy CRs in ``average'' clouds, irradiated only by ambient, galactic CRs (cf. Fig.~\ref{w28}).


For more distant sources, the ``magnifying-glass'' effect of {\spicapol}  on small spatial scales would of course decrease, 
but an interesting link could then be established with the \emph{\v{C}erenkov Telescope Array} (\emph{CTA}, presently under construction; \citealt{Actis+2011}), 
which operates in the 20 GeV--300 TeV regime. Until \emph{CTA} actually observes, it is difficult to make accurate predictions on how far SNR-MC systems 
such as W28 could be detected in $\gamma$-rays. 
For CR studies, the main point is not simply the detection, but the location of the emission with respect to the shock. For the moment, only two cases are known in 
which the $\gamma$-rays are clearly upstream of the shock: W28 (detected by \emph{CGRO}, \emph{Fermi} and \emph{HESS}, so from GeV to TeV energies), 
and W44 (detected by \emph{Fermi} only, so at GeV energies only). About 20\% of \emph{HESS} sources, and most of the SNRs detected by \emph{Fermi} 
are SNR-MC systems (\citealt{HESS2018}; \citealt{Ackermann+2011}), but apart from W28 and W44 it is difficult to distinguish between upstream and 
downstream  $\gamma$-ray emissions (or both). Taking into account that \emph{CTA} is $\sim$ 10 times more sensitive than \emph{HESS}, and 
taking W28 as a template, representative of a Galactic disk SNR-MC population, we estimate that $\approx$ ten W28-like sources could be possibly 
detected and be sufficiently resolved by \emph{CTA}, hence be good targets for future {\spicapol} observations, 
which would in turn spur new theoretical work on CR acceleration on very small spatial scales not considered at present.

\begin{figure*}
\begin{center}
\begin{subfigure}{.4\textwidth}
        \centering
        \includegraphics[width=\linewidth]{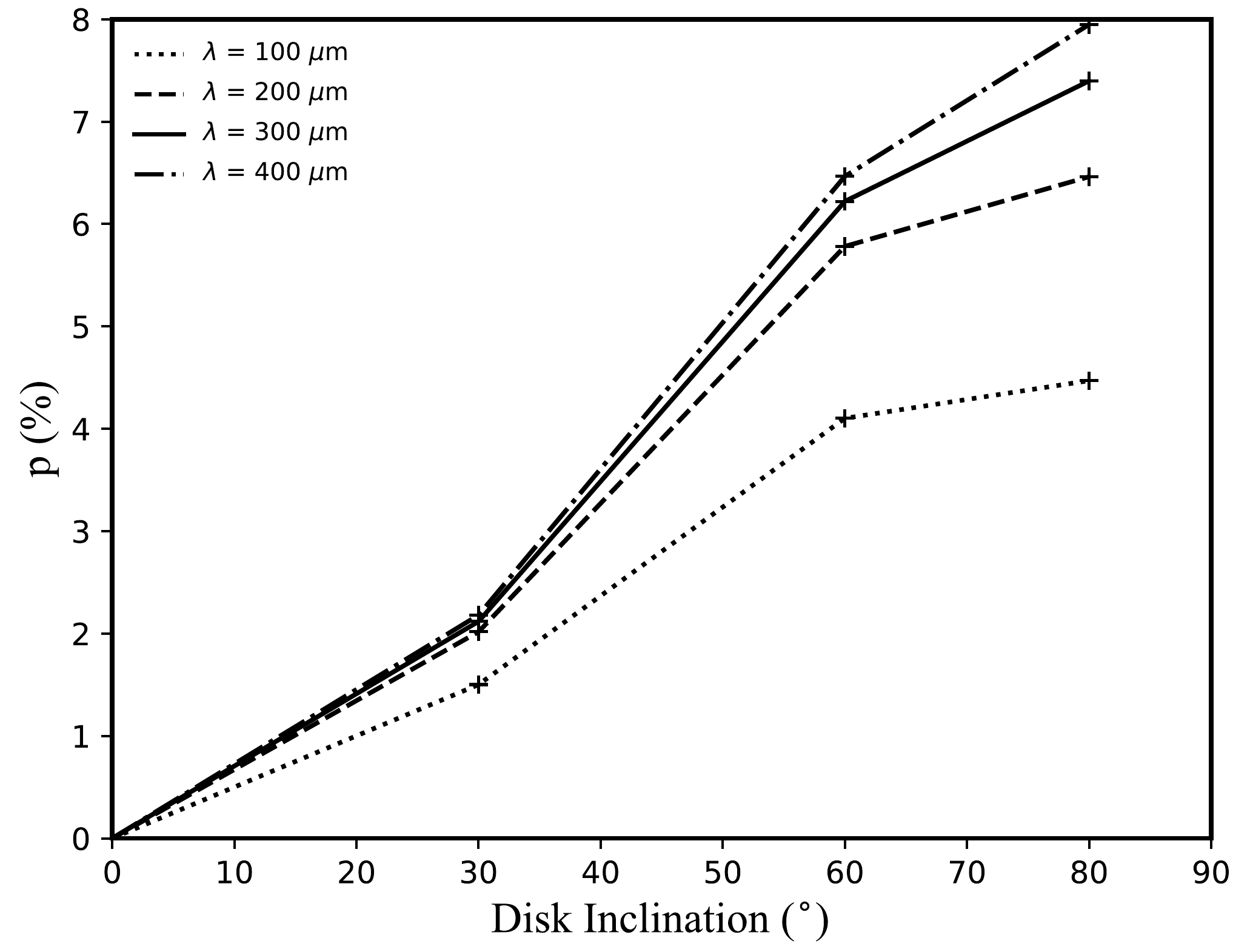}
        \caption{}\label{fig:fig_a}
    \end{subfigure} %
\begin{subfigure}{.42\textwidth}
        \centering
       \includegraphics[width=\linewidth]{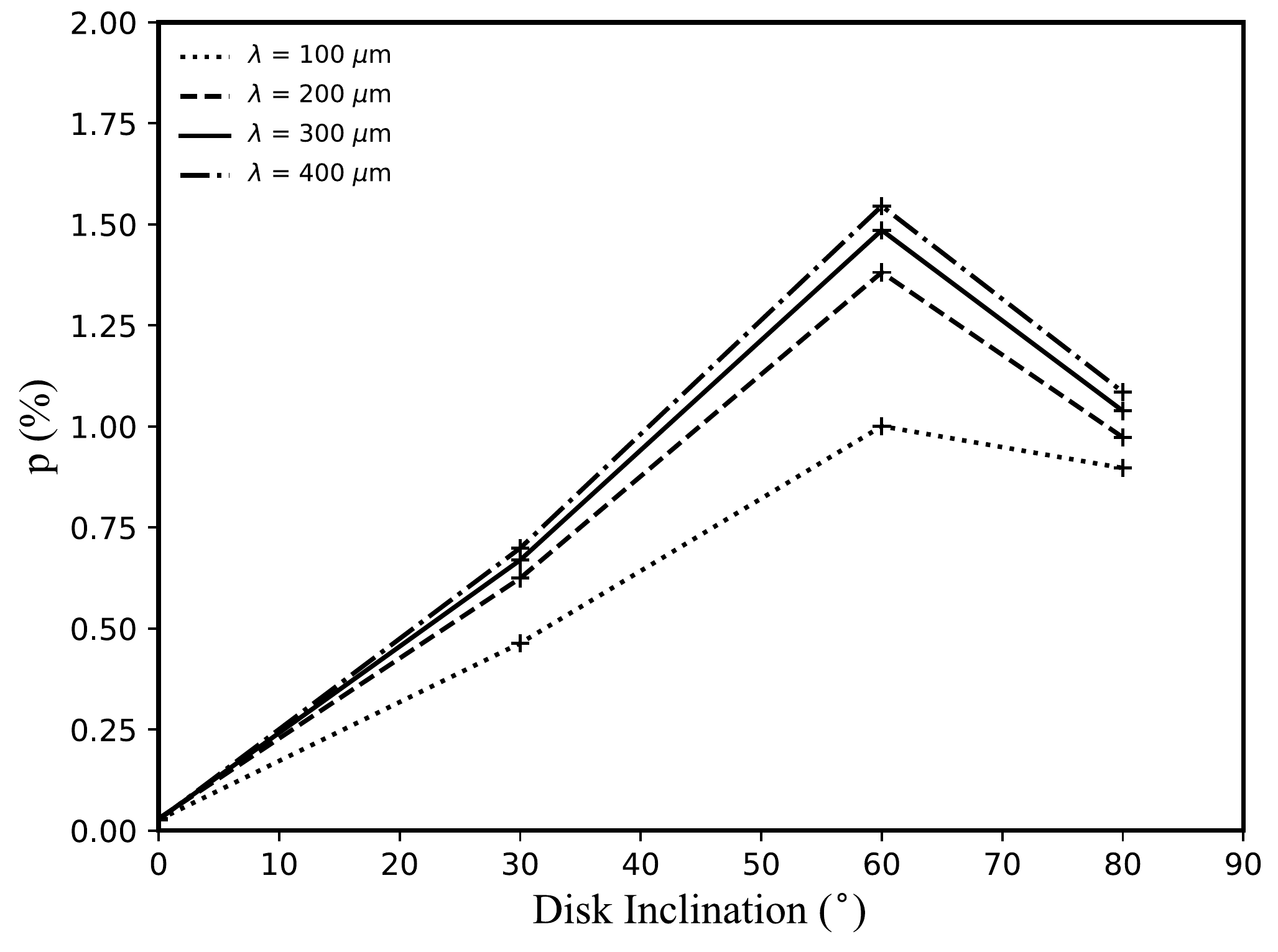}
        \caption{}\label{fig:fig_b}
    \end{subfigure} %
    \caption{Spatially integrated polarization levels expected at 100--400$\, \mu$m 
    from magnetized protoplanetary disks as a function of disk inclination
      (0$^\circ$ refers to a pole-on disk configuration). The left panel (a)
      considers a poloidal magnetic field topology, and the right panel (b) a
      toroidal one. See \cite{Li+2016} for more details
      about the assumptions of the disk model and the adopted parameter 
      values.}
\label{diskspol}
\end{center}
\end{figure*}

\section{Polarized dust emission from protoplanetary disks}
\label{sec:disks} 

As already mentioned in Sect.~\ref{sec:filaments}, magnetic fields 
may regulate the gravitational collapse 
and fragmentation of prestellar dense cores, thereby influencing 
the overall star formation efficiency  \citep[][]{MouschoviasCiolek1999,Dullemond+2007,Crutcher2012}.
%
It is thus natural to expect that, during core collapse, magnetic fields can be
dragged inward, leaving a remnant field in the protoplanetary disk
formed subsequently. 

If protoplanetary disks are indeed (weakly) magnetized, then the MHD
turbulence arising from magneto-rotational instability (MRI) is
thought to be the primary source of disk viscosity, a crucial driving
force for disk evolution (e.g. disk accretion) and planet formation
\citep{Balbus+1998, Turner+2014}.  Despite this general consensus, our
knowledge about magnetic fields in disks is actually very limited and
incomplete at this stage, largely due to the lack of observational
constraints on magnetic field properties (geometry and strength) in
protoplanetary disks.  
Polarimetric observations of thermal dust
emission at centimeter or millimeter wavelengths with single-dish
telescopes, such as the Caltech Submillimeter Observatory (CSO) and JCMT,  
or interferometric arrays, such as the Karl G. Jansky Very Large Array (JVLA), the SMA, 
the Berkeley-Illinois-Maryland Array (BIMA), and the Combined Array for Research in Millimeter-wave Astronomy 
(CARMA), 
have been used extensively to map magnetic field structure in YSOs at scales from $\sim 50\,
$AU to thousands of AU (see \citealp{Crutcher2012} for a review).
However, due to the limited sensitivity and angular resolution offered
by current facilities and the nature of centimeter/millimeter observations, most of
these studies have been focused on magnetic fields in molecular clumps/cores (cf. \S~\ref{sec:protostars}), 
or Class~0/Class~I objects \citep[e.g.][]{Qiu+2013,Rao+2014,Liu+2016},
rather than classical protoplanetary disks around Class~II objects.
Using CARMA, \citet{Stephens+2014} spatially resolved the HL Tau disk
in polarized light at 1.3 mm, and their best-fit model suggested that
the observation was consistent with a highly tilted (by
$\sim$50$^\circ$ from the disk plane), toroidal magnetic field threading the
disk.  This conclusion was challenged by follow-up studies, which
showed that the 1.3 mm polarization of HL Tau could also arise solely
from dust scattering as opposed to dichroic emission from elongated grains aligned with the magnetic field \citep{Kataoka+2015,Yang+2016}.  
However, more
recent ALMA observations at 0.87, 1.3, and 3.1~mm indicated that dust
scattering alone may not be able to explain all of the
multi-wavelength polarization data
\citep[][]{Kataoka+2017,Stephens+2017}.

Recently, \citet{Li+2016} have been able to highlight the signature of
a magnetic field in the AB Aur protoplanetary disk at mid-IR 
wavelengths.  Using observations of the AB Aur protoplanetary disk at
10.5 $\mu$m with the GTC/CanariCam imager and polarimeter, they
detected a polarization pattern in the inner regions of the disk
compatible with dichroic emission polarization produced by elongated
grains aligned by a tilted poloidal magnetic field. The observed polarization
level (2-3 \%) was somewhat lower than that predicted by theory
\citep{Cho_Lazarian2007}, although this is something naturally
expected since the modeling assumes alignment efficiencies and
intrinsic particles polarizability which are probably overestimated.
At a wavelength of 10.5 $\mu$m where protoplanetary disks are
optically thick, the observations are probing the disk properties down
to depths corresponding to the $\tau$=1 optical depth surface.  This
depth is relatively small (less than $\sim$10\%) compared to the disk
scale height. Longer wavelengths, up to about 200 $\mu$m where the
disk can still be moderately optically thick up to large distances
from the star, are emitted by cooler material located deeper within
the disk.  Thus, by measuring the level of polarization at wavelengths
in the range 100--300 $\mu$m, we expect to be able to compare the
levels of polarization as a function of optical thickness, thereby
getting an indirect signature of the magnetic field at different depths
within protoplanetary disks. This will provide constraints on the importance
of MRI-induced turbulence.  Together with 10 $\mu$m and ALMA similar
types of observations, this will allow us to build a tomographic view of the
magnetic field along the vertical profile of the disks.  Such measurements
would also have large impacts on our understanding of planet formation
processes.

For the purpose of this paper, we used the same modeling approach as
described in \citet{Li+2016}, with also the same disk parameters based
on the example of AB Aur, in order to predict expected polarization
levels levels at 100 $\mu$m, 200 $\mu$m, and 350 $\mu$m if observed by
SPICA. Given the angular resolution of SPICA at these wavelengths, we
do not expect, apart from exceptional cases, to angularly resolve the
polarized emission, therefore we computed an integrated value,
considering the object as unresolved.

In Fig.~\ref{diskspol}, we show the predictions of the model,
integrated over the full spatial extent of the disk, for different disk
inclinations with respect to the line of sight and for observing
wavelengths within the {\spicapol} range.  Two types of magnetic field
configuration are considered, poloidal and toroidal, which are the
simplest ones, and those also widely discussed in the literature.  We
can see from the simulations that the poloidal magnetic field configuration
produces stronger integrated polarization signatures compared to the
toroidal configuration. 

As mentioned earlier, the origin of the dust continuum polarization 
on the disk scale is still uncertain, with potential contributions from scattering 
by large grains in addition to that from emission by magnetically (or radiatively) aligned grains.
The {\spicapol} instrument will generally not (or barely be able to) 
resolve protoplanetary disks, which poses the problem of
disentangling these various mechanisms.
Fortunately, polarization by aligned dust grains and polarization by dust self-scattering
have different dependences on wavelength and optical depth \citep{HYang+2017}.
SPICA will greatly extend the wavelength coverage of ALMA (from 870$\, \mu$m to 100$\, \mu$m), 
which will make it easier to disentangle the contributions from the competing mechanisms. 
Such an effort is a pre-requisite for using dust  
continuum polarization to probe both disk magnetic fields and grain growth, the crucial first step 
toward the formation of planetesimals and ultimately planets.  
Moreover, given the plan to image the whole extent 
of  nearby star-forming regions with {\spicapol} (cf. end of \S~\ref{subsec:filaments}), 
several tens of protoplanetary disks will be detected in Stokes I, Q, U. 
It will therefore be possible to derive statistical trends about the presence of
magnetic fields, and any bias can be controlled providing that the inclination
and position angles of the disks are known.

\section{Variability studies of protostars in the far-infrared}
\label{sec:proto-var} 

At the core ($\leq 0.1\,$pc) scale, 
the formation of a solar-type star is well understood as a continual mass assembly process whereby material in the protostellar envelope is accreted 
onto a circumstellar disk and then transported inward and onto the protostar via accretion columns \citep{Hartmann+2016}.
The observational evidence 
for the mass assembly rate of low-mass stars is provided by the lifetimes of the various stages and the bolometric luminosities, 
which are dominated by accretion energy at early times \citep[e.g.][and references therein]{Dunham+2014}.
These two quantifiable measures are in significant disagreement and circumstantial 
evidence exists for the episodic nature of mass assembly -- bullets in outflows \citep[e.g.][]{Plunkett+2015}, 
FU and EX Ori phenomenon \citep{Hartmann+1996,Herbig2008}, 
numerical calculations 
of disk transport \citep[e.g.][]{Armitage2015}.
Moreover, high spatial resolution images of young disks reveal macroscopic structure including rings \citep{ALMA+2015}, 
spirals \citep{Perez+2016}, 
and fragmentation \citep[][]{Tobin+2016}, 
suggesting that the transport of material through the disk is not a smooth and steady process.

\begin{figure}
\begin{center}
\includegraphics[width=\columnwidth]{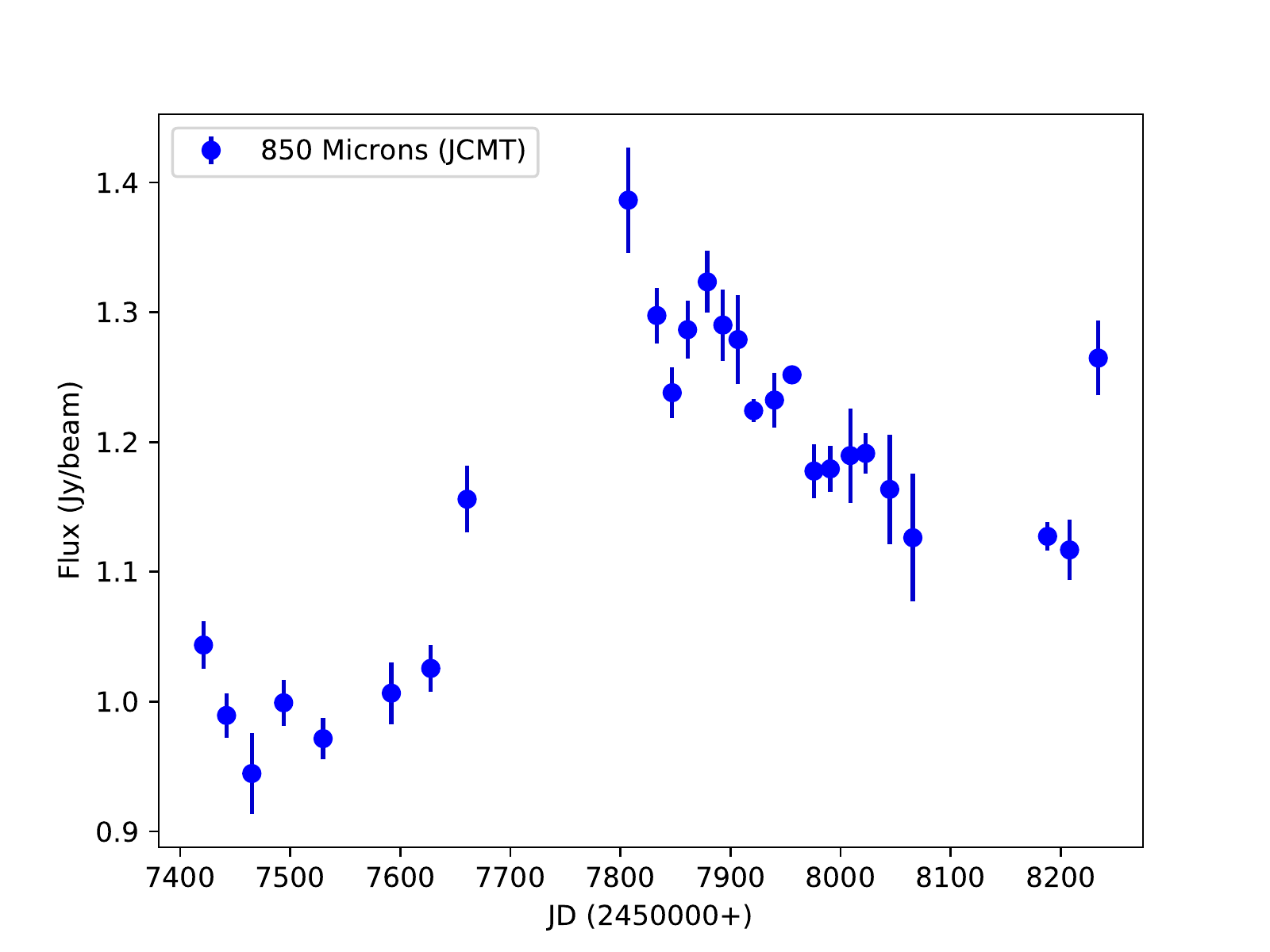}
\caption{Time variation observed over 27 epochs at $850\, \mu$m for the Class~I protostar EC~53  in the Serpens Main star-forming region 
as part of the ``JCMT Transient Survey''  \citep[][]{Yoo+2017}. 
The typical uncertainty in a single measurement is $\sim 20\,$mJy (S/N $\sim$ 50) and the peak to peak brightness variation is almost 500\,mJy. 
Figure adapted from \citep{Johnstone+2018b}. 
}
\label{proto_var_fig} 
\end{center}
\end{figure}

Recently, significant variability has been detected in the submillimeter continuum emission of several nearby protostars 
\citep[see, e.g.,][]{Mairs+2017b,Yoo+2017, Johnstone+2018},  
through an ongoing multi-year monitoring survey of eight nearby star-forming regions with JCMT at $850\, \mu$m 
\citep[][]{Herczeg+2017}. While uncertain due to small number statistics, it appears that roughly 10\% of deeply embedded protostars vary 
over year timescales by around 10\% at submillimeter wavelengths \citep{Johnstone+2018}. The dominant mode of variability uncovered by the survey is quasi-secular, with the protostellar brightness increasing 
or decreasing for extended - multi-year - periods \citep[][]{Mairs+2017b, Johnstone+2018} 
and suggesting a link to non-steady accretion processes taking place within the circumstellar disk where the orbital timescales match those of the observed variability. These long timescales also allow for significant amplification of the overall change in submillimeter brightness after many years. One source, EC 53 in Serpens Main, has an eighteen-month quasi-periodic light curve \citep[][and Fig.~\ref{proto_var_fig}]{Yoo+2017}, previously identified through near-IR observations \citep[][]{Hodapp+2012}, 
which is likely due to periodic forcing by a long-lived structure within the inner several AU region of the disk. 

Stronger variability is expected at far-IR wavelengths where Class~0 and Class~I YSOs have the peak of their SEDs 
and the envelope emission directly scales with the internal (accretion) luminosity of the underlying protostar \citep[][]{Dunham+2008,Johnstone+2013}. At submillimeter wavelengths the emission typically scales with the envelope temperature and thus the far-IR signal is expected to be around four times larger such that a 10\% variability in the submillimeter relates to a 40\% variability in the far-IR.  Contemporaneous monitoring of EC 53 in the near-IR and at 850/450$\, \mu$m has confirmed that the longer submillimeter wavelength shows just such a diminished response to the underlying change in internal luminosity 
as proxied by near-IR observations (Yoo et al., in prep.). 
Thus, carefully calibrated monitoring of protostars with SPICA should uncover a significantly larger fraction of variables than the 10\% obtained by the JCMT survey.

Thanks to its high continuum sensitivity and mapping speed at wavelengths around the peak of protostellar SEDs, {\spicapol}, used as a total-power imager, will be ideal for monitoring hundreds of forming stars over multi-year epochs, allowing an unprecedented statistical determination of the variation in accretion on these timescales. Typical nearby deeply embedded protostars have far-IR brightnesses greater than $\sim10\, $mJy and thus will be observed to a S/N $\simgt $ 100 by {\spicapol} in a fast scanning mode. 
As demonstrated for ground-based submillimeter observations \citep[][]{Mairs+2017a}, 
instrument stability will need to be carefully monitored in order to achieve precise relative flux calibration of a few percent between epochs. Additional critical requirements for {\spicapol} will be a large, $10^5$ or higher, dynamic range and instrument robustness against extremely bright sources within the field. 

Three interconnected monitoring surveys are envisioned. First, the bulk of the $\sim1000$ nearby, Gould Belt, deeply embedded protostars will be observed every six months while SPICA is in orbit, requiring coverage of $\sim 20$~deg$^2$ (a modest twenty hours of observing per epoch). This will allow for a detailed statistical characterization of variability across multiple years. Additionally, a few carefully chosen nearby star formation fields, each roughly a square degree, will be observed weekly during their expected few month continuous observing window 
\citep[for information on observing strategies for SPICA, see][]{Roelfsema+2018}.
For both of these nearby samples, an even larger number of Class~II YSOs 
will be observable within each field. While these sources will be fainter at far-IR wavelengths as the emission probes the disks directly, the enhanced numbers will allow for a determination of the importance of variability throughout the evolution of a protostar. 
Finally, a sample of more distant high-mass star-forming regions should be observed yearly to search for 
rare, but extremely bright, bursts such as FU Ori events. While SPICA will not have the spatial resolution to separate individual protostars 
within these regions, evidence of a significant brightening 
can be easily followed-up with ground-based telescopes such as ALMA (see \citealp{Hunter+2017}  
for an example of a brightening in a high-mass star-forming region).

\section{Concluding remarks}
\label{sec:conclusions} 

Magnetic fields are a largely unexplored ``dimension'' of the cold Universe. 
While they are believed to be a key dark ingredient 
of the star formation process through most of Cosmic time, 
they remain very poorly constrained observationally, especially in the cold ISM of galaxies 
\citep[e.g.][]{Crutcher2012}. 

Benefiting from a cryogenic telescope, SPICA-POL or  {\spicapol} will be two to three orders of magnitude
more sensitive than existing or planned far-IR/submillimeter polarimeters (cf. Fig.~\ref{spica_sensitivity}) 
and will therefore lead to a quantum step forward in the area of far-IR dust polarimetric imaging, 
one of the prime observational techniques to probe the topology of magnetic fields in cold, mostly neutral environments.
In particular, systematic polarimetric imaging surveys of Galactic molecular clouds and nearby galaxies
with {\spicapol} have the potential to revolutionize our understanding of the origin and role of magnetic fields
in the cold ISM of Milky-Way-like galaxies on scales from $\sim 0.01\, $pc to a few kpc.
The three main science drivers for {\spicapol} are 1) probing how magnetic fields control the formation, 
evolution, and fragmentation of dusty molecular filaments (Sect.~\ref{sec:filaments}), thereby setting the initial conditions for individual 
protostellar collapse (Sects.~\ref{sec:protostars} and ~\ref{sec:massive-sf}); 
2) characterizing the structure of both turbulent and regular magnetic fields in the cold ISM of nearby galaxies, including the Milky Way, 
and constraining galactic dynamo models (Sect.~\ref{sec:turbulence} and Sect.~\ref{sec:galaxies}); 
and 3) testing models of dust grain alignment and informing dust physics (Sect.~\ref{sec:dust-physics}). 
Other science areas can be tackled with, or uniquely informed by, {\spicapol} observations, 
including the problem of the interaction of cosmic rays with molecular clouds (Sect.~\ref{sec:cosmic-rays}),
the study of the magnetization of protoplanetary disks (Sect.~\ref{sec:disks}),  
and the characterization of variable accretion in embedded protostars (Sect.~\ref{sec:proto-var}). 
Last, but not least, the leap forward provided by {\spicapol} in far-IR imaging polarimetry 
will undoubtedly lead to unexpected discoveries, such as the potential detection 
of polarization from the CIB (Sect.~\ref{subsec:cib}).

\begin{acknowledgements}
This paper is dedicated to the memory of Bruce Swinyard, who initiated the SPICA project in Europe, 
but unfortunately died on 2015 May 22 at the age of 52. He was ISO-LWS calibration scientist, Herschel-SPIRE instrument scientist, 
first European PI of SPICA and first design lead of SAFARI. 
This work has received support from the European Research Council 
under the European Union's Seventh Framework Programme 
(ERC Advanced Grant Agreement no. 291294 --  `ORISTARS', and Starting Grant Agreement no. 679937 -- `MagneticYSOs'). 
D.$\, $J. is supported by the National Research Council of Canada and by an NSERC Discovery Grant. 
H.$\, $B. acknowledges support from the ERC Consolidator Grant CSF-648505 
and financial help from the DFG via the SFB881 ``The Milky Way System'' (subproject B1).
The National Radio Astronomy Observatory is a facility of the National Science Foundation operated under 
cooperative agreement by Associated Universities, Inc.

\end{acknowledgements}

\begin{appendix}

\section*{Affiliations}

\affil{$^1$Laboratoire d'Astrophysique (AIM), CEA, CNRS, Universit\'e Paris-Saclay, Universit\'e Paris Diderot, Sorbonne Paris Cit\'e, 91191 Gif-sur-Yvette, France}%
\affil{$^2$Institut de Recherche en Astrophysique et Plan\'etologie (IRAP), CNRS, 9 Av. Colonel Roche, BP 44346, 31028 Toulouse, France}%
\affil{$^3$Institut d'Astrophysique Spatiale (IAS), CNRS (UMR 8617) Universit\'e Paris-Sud 11, B\^atiment 121, 91400 Orsay, France}%
\affil{$^4$Laboratoire Univers et Particules de Montpellier, Universit\'{e} de Montpellier, CNRS/IN2P3, CC 72, Place Eug\`{e}ne Bataillon, 34095 Montpellier Cedex 5, France}%
\affil{$^{5}$LERMA/LRA - ENS Paris - UMR 8112 du CNRS, 24 rue Lhomond, 75231, Paris Cedex 05, France}%
\affil{$^6$Nordita, KTH Royal Institute of Technology and Stockholm University, Roslagstullsbacken 23, 10691 Stockholm, Sweden}%
\affil{$^{7}$Department of Physics and ITCP, 
University of Crete, Voutes, GR-71003 Heraklion, Greece}%
\affil{$^8$Department of Physics, Graduate School of Science, Nagoya University, Furo-cho, Chikusa-ku, Nagoya 464-8602, Japan}
\affil{$^9$Harvard-Smithsonian Center for Astrophysics, Cambridge, MA02138, USA}
\affil{$^{10}$Laboratoire d'Astrophysique de Bordeaux, Univ. de Bordeaux -- CNRS/INSU, BP 89, 33271 Floirac Cedex, France}%
\affil{$^{11}$Institut de Ci\`encies de l'Espai (ICE, CSIC), Can Magrans s/n, 08193 Cerdanyola del Vall\`es, Catalonia, Spain; 
Institut d'Estudis Espacials de Catalunya (IEEC), E-08034, Barcelona, Catalonia, Spain}%
\affil{$^{12}$Univ. Grenoble Alpes, CNRS, Institut de Plan\'etologie et d'Astrophysique de Grenoble, 38000 Grenoble, France}%
\affil{$^{13}$Institut d'Astrophysique de Paris (IAP), 98bis Bd Arago, 75014 Paris, France}%
\affil{$^{14}$NRC Herzberg Astronomy and Astrophysics, 5071 West Saanich Road, Victoria, BC, V9E 2E7, Canada; 
Department of Physics and Astronomy, University of Victoria, Victoria, BC, V8P 5C2, Canada}%
\affil{$^{15}$APC, AstroParticule et Cosmologie, Universit\'e Paris Diderot, CNRS, CEA, Observatoire de Paris, Sorbonne Paris, 75205 Paris, France}%
\affil{$^{16}$School of Sciences, European University Cyprus, 1516 Nicosia, Cyprus}%
\affil{$^{17}$Department of Physics and Astronomy, The University of Western Ontario, London, ON N6A 3K7, Canada}%
\affil{$^{18}$Aix-Marseille Univ., CNRS, LAM, Laboratoire d'Astrophysique de Marseille, 13013, Marseille, France}%
\affil{$^{19}$Max Planck Institute for Astronomy, K\"onigstuhl 17, 69117 Heidelberg, Germany}%
\affil{$^{20}$School of Mathematics, Statistics and Physics, Newcastle University, Newcastle upon Tyne, NE1 7RU, UK}%
\affil{$^{21}$National Radio Astronomy Observatory, 520 Edgemont Road, Charlottesville, VA 22903-2475, USA}%
\affil{$^{22}$Institute of Space and Astronautical Science, Japan Aerospace Exploration Agency, 
Kanagawa 252-5210, Japan}%
\affil{$^{23}$National Optical Astronomy Observatory, 950 N. Cherry Ave., Tucson, AZ 85726, USA}%
\affil{$^{24}$Department of Astronomy, University of Virginia, Charlottesville, VA 22901, USA}%
\affil{$^{25}$Max Planck Institute for Radioastronomy, Auf dem H\"ugel 69, 53111 Bonn, Germany}%
\affil{$^{26}$Department of Astronomy, Graduate School of Science, University of Tokyo, Tokyo, Japan}%
\affil{$^{27}$INAF-Istituto di Radioastronomia, via P. Gobetti, 101, 40129, Bologna, Italy}%
\affil{$^{28}$School of Physics \& Astronomy, Cardiff University, Queen's Buildings, The Parade, Cardiff, CF24 3AA, UK}%
\affil{$^{29}$Max-Planck-Institut f\"ur extraterrestrische Physik, Garching, Germany}%
\affil{$^{30}$Istituto di Astrofisica e Planetologia Spaziali, INAF, Via Fosso del Cavaliere 100, I-00133 Roma, Italy}%
\affil{$^{31}$School of Astronomy, Institute for Research in Fundamental Sciences, P.O. Box 19395-5531, Tehran, Iran}%
\affil{$^{32}$Research School of Astronomy and Astrophysics, Australian National University, Canberra, ACT 2611, Australia}
\affil{$^{33}$SRON Netherlands Institute for Space Research, Groningen, The Netherlands; Kapteyn Astronomical Institute, University of Groningen, The Netherlands}%
\affil{$^{34}$Jeremiah Horrocks Institute, University of Central Lancashire, Preston PR1 2HE, UK}%
\affil{$^{35}$Department of Astronomy, University of Florida, Gainesville, FL 32611, USA}%

\end{appendix}

\bibliographystyle{pasa-mnras}
\bibliography{spica-pol,Refs_FB}

\end{document}